\documentclass[letterpaper,12pt]{article}
\usepackage{jheppub2} 
\usepackage[T1]{fontenc}

\usepackage{natbib}
\usepackage{graphicx}
\usepackage{hyperref}
\usepackage{amsmath}
\usepackage{amsfonts}
\usepackage{amssymb}
\usepackage{enumitem}
\usepackage{tocvsec2}

\newcommand{\ZF}{(\mathbb{Z}_2)_{\rm F}}
\newcommand{\ZC}{(\mathbb{Z}_2)_{\rm C}}
\newcommand{\ZZ}{{\mathbb Z_2}}

\def\tr{\text{tr}\,}
\def\Tr{\text{Tr}\,}
\def\Re{\mathrm{Re}\,}
\def\half{\tfrac{1}{2}}
\def\LambdaQCD{\Lambda_{\rm QCD}}
\def\Seff{S_{\rm eff}}
\def\UG{U(1)_{\rm G}}
\def\UX{U(1)_{X}}
\def\UY{U(1)_{Y}}
\def\SI{S_{\rm I}}
\def\Vm{V_{\rm m}}
\def\muBs{\mu_B^{\rm sat}}

\hypersetup{
    colorlinks=true,   % false: boxed links; true: colored links
    linkcolor=red,     % color of internal links (box color = linkbordercolor)
    citecolor=blue,    % color of links to bibliography
    filecolor=magenta, % color of file links  
    urlcolor=blue      % color of external links 
}

\makeatletter
\def\l@subsubsection#1#2{}
\makeatother    

\advance\footnotesep 1.5pt

\begin{document}

\title{Higgs-confinement phase transitions \\ with fundamental representation matter}

\author[1]{Aleksey Cherman,}
\emailAdd{acherman@umn.edu}

\author[1]{Theodore Jacobson,}
\emailAdd{jaco2585@umn.edu}
\affiliation[1]{School of Physics and Astronomy, University of Minnesota, Minneapolis MN 55455, USA}

\author[2]{Srimoyee Sen,}
\emailAdd{srimoyee08@gmail.com}
\affiliation[2]{Department of Physics and Astronomy,  Iowa State University, Ames IA 50011, USA}

\author[3]{Laurence G. Yaffe}
\emailAdd{yaffe@phys.washington.edu}
\affiliation[3]{Department of Physics,  University of Washington, Seattle WA 98195-1560, USA}

\abstract{
    We discuss the conditions under which Higgs and confining regimes
    in gauge theories with fundamental representation matter fields can be sharply
    distinguished. It is widely believed that these regimes are
    smoothly connected unless they are distinguished by the
    realization of global symmetries.  However, we show that when a
    $U(1)$ global symmetry is spontaneously broken in \emph{both} the
    confining and Higgs regimes, the two phases can be separated by a phase boundary.  
    The phase transition between the two regimes may be
    detected by a novel topological vortex order parameter.
    We first illustrate these ideas by explicit calculations in gauge theories
    in three spacetime dimensions.
    Then we show how our analysis generalizes to four dimensions, 
    where it implies that nuclear matter
    and quark matter are sharply distinct phases of QCD with an
    approximate $SU(3)$ flavor symmetry.     
} 

\maketitle

\maxtocdepth{subsection} 
 
\section{Introduction}

In gauge theories with fundamental representation matter fields, one
can often dial parameters in a manner which smoothly interpolates
between a Higgs regime and a confining regime without undergoing any
change in the realization of global
symmetries~\cite{tHooft:1979yoe,Osterwalder:1977pc,Fradkin:1978dv,Banks:1979fi}.
In the Higgs regime gauge fields become massive via the usual Higgs
phenomenon, while in the confining regime gauge fields also become
gapped (or acquire a finite correlation length) due to the
non-perturbative physics of confinement, with an approximately linear
potential appearing between heavy fundamental test charges over a
finite range of length scales which is limited by the lightest meson
mass. In this paper we examine situations in which the Higgs and
confining regimes of such theories can be sharply distinguished.

This is, of course, an old and much-studied issue.  In specific
examples, when both regimes have identical realizations of global
symmetries, it has been shown that confining and Higgs regimes can be
smoothly connected with no intervening phase transitions~\cite{Fradkin:1978dv,Banks:1979fi}. These examples, which we will refer to as
as the ``Fradkin-Shenker-Banks-Rabinovici theorem,'' have inspired a widely held
expectation that there can be no useful gauge-invariant order
parameter distinguishing Higgs and confining phases in any gauge
theory with fundamental representation matter fields.%
\footnote
    {%
    By a useful order parameter we mean an expectation value of
    a physical observable whose non-analytic change also
    indicates non-analytic behavior in thermodynamic observables and
    correlation functions of local operators.
    For a rather different take on these issues,
    see Refs.~\cite{Fredenhagen:1985ft,Greensite:2017ajx,Greensite:2018mhh,Greensite:2020nhg}.
    }
But there are physically
interesting situations in which the Fradkin-Shenker-Banks-Rabinovici theorem does not
apply. We are interested in systems where no local order parameter can
distinguish Higgs and confining regimes and yet the conventional
wisdom just described is incorrect. We will analyze model theories,
motivated by the physics of dense QCD, where Higgs and confining
regimes cannot be distinguished by the realization of global
symmetries and yet these are sharply distinct phases necessarily
separated by a quantum phase transition in the parameter space of the
theory.

We will consider a class of gauge theories with two key features. The
first is that they have fundamental representation scalar fields which
are charged under a $U(1)$ global symmetry. Second, this $U(1)$ global
symmetry is spontaneously broken in \emph{both} the Higgs and
confining regimes of interest. In this class of gauge theories, we
argue that one can define a natural non-local order parameter which
does distinguish the Higgs and confinement regimes. This order
parameter is essentially the phase of the expectation value of the
holonomy (Wilson loop) of the gauge field around $U(1)$ global
vortices; its precise definition is discussed below. We will find that
this vortex holonomy phase acts like a topological observable; it is
constant within each regime but has differing quantized values in the
two regimes.%
\footnote
    {%
    This statement assumes a certain global flavor symmetry.
    In the absence of such a symmetry,
    the phase of the vortex holonomy is constant in the $U(1)$-broken
    confining regime and changes non-analytically at the onset of the
    Higgs regime.
    }
We present a general
argument --- verifying  it by explicit calculation where possible ---
that implies that non-analyticity in our vortex holonomy observable
signals a genuine phase transition separating the $U(1)$-broken Higgs
and $U(1)$-broken confining regimes.

The Higgs-confinement transition we discuss in this paper does not map
cleanly onto the classification of topological orders which is much
discussed in modern condensed matter
physics~\cite{Wegner:1984qt,Wen:1989zg,Wen:1989iv,Wen:2012hm}.  The
basic reason is that the topological order classification is designed
for gapped phases of matter, while here we focus on gapless phases.
Some generalizations of topological order to gapless systems have been
considered in the condensed matter literature, see
e.g. Refs.~\cite{SACHDEV200258,Kitaev:2006lla,Sachdev:2018ddg}, but these examples differ
in essential ways from the class of models we consider here.  Our
arguments also do not cleanly map onto the related idea of classifying
phases based on realizations of higher-form global
symmetries~\cite{Gukov:2013zka,Kapustin:2013uxa,Kapustin:2014gua,
Gaiotto:2014kfa,Metlitski:2017fmd,Lake:2018dqm,Wen:2018zux},
because the models we consider do not have any obvious higher-form
symmetries.  But there is no reason to think  that existing
classification ideas can detect all possible phase transitions.  We
argue that our vortex order parameter provides a new and useful way to
detect certain phase transitions which are not amenable to standard
methods.

Let us pause to explain in a bit more detail why the Fradkin-Shenker-Banks-Rabinovici
theorem does not apply to theories of the sort we consider.
The Fradkin-Shenker-Banks-Rabinovici theorem presupposes that Higgs fields are uncharged
under any global symmetry.  This assumption may seem innocuous.
After all, if Higgs fields are charged under a global symmetry,
it is tempting to think that this global symmetry will be spontaneously
broken when the Higgs fields develop an expectation value, 
implying a phase transition associated with a change in symmetry
realization and detectable with a local order parameter.
In other words, a typical case lying within the Landau paradigm of
phase transitions.

But such a connection between Higgs-confinement transitions and
a change in global symmetry realization is model dependent.
In the theories we consider in this paper, as well as in dense QCD,
these two phenomena are unrelated.
Our scalar fields will carry a global $U(1)$ charge,
but crucially, the realization of all global symmetries will be the
same in the confining and Higgs regimes of interest.
Consequently,
the Fradkin-Shenker-Banks-Rabinovici theorem does not apply to these models and yet
the confining and Higgs regimes are not distinguishable within the
Landau classification of phases.
Nevertheless,  we will see that they are distinct.
%% the behavior of Wilson loops that encircle $U(1)$ global vortices shows
%% that the Higgs and confinement regimes are separated by a phase
%% transition. 

The basic ideas motivating this paper were introduced by three of us in
an earlier study of cold dense QCD matter~\cite{Cherman:2018jir}.
We return to this motivation at the end of this paper 
in Sec.~\ref{sec:QCD},
where we generalize our analysis to cover
non-Abelian gauge theories in four spacetime dimensions
and explain why it provides compelling evidence against the
Sch\"afer-Wilczek conjecture of quark-hadron continuity in dense QCD~\cite{Schafer:1998ef}.
The bulk of our discussion is focused on a simpler set of model theories
which will prove useful to refine our understanding of Higgs-confinement
phase transitions. 

We begin, in Sec.~\ref{sec:our_model}, by introducing a simple Abelian
gauge theory in three spacetime dimensions in which Higgs and
confinement physics can be studied very explicitly.  In
Sec.~\ref{sec:vortices_and_holonomies} we introduce our vortex order
parameter and use it to infer the existence of a Higgs-confinement
phase transition. Sec.~\ref{sec:QCD} discusses the application of our
ideas to four-dimensional gauge theories such as QCD, while
Sec.~\ref{sec:conclusion} contains some concluding remarks. 
Finally, in Appendices \ref{sec:EoMAppendix}--\ref{sec:gaugingU1}
we collect some technical results on vortices,
discuss embedding our Abelian model within a non-Abelian theory, and
consider the consequences of gauging of our $U(1)$ global symmetry to
produce a $U(1) \times U(1)$ gauge theory. 
%  in which conventional
% topological order ideas may be applied.

%%%%%%
\section{The model}
\label{sec:our_model}
%%%%%%

We consider compact $U(1)$ gauge theory in three Euclidean spacetime
dimensions.
Let $A_{\mu}$ denote the (real) gauge field. Our analysis assumes $SO(3)$
Euclidean rotation symmetry, together with a parity (or time-reversal)
symmetry. Parity symmetry precludes a Chern-Simons term, so the gauge
part of the action is just a photon kinetic term,
\begin{align}
    S_{\rm \gamma} = \int d^{3}x \> \frac{1}{4e^2} \, F_{\mu\nu}F^{\mu\nu} \,.
\label{eq:S_gamma}
\end{align}
The statement that the gauge group is compact (in this continuum description)
amounts to saying that the Abelian description (\ref{eq:S_gamma}) is valid
below some scale $\Lambda_{\rm UV}$, and that the UV completion of the theory above
this scale allows finite action monopole-instanton field configurations
whose total magnetic flux is quantized
\cite{Polyakov:1976fu}.
Specifically, we demand that the flux through any 2-sphere is an integer,
\begin{align}
    \int_{S^2} F = 2\pi k \,, \qquad k \in \mathbb{Z} \,,
\label{eq:flux_quant}
\end{align}
where $F \equiv \half F_{\mu\nu} \> dx^{\mu} \wedge dx^{\nu}$
is the 2-form field strength.
Condition (\ref{eq:flux_quant}) implies charge quantization
and removes the freedom to perform arbitrary
field rescalings of the form $A \to A' \equiv (q'/q) \, A$.
As shown by Polyakov, the presence of 
monopole-instantons, regardless of how dilute,
leads to confinement on sufficiently large distance scales
\cite{Polyakov:1976fu}.
%%%%%%%%
\subsection{Action and symmetries}
%%%%%%%%%
We choose the matter sector of our model to be comprised of two
 oppositely-charged scalar fields, $\phi_+$ and $\phi_-$, plus one
 neutral scalar $\phi_0$. We assign unit gauge charges $q = \pm 1$ to
 the charged fields, making them analogous to
 fundamental representation matter fields in a non-Abelian gauge
 theory.%
 \footnote
     {%
     The fact that our charged matter fields have minimal charges of $\pm 1$
     is an essential difference from a similar model
     studied by Sachdev and Park~\cite{SACHDEV200258} 
     in a condensed matter context, see also \cite{Sachdev:2018ddg}.
     The model of Ref.~\cite{SACHDEV200258} has a $U(1)$ global symmetry
     and fields with charges $-1$ and $+2$ under an emergent $U(1)$
     gauge symmetry.  The existence of non-minimally charged matter
     fields allowed Sachdev and Park to use topological order ideas to delineate
     distinct phases. That approach does not work in our model.
     }
 We require the theory to have a single zero-form
 global $U(1)$ symmetry%
 \footnote
     {%
     A zero-form global symmetry is just an ordinary global symmetry which
     acts on local operators.
     }
under which the
fields $\phi_\pm$ both have charge assignments of $-1$ while $\phi_0$
has a charge assignment of $+2$.
These charge assignments, summarized here:
\begin{align}
\begin{array}{c|ccc}
 & \phantom{+}\phi_+ &  \phantom{+}\phi_- & \phantom{+}\phi_0 \\ \hline
U(1)_{\rm gauge} & +1 & -1 & \phantom{+}0 \\
U(1)_{\rm global} & -1 & -1 & +2
\end{array}
\label{eq:chargeTable}
\end{align}
are chosen in a manner which will allow independent control of the
Higgsing of the $U(1)$ gauge symmetry (or lack thereof) and the
realization of the $U(1)$ global symmetry by adjusting suitable mass
parameters.  This is the essential structure needed to
examine the issues motivating this paper in the context of a model
Abelian theory.

The complete action of our model consists of the gauge action
(\ref{eq:S_gamma}), standard scalar kinetic terms, 
plus a scalar potential containing interactions
consistent with the above symmetries,
\begin{align}
\label{eq:the_model}
    S =  \int d^{3}x \, &\left[ \frac{1}{4e^2} \, F_{\mu \nu}^2 
	+ |D_{\mu}\phi_{+}|^2
	+ |D_{\mu} \phi_{-}|^2
	+ m_{c}^2 \, \big(|\phi_{+}|^2+|\phi_{-}|^2 \big)
	+ |\partial_{\mu} \phi_{0}|^2  \right.
	+ m_{0}^2 \, |\phi_{0}|^2
\nonumber\\ &\left. \vphantom{\int}
	- \epsilon \, \big(\phi_{+} \phi_{-} \phi_{0} + \mathrm{h.c.} \big)\right.
	+ \lambda_{c} \big(|\phi_{+}|^4+|\phi_{-}|^4 \big)
	+ \lambda_{0} |\phi_{0}|^4 
\nonumber\\ &\left. \vphantom{\int}
	+ g_{c} \big(|\phi_{+}|^6+|\phi_{-}|^6 \big)
	+ g_{0} |\phi_{0}|^6
	+ \cdots
	+ \Vm(\sigma)
    \right]  .
\end{align}
The mass dimensions of the various couplings are
$[e^2] = [\lambda_{c}] =[\lambda_{0}]=1$, 
$[\epsilon] = 3/2$, and $[g_c] = [g_0] = 0$.  
The ellipsis ($\cdots$) represents possible further
scalar self-interactions, 
consistent with the imposed symmetries,
arising via renormalization.
The term $\Vm(\sigma)$ describes the effects of monopole-instantons,
and is given explicitly below.

The cubic term $\epsilon \, \phi_+ \phi_- \phi_0$ 
ensures that the model has a single $U(1)$ global symmetry,
not multiple independent phase rotation symmetries.  From here onward, we will denote the $U(1)$
global symmetry by $\UG$. 
The simplest local order parameter for the $\UG$ symmetry
is just the neutral field expectation value $\langle \phi_0 \rangle$.
This order parameter has a charge assignment (\ref{eq:chargeTable})
of +2 under the $\UG$ symmetry;
there are no gauge invariant local order parameters with 
odd $\UG$ charge assignments.

In addition to the $U(1)$ gauge redundancy and the $\UG$
global symmetry, this model has two internal $\ZZ$ discrete symmetries.
One is a conventional (particle $\leftrightarrow$ antiparticle)
charge conjugation symmetry,
\begin{align}
    \ZC :\quad
    \phi_\pm \to \phi_\pm^{*} \,,\quad
    \phi_0 \to \phi_0^* \,,\quad
    A_{\mu} \to -A_{\mu} \,.
\label{eq:ZC}
\end{align}
The other is a charged field permutation symmetry,
\begin{align}
    \ZF:\quad
    \phi_+ \leftrightarrow \phi_- \,,\quad
    A_{\mu} \to -A_{\mu} \,.
\label{eq:ZP}
  \end{align} 

A conserved current
$j_{\rm mag}^{\mu} \equiv \epsilon^{\mu\nu\lambda} F_{\nu \lambda}$
associated with a $U(1)$ magnetic global symmetry
is also present if monopole-instanton effects are neglected.
But for our \emph{compact} Abelian theory this
symmetry is not present.
The functional integral representation of the theory 
includes a sum over finite-action magnetic monopole-instanton configurations
with all integer values of total magnetic charge.
These induce corrections to the effective potential
(below the scale $\Lambda_{\rm UV})$
of the form \cite{Polyakov:1976fu}
\begin{align}
    \Vm(\sigma) = - 
    \mu_{\rm UV}^3 \, e^{-\SI} \, \cos(\sigma) \,.
\label{eq:S_monopole}
\end{align}
Here $\SI$ is the minimal action of a monopole-instanton,
% \footnote{Expression (\ref{eq:S_monopole}) relies on a dilute gas
% approximation, valid when the instanton action is large, $\SI \gg 1$.}
%
and $\sigma$ is the dual photon field, related to the original gauge
field by the Abelian duality relation%
\footnote
    {%
    Expression (\ref{eq:S_monopole}) relies on a dilute gas
    approximation, valid when the instanton action is large, $\SI \gg 1$.
    The duality relation (\ref{eq:duality_relation}) appears
    when one imposes the Bianchi identity for $F_{\mu\nu}$  by adding a
    Lagrange multiplier term $i\int d^{3}x\, \frac{\sigma}{4\pi}
    \epsilon^{\mu\nu\lambda} \, \partial_{\mu} F_{\nu\lambda}$ to the
    Euclidean action. Relation \eqref{eq:duality_relation} is the resulting
    equation of motion for
    $F_{\mu\nu}$, and integrating out $F_{\mu\nu}$ gives the Abelian dual
    representation of Maxwell theory.}
\begin{align}
    F_{\mu\nu}
    = \frac{ie^2 }{2\pi}\, \epsilon_{\mu \nu \lambda} \, \partial^{\lambda} \sigma \,.
    \label{eq:duality_relation}
\end{align}
With this normalization the dual photon field is a periodic scalar,
$\sigma \equiv \sigma + 2\pi$, 
with the Maxwell action becoming the kinetic term
$\frac{1}{2}\left(\frac{e}{2\pi}\right)^2 (\partial\sigma)^2$.
The parameter
$\mu_{\rm UV}$ is a short-distance scale associated with the inverse
core size of monopole-instantons.
The $U(1)$ magnetic transformations act as arbitrary shifts on the
dual photon field,
$\sigma \to \sigma + c$.
Such shifts are clearly not a symmetry, except for integer multiples
of $2\pi$.
Consequently, the $\UG$ phase rotation symmetry is the only
continuous global symmetry in our model.

In summary, the faithfully-acting internal global symmetry group of our model is
\begin{align}
    G_{\rm internal} = \frac{\left[ \UG  \rtimes \ZC \right] \times \ZF }{\mathbb{Z}_2} \,.
\end{align}
The quotient by  $\mathbb{Z}_2 \subset \UG:  \phi_{\pm} \to - \phi_{\pm}$ is necessary because it also lies in the gauge group $U(1)$.

When the charged scalar mass squared,  $m_c^2$, is sufficiently
negative this theory has a Higgs regime in which the charged scalar
fields are ``condensed.''  In this regime gauge field fluctuations are
suppressed since the photon acquires a mass term,
\begin{align}
    \big(|\langle\phi_+\rangle|^2+|\langle \phi_- \rangle|^2 \big)
    A_{\mu}A^{\mu}
    \equiv
   \frac{m_A^2}{2e^2} \, A_\mu A^\mu \,,
\label{eq:m_A}
\end{align}
(to lowest order in unitary gauge).%
\footnote
{%
    Our charged scalar fields may be viewed as analogs of the
    electron pair condensate in a Ginsburg-Landau treatment of
    superconductivity, in which case $m_A$ is the Meissner mass whose
    inverse gives the penetration length of magnetic fields.
} 
Monopole-instanton--antimonopole-instanton pairs become
bound by flux tubes with a positive action per unit length $T_{\rm mag}$.% $\sim m_A^2/e^2$.%
\footnote
    {%
    On sufficiently long length scales
    when the flux tube length $L \gtrsim 2 \SI /T_{\rm mag}$,
    these magnetic flux tubes can break due to production of
    monopole-instanton--antimonopole-instanton pairs.
    This is completely analogous to the situation
    in the confining regime,
    discussed next,
    where electric flux tubes exist over a limited range of scales
    controlled by the mass of fundamental dynamical charges.
    }

In contrast,
for sufficiently positive $m_c^2$ our model should be regarded
as a confining gauge theory.
Recall that in the context of QCD,
the confining regime is characterized by a 
static test quark--antiquark potential which rises linearly with separation,
$V_{q\bar q} \sim \sigma r$,
for separations large compared to the strong scale,
$r \gg \LambdaQCD^{-1}$.
But such a linear potential is only present for separations where
the confining string cannot break, which requires that
$\sigma r < 2 m_q$, with $m_q$ the mass of dynamical quarks.
So confinement is only a sharply-defined criterion in the heavy quark
limit, $m_q \gg \sigma/\LambdaQCD =\mathcal O(\LambdaQCD)$.
Nevertheless, it is conventional to speak of QCD as
a confining theory even with light quarks, as this
is a qualitatively useful picture of the relevant dynamics.
This summary applies verbatim to our compact $U(1)$ 3D gauge theory
with massive unit-charge matter, with $\Lambda_{\rm QCD}$ replaced by
an appropriate non-perturbative scale which depends
(exponentially) on the monopole-instanton action $\SI$ \cite{Polyakov:1976fu}.

Finally, we note that
in the absence of monopole-instanton effects,
oppositely charged static test particles in 3D Abelian gauge theory
would experience logarithmic Coulomb interactions which 
grow without bound with increasing separation.
Such a phase could be termed
``confined,'' but for our purposes
this terminology is not helpful.
We find it more appropriate to reserve the term ``confinement'' for
situations where the potential between test charges is linear over
a significant range of distance scales.  With this terminology,
3D compact $U(1)$ gauge theory with finite-action monopole-instantons
and very heavy charged matter is confining, while the non-compact
version of the theory, which does not have a regime with a linear
potential between test charges, is not confining.

%%%%%%%%%   
\subsection{Analogy to dense QCD}
%%%%%%%%%

Our 3D Abelian model is designed to mimic many features of real 4D QCD
at non-zero density.
Explicitly,
\begin{enumerate}
\item
    Both theories contain fundamental representation matter fields
    and are confining in the sense described above.
    Of course, the gauge groups are completely different: $SU(N)$ versus
    $U(1)$. 

\item{
    QCD with massive quarks of equal mass has a vector-like
    $U(N_f)/{\mathbb{Z}_N}$ internal global symmetry.  The
    quotient arises because $\mathbb{Z}_N$
    transformations are part of the $SU(N)$ gauge symmetry.
    In our model, the corresponding global symmetry is
    $[\ZF \times \UG ]/{\mathbb{Z}_2}$.
    The $\ZF \times \UG$ symmetry is analogous to $U(N_f)$},
    while the discrete quotient arises for the same reason
    as in QCD.

\item
    The scalar fields in the 3D Abelian model may be regarded
    as playing the role of color anti-fundamental diquark
    operators which acquire non-zero vacuum expectation values
    in high density QCD, see Ref.~\cite{Alford:2007xm} for a review.
    The symmetry group $\UG$ is analogous to quark number
    $U(1) \subset U(N_f)$,
    while $\UG/\mathbb{Z}_2$ is analogous to baryon number $U(1)_B$.
    Note one distinction in the transformation properties
    of the scalar fields in our Abelian model and the diquark
    condensates in QCD; the former have charge $1$ under our $\UG$
    group whereas the latter have
    charge $2$ under quark number.
    The $\ZF$ permutation symmetry of our 3D Abelian model is
    analogous to the $\mathbb{Z}_{N_f} \subset U(N_f)$ cyclic
    flavor permutation symmetry of 4D QCD.

\item
    Since the charged scalars $\phi_{\pm}$ are analogous
    to anti-fundamental
    diquarks in three-color QCD, $\phi^{\dag}_{+}\phi^{\dag}_{-}$ is
    akin to a dibaryon.  This means that $\phi_0$ can also be
    interpreted as a dibaryon interpolating operator, and the condensation of $\phi_0$ in our model is directly
    analogous to the dibaryon condensation which occurs in dense QCD.

\item
    In QCD, the Vafa-Witten theorem~\cite{Vafa:1983tf} implies
    that phases with spontaneously broken $U(1)_B$ symmetry
    can only appear at non-zero baryon density,
    while in our Abelian model $\UG$-broken phases can appear
    at zero density. This difference reflects the fact that QCD contains only
    fermionic matter fields, while our Abelian model has fundamental scalar fields.

\end{enumerate}

%%%%%%%%%%%%%%%
\subsection{Symmetry constraints on the phase structure}
\label{sec:Landau_constraints}
%%%%%%%%%%%%%%%

\begin{figure}[t]
    \centering
    \includegraphics[width=0.6\textwidth]{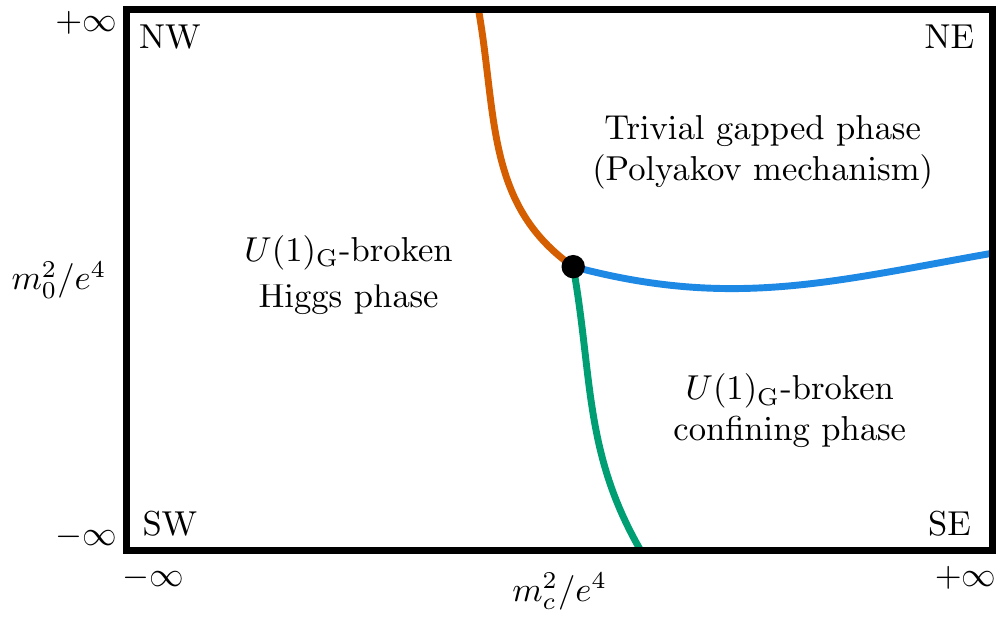}
    \caption{A sketch of the simplest consistent phase diagram of our
        model as a function of the charged and neutral scalar mass
        parameters $m_c^2$ and $m_0^2$. The four corners correspond to
        weakly-coupled regimes in parameter space;
	curves in the interior of the figure
        represent phase transitions.  These phase transition curves
        are robust: they cannot be evaded by varying any parameters of
        the model which are consistent with its symmetries.
    \label{fig:3D_phase_diagram} 
    }
 \end{figure}

We begin analyzing the phase structure of the model
\eqref{eq:the_model} using the Landau paradigm based on realizations
of symmetries with local order parameters.   We
will consider the phase diagram as a function of the charged and
neutral scalar masses, $m_c^2$ and $m_0^2$. 
We focus on the regime
where quartic and sextic scalar self-couplings are positive, the cubic, quartic
and gauge couplings are comparable, $\epsilon/e^3$, $|\lambda_c|/e^2$ and
$|\lambda_0|/e^2$ are all $\mathcal O(1)$, and the dimensionless sextic couplings
are small, $g_c$, $g_0 \ll 1$.
% \footnote{The phase structure has no
% qualitative dependence on the sign of $\epsilon$. However, an extra
% $U(1)$ global symmetry appears at the point $\epsilon = 0$, so this
% point in parameter space is discussed at the end of this section.} 
The simplest phase diagram
consistent with our analysis is sketched in
Fig.~\ref{fig:3D_phase_diagram}.

Interpreting Fig.~\ref{fig:3D_phase_diagram} as if it were a map,
let us refer to the four weakly-coupled corners of parameter
space by their compass directions:
\begin{subequations} \label{eq:corners} 
\begin{align}
\textrm{NW}  &: \{-m_c^2 \gg e^4,\, m_0^2 \gg e^4 \}, &\textrm{NE}   &: \{m_c^2 \gg e^4,\, m_0^2 \gg e^4 \}, \\
\textrm{SW} &: \{-m_c^2 \gg e^4,\, -m_0^2 \gg e^4 \}, &\textrm{SE}  &: \{m_c^2 \gg e^4,\, -m_0^2 \gg e^4 \},
\end{align}
\end{subequations}
each of which  we discuss in turn.   In this section we explain the origin of
the phase transition curve (orange) separating the NE region from the W
side of Fig.~\ref{fig:3D_phase_diagram}, as well as the 
(blue) curve separating the NE and SE regions.  The bulk of the paper is
dedicated to understanding the origin of the phase transition curve
(green) separating the  SE region from the W side of
Fig.~\ref{fig:3D_phase_diagram}.

First, consider region NE where $m_{c}^2$, $m_0^2 \gg e^4$.
In this regime our model has a
unique gapped vacuum state and no broken symmetry.
To see this, one may integrate out all the matter fields
and observe that the resulting tree-level effective action is
\begin{align}
    \Seff = \int d^3x\, \left[ \frac{1}{4e^2} \, F_{\mu\nu}^2 + \Vm(\sigma)\right]\,.
\end{align}
The monopole potential $\Vm(\sigma)$ has a unique minimum
for the dual photon $\sigma$ and induces a non-zero photon mass,
\begin{equation}
    m_{\gamma}^2 = 4\pi^2({\mu_{\rm UV}^3}/{e^2}) \, e^{-\SI} \,.
\label{eq:mgamma}
\end{equation}
Hence, the vacuum is gapped and unique.
Both the continuous $\UG$ and the discrete
$\ZC$ and $\ZF$ global symmetries are unbroken,
and hence
region NE may be termed ``confining and unbroken.''

Now consider the entire E side where $m_c^2 \gg e^4$ while the neutral mass
$m_0^2$ is arbitrary.
Then one may integrate out the charged fields
and the effective action becomes
\begin{align}
    \Seff = \int d^3x\, \left[
	\frac{1}{4e^2}\, F_{\mu\nu}^2 + \Vm(\sigma)
	+ |\partial_{\mu} \phi_0|^2
	+ m_0^2 |\phi_0|^2
	+ \lambda_0 |\phi_0|^4
	+ g_0 |\phi|^6
	+\cdots
    \right]\,.
%\nonumber\\
\end{align}
This is a 3D XY model plus a decoupled compact $U(1)$ gauge theory.
The photon is still gapped by the Polyakov mechanism. If we take
$m_0^2 \gg |\lambda_0|^2$, then we come back to the discussion of the
previous paragraph. If we take $-m_0^2 \gg |\lambda_0|^2$, then
$\phi_0$ develops a non-vanishing expectation value, the
$\UG/\mathbb{Z}_2$ symmetry is spontaneously broken, and there is a
single massless Nambu-Goldstone boson. So region SE is ``confining and
$\UG$ symmetry broken.'' The discrete $\ZF$ symmetry is unbroken in
this region, as is a redefined $\ZC$ symmetry which combines the basic
$\ZC$ transformation (\ref{eq:ZC}) with a $\UG$ transformation that
compensates for the arbitrary phase of the condensate $\langle \phi_0
\rangle$. This symmetry-broken regime must be separated from the
symmetry-unbroken regime by a phase transition depending on the value
of $m_0^2/\lambda_0^2$. If we take our quartic and sextic couplings to
be positive, this is just the well-known XY model phase transition,
which is second order in three spacetime dimensions.

Next, consider what happens on the W side where $-m_c^2 \gg e^4$ while
$m_0^2$ is arbitrary.
In this case the charged scalar fields $\phi_\pm$ will acquire
non-zero expectation values
(using gauge-variant language), with
$v_c \equiv |\langle \phi_\pm \rangle| = \mathcal O(|m_c \, \lambda_c^{-1/2}|)$.%
\footnote
    {%
    In this and subsequent parametric estimates, we neglect the
    cubic coupling and sextic couplings.  For the sextic couplings
    this is justified by our assumption that they are small.
    We have dropped $\epsilon$-dependence purely for simplicity:  
    taking it into account is straightforward but results in much
     more cumbersome expressions.
    } 

This has several effects. First, since these fields transform
non-trivially under the $\UG$ symmetry, this global symmetry is
spontaneously broken leading to a massless Nambu-Goldstone excitation.
Second, the $U(1)$ gauge field becomes Higgsed, as discussed above,
with the photon acquiring a mass $m_A$. Writing $\phi_\pm = (v_c +
H_\pm/\sqrt 2) \, e^{-i\chi}$, up to an arbitrary $U(1)$ gauge
transformation, the resulting effective action has the form
\begin{align}
    S = \int & d^{3}x \, \left[
    \frac{1}{4e^2} \, F_{\mu\nu}^2
    + \half m_A^2 \, A_{\mu}A^{\mu} 
    + |\partial_\mu \phi_0|^2
    + m_0^2 |\phi_0|^2
    + \lambda_0 |\phi_0|^4 \right.
\nonumber\\ 
    &\left. \vphantom{\int d^3x} 
    + 2 v_c^2 (\partial_{\mu}\chi)^2
    - 2\epsilon v_c^2 \, \Re( e^{-2i \chi} \phi_0)
    + \sum_{i = \pm} \Big[ \half ( \partial_{\mu} H_{i})^2
    + \half m_H^2 H_i^2 \Big]
    + \cdots
     \right] ,
\end{align}
where $\chi$ is the $\UG$ Nambu-Goldstone boson
and $H_\pm$ are real Higgs modes with mass $m_H$.
This regime, extending inward from the W boundary of the phase diagram,
may be termed ``Higgsed and $\UG$ symmetry broken.''
The discrete symmetries remain unbroken
in the same manner as in region SE.
Regardless of the sign of $m_0^2$,
the neutral scalar $\phi_0$ acquires a non-zero vacuum expectation value
whose phase, $2\chi$, is set by the phase of the Higgs condensate.
As $m_0^2$ is varied from large positive to large negative values,
the magnitude
$|\langle \phi_0 \rangle|$ varies from
a small $\mathcal O(\epsilon v_c^2 \, m_0^{-2})$ value to
a large $\mathcal O(m_0 \lambda_0^{-1/2})$ value,
while always remaining non-zero.
Throughout this Higgs regime
% ,
% the action of a single monopole-instanton diverges with
% the spacetime volume, and 
monopole-instanton--antimonopole-instanton pairs become
linearly confined by magnetic flux tubes as noted earlier.

The fact that the $\UG$ symmetry is spontaneously broken in this
Higgs regime means that the entire W region of parameter space with
$-m_c^2 \gg e^4$
must be separated by a phase transition from the trivially gapped
region NE where $m_c^2$ and $m_0^2$ are large and positive.
But the pattern of global symmetry breaking throughout the W side Higgs regime
of $-m_c^2 \gg e^4$ is identical to that in region SE where
$m_c^2 \gg e^4$ and $-m_0^2 \gg \lambda_0^2$.
This raises the central question in this paper:

\bigskip
\emph{Are the Higgs and confining $\UG$-breaking regimes
smoothly connected, or are they distinct phases?}
\bigskip

\noindent
As summarized in the introduction and sketched in
Fig.~\ref{fig:3D_phase_diagram}, we will find that the Higgs and
confining $\UG$-breaking regimes must be distinct phases, separated by
at least one phase transition, even though there are no distinguishing
local order parameters.

Before leaving this section, we pause to consider two further issues:
the realization of the $\ZF$ symmetry and the nature of the $\epsilon
\to 0$ limit.  In our discussion below we will assume that the $\ZF$
symmetry is not spontaneously broken. It is
possible to tune the scalar potential to break $\ZF$ spontaneously,
but this results in a Higgs phase which is separated by an obvious
phase boundary from both the $\UG$-broken confining phase and the
$\ZF$-invariant Higgs phase.  This makes the $\ZF$-broken regime
uninteresting for the purposes of this paper.

Next, one should observe that
$\epsilon \to 0$ is a non-generic limit of the model. An additional
global symmetry which purely phase rotates the charged fields,
$\phi_{\pm} \to e^{i\alpha} \, \phi_{\pm}$, is present when $\epsilon
= 0$; we denote this symmetry as $U(1)_{\rm extra}$. The $\epsilon=0$
theory has four distinct phases distinguished by realizations of the
$\UG$ and $U(1)_{\rm extra}$ symmetries. There is a phase where only
the $U(1)_{\rm extra}$ symmetry is spontaneously broken, with one
Nambu-Goldstone boson. This phase is not present at non-zero
$\epsilon$. At $\epsilon = 0$, the $\UG$-broken Higgs phase in
Fig.~\ref{fig:3D_phase_diagram} becomes a phase with two spontaneously
broken continuous global symmetries, $\UG$ and $U(1)_{\rm extra}$, and
has two Nambu-Goldstone bosons. This is a distinct symmetry
realization from the $\UG$-broken confining regime with only a single
Nambu-Goldstone boson implying, by the usual Landau paradigm
reasoning, at least one intervening separating phase transition.

When $\epsilon$ is non-zero but very small compared to all other
scales, there is a parametrically light pseudo-Nambu-Goldstone
boson with a mass $m_{\rm pNGB} \propto \sqrt{\epsilon}$ in the
Higgs regime.
Determining whether the $\UG$-broken Higgs and confining
regimes remain distinct for non-zero values of $\epsilon$
is the goal of our next section in which
we examine the long-distance behavior of
holonomies around vortices.
In this analysis, it will be important that the
holonomy contour radius be large compared to
microscopic length scales --- which include the
Compton wavelength of the pseudo-Goldstone boson, $m_{\rm pNGB}^{-1}$.
There is non-uniformity between the large distance limit of the
holonomy and the $\epsilon\to 0$ limit, and consequently
the physics of interest must be studied directly in the
theory with $\epsilon \ne 0$.

%%%%%%%%%%%
\section{Vortices and holonomies}
\label{sec:vortices_and_holonomies}
%%%%%%%%%%%
% This section introduces a novel topological order parameter for
% systems with a spontaneously broken $U(1)$ global symmetry, and then
% applies this order parameter to infer that the Higgs and confining
% $\UG$-breaking regimes of the model \eqref{eq:the_model} are separated
% by at least one phase transition.  Our order parameter is introduced
% in Sec.~\ref{sec:order_param_def}, and its behavior in the Higgs
% regime and the confining $\UG$-broken regimes is studied in
% Sec.~\ref{sec:Higgs_holonomy_Abelian} and
% Sec.~\ref{sec:holonomy_confining} respectively.
% Sec.~\ref{sec:ColemanWeinberg} shows that non-analyticies in our
% topological order parameter are associated with genuine phase
% transitions of the theory, while Sec.~\ref{sec:broken_permutations} of
% \eqref{eq:the_model} shows that explicitly breaking the $\ZF$ symmetry
% does not remove the phase transition between the Higgs and confining
% $\UG$ broken regimes.

%%%%%%%%%
\subsection{The order parameter $O_{\Omega}$}
\label{sec:order_param_def}
%%%%%%%%%

Consider the portion of the phase diagram in which the $\UG$ symmetry
is spontaneously broken. Then the field $\phi_0$ has a non-vanishing
expectation value and the spectrum contains a Nambu-Goldstone boson.
The Goldstone manifold has a non-trivial first homotopy group,
$\pi_1(\UG) = \mathbb{Z}$. This implies that there are stable global
vortex excitations, which are particle-like excitations in two spatial
dimensions.  

Vortex excitations may be labeled by an integer winding number $w$
indicating the number of times the phase of $\langle \phi_0 \rangle$
wraps the unit circle as one encircles a vortex.
More explicitly, one may write the winding number as a contour integral
of the gradient of the phase,
\begin{equation}
    w = \frac 1{2\pi}
    \oint_C dx^\mu \, u_\mu \,,
\end{equation}
where 
$ u_\mu \equiv -i \partial_\mu \left( \langle \phi_0 \rangle / |
\langle \phi_0 \rangle| \right)$. 
Using the language of a
superfluid, $u_\mu$ is the superfluid flow velocity, and the
winding number $w$ is the quantized circulation around a vortex.

As with vortices in superfluid films, vortex excitations have
logarithmic long range interactions, 
 with a $1/r$ force between vortices separated by distance $r$.
A single vortex in
infinite space has a logarithmically divergent long distance
contribution to its self-energy. Nevertheless, vortices are important
collective excitations and, in any sufficiently large volume, a
non-zero spatial density of vortices and antivortices will be present
due to quantum and/or thermal fluctuations.  From a spacetime
perspective, vortex/antivortex world lines, as they appear and
annihilate, form a collection of closed loops, with an action scaling
as  $L \log L$ for loops with characteristic size $L$.%
\footnote
    {This is only a logarithmic enhancement
    over the linear scaling of a vortex loop action in
    superconductors (or simple Abelian Higgs models).}

%%%%%%%%%%%%%%
\begin{figure}
\centering
\includegraphics[width=.4\textwidth]{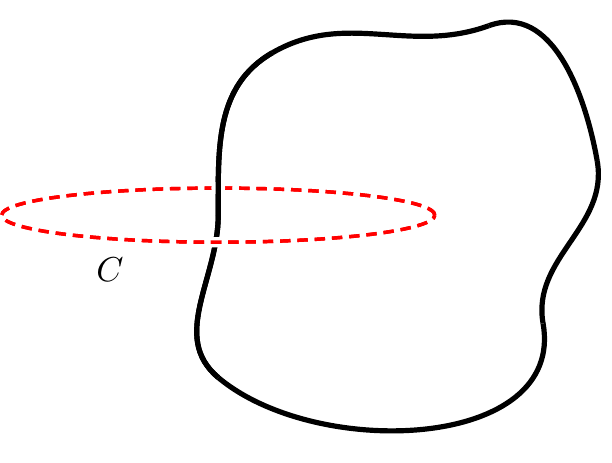}
\caption
    {%
    A contour $C$ (red dashed curve) which links a vortex world-line
    (solid black curve).
    Of interest is the gauge field holonomy $\Omega \equiv e^{i\oint_C A}$
    for contours $C$ far from the vortex core.
    \label{fig:vortex_holonomy}
    }
\end{figure}
%%%%%%%%%%%%%%

Consider the gauge field holonomy,
$\Omega \equiv e^{i\oint_C A}$,
evaluated on some large circular contour $C$
surrounding a vortex of non-zero winding number $k$,
illustrated in Fig.~\ref{fig:vortex_holonomy},
which we denote by $\langle \Omega(C) \rangle_k$.
Let $r$ denote the radius
of the contour $C$ encircling the vortex.
We are interested in the phase of the holonomy,
but as the size of the contour $C$ grows, short distance quantum fluctuations
will cause the magnitude of the expectation
$\langle \Omega (C) \rangle_k$ to decrease
(with at least exponential perimeter-law decrease).
To compensate, we consider the large distance limit of a
ratio of the holonomy expectation values
which do, or do not, encircle a vortex of minimal non-zero
winding number,
\begin{align}
    O_{\Omega}
    \equiv
    \lim_{r \to \infty} %%% \lim_{V\to\infty}
    \frac{\langle \Omega(C) \rangle_1}{\langle \Omega (C) \rangle} \,.
\label{eq:order_parameter}
\end{align}
Here, the numerator should be understood as an expectation value
defined by a constrained functional integral in which there is a prescribed
vortex loop of characteristic size $r$ and winding number 1
linked with the holonomy loop of size $r$,
with both sizes, and the minimal separation between the two loops,
scaling together as $r$ increases.
The denominator is the ordinary unconstrained vacuum expectation value.

The quantity $O_{\Omega}$ measures the phase acquired by a particle
with unit gauge charge when it encircles a minimal global vortex. Or
equivalently, it is the phase acquired by a minimal global vortex when
it is dragged around a particle with unit gauge charge.

Our analysis below will demonstrate that $O_{\rm \Omega}$ cannot be a
real-analytic function of the charged scalar mass parameter
$m_c^2/e^4$.  We will also argue that non-analyticities in the
topological order parameter $O_{\Omega}$ are associated with
genuine thermodynamic phase transitions.  

A quick sketch of the argument is as follows.  Since 
the vacuum is invariant under the
$\ZC$ charge conjugation symmetry,
the denominator of
$O_{\Omega}$ must be real and at sufficiently weak
coupling is easily seen to be positive.%
\footnote
    {%
    One may equally well appeal to reflection symmetry,
    as this reverses the orientation of a reflection symmetric
    contour like a circle, and hence maps the holonomy on a
    circular contour to its complex conjugate.
    This alternative will be relevant for our later discussion
    in Sec.~\ref{sec:QCD} of dense QCD and related models
    with non-zero chemical potential, where
    charge conjugation symmetry is explicitly broken by the chemical
    potential but the ground state remains invariant under
    reflections.
    \label{fn:reflect}
    }
In the constrained expectation value in the numerator of
$O_{\Omega}$, the $\ZC$ symmetry is explicitly broken by the
unit-circulation condition that enters the definition of
$\langle \Omega(C) \rangle_1$.  But the unit-circulation condition does
not break $\ZF$ permutation symmetry \eqref{eq:ZP},
which also flips the sign of the gauge field.%
\footnote
    {%
    The $\ZF$ symmetry
    cannot be spontaneously broken due to the presence of a vortex because the vortex worldvolume is one-dimensional, and discrete symmetries
    cannot break spontaneously in one spacetime dimension.
    (The exception to this statement involving
    mixed 't Hooft anomalies~\cite{Gaiotto:2017yup}
    is irrelevant in our case.)
    }
Therefore the numerator of $O_{\Omega}$ must be invariant
under $\ZF$, and hence real.  We will see below that
it is negative deep in the Higgs regime, but is positive deep in
the $\UG$-broken confining regime.
In the large-$r$ limit defining our vortex observable $O_\Omega$,
the magnitudes of the holonomy expectations in numerator and denominator
will be identical.
%%% This means that when our system has an unbroken $\ZF$ symmetry,
Hence, our vortex observable $O_{\Omega}$ obeys
\begin{align}
    O_{\Omega} = 
    \begin{cases}
        -1 \,, & \UG\textrm{-broken Higgs regime;} \\
        +1 \,, & \UG\textrm{-broken confining regime,} 
    \end{cases}
\end{align}
and therefore cannot be analytic as a function of $m_c^2/e^4$.

In the remainder of this section we support the above claims.
We study the properties of vortices in the Higgs and
confining $\UG$-broken regimes in Secs.
\ref{sec:Higgs_holonomy_Abelian} and \ref{sec:holonomy_confining},
respectively.   Then in Sec.~\ref{sec:ColemanWeinberg} we argue that
non-analyticities in our topological order parameter are associated
with genuine thermodynamic phase transitions.  Finally, in
Sec.~\ref{sec:broken_permutations} we extend the treatment
and consider the effects of perturbations which explicitly break
the $\ZF$ symmetry.
We find that $O_{ \Omega}$ remains a non-analytic function of the
charged scalar mass parameter(s) even in the presence of such perturbations.
This shows that the phase transition line separating the 
Higgs and confining $\UG$-broken regimes is robust against
sufficiently small $\ZF$-breaking perturbations.

%%%%%%%%%%
\subsection{$O_{\Omega}$ in the Higgs regime}
\label{sec:Higgs_holonomy_Abelian}
%%%%%%%%%%

We first consider $O_{\Omega}$ deep in the Higgs regime,
$-m_c^2 \gg e^4$ and, to begin, neglect quantum fluctuations altogether.
So the holonomy expectation values in the definition (\ref{eq:order_parameter})
of $O_\Omega$ just require evaluation of the holonomy in the appropriate
energy-minimizing classical field configurations.

As always, the holonomy $\Omega(C)$ is the exponential of the line integral
$\oint_C A$ (times $i$) which, in our Abelian theory,
is just the magnetic flux passing through a surface spanning the curve $C$.
For the ordinary vacuum expectation value in the denominator of $O_\Omega$,
vacuum field configurations have everywhere vanishing magnetic field
and hence $\langle \Omega(C) \rangle = 1$.

For the constrained expectation value in the numerator, one needs to
understand the form of the minimal vortex solution(s).
Choose coordinates such that the vortex lies at the origin of space
and let $\{r,\theta\}$ denote 2D polar coordinates.
For a vortex configuration with winding number $k$,
the phase of the neutral scalar $\phi_0$ must wrap
$k$ times around the unit circle as one encircles the origin.

There exist classical solutions which preserve rotation invariance,
and we presume that these rotationally invariant solutions capture the
relevant global energy minima. Such field configurations may be
written in the explicit form
\begin{subequations}%
\label{eq:ansatz}%
\begin{align}
    \phi_{+}(r,\theta) &= v_c \, f_+(r) \, e^{i \nu_+ \theta } \,, &
    \phi_{0}(r,\theta) &= v_0 \, f_0(r) \, e^{i k \theta } \,, \\
    \phi_{-}(r,\theta) &= v_c \, f_-(r) \, e^{i \nu_- \theta} \,, &
    A_{\theta}(r) &= \frac{\Phi\, h(r)}{2\pi r} \,.
\end{align}
\end{subequations}
Here $v_0$ and $v_c$ are the magnitudes of the vacuum expectation
values of $\phi_0$ and $\phi_\pm$, determined by minimizing the
potential terms in the action. The angular wavenumbers $\nu_+$,
$\nu_-$, and $k$ must be integers to have single valued configurations
and $k$, by definition, is the winding number of the vortex
configuration. For non-zero values of $k$ and $\nu_\pm$ the radial
functions $f_0(r)$ and $f_\pm(r)$ interpolate between 0 at the origin
and 1 at infinity. Similarly, to minimize energy the gauge field must
approach a pure gauge form at large distance, implying that $h(r)$ may
also be taken to interpolate between 0 and 1 as $r$ goes from the
origin to infinity. The associated magnetic field is
\begin{equation}
    B(r) = \frac{(r A_\theta(r))'}{r} = \frac{ \Phi \, h'(r)}{2\pi r} \,.
\end{equation}
The gauge field in ansatz (\ref{eq:ansatz}) is written in a form which
makes the coefficient $\Phi$ equal to the total magnetic flux,
\begin{equation}
    \Phi_B \equiv \int d^2x \> B
    = 2\pi \int_0^\infty r \, dr \> B(r)
    = \Phi \int_0^\infty dr \> h'(r)
    = \Phi   \,.
\label{eq:flux}
\end{equation}

To avoid having an energy which diverges linearly with volume
(relative to the vacuum), the phases of $\phi_0$, $\phi_+$ and
$\phi_-$ must be correlated in a fashion which minimizes the cubic
term in the action.   Below we will suppose that the  coefficient of
the cubic term $\epsilon>0$, but essentially the same formulas would
result if $\epsilon < 0$.  (The singular point $\epsilon=0$ must be
handled separately, see the discussion at the end of Sec.~\ref{sec:Landau_constraints}.)
Minimizing the cubic term in the action forces the product
$\phi_0 \, \phi_+ \, \phi_-$ to be real and positive, implying that
\begin{equation}
    \nu_+ = n - k \,, \qquad \nu_- = -n \,,
\label{eq:nupm}
\end{equation}
for some integer $n$.

After imposing condition (\ref{eq:nupm}),
there remains a logarithmic dependence on the spatial volume
caused by the scalar kinetic terms which, due to the
angular phase variation of the scalar fields,
generate energy densities falling as $1/r^2$.
Explicitly, this long-distance energy density is
\begin{equation}
    \mathcal E(r)
    =
    \frac {v_c^2}{r^2}
    \left[
	\left(n-k - \frac \Phi{2\pi}\right)^2 + \left(-n + \frac \Phi{2\pi}\right)^2
    \right]
    +
    \frac {v_0^2 \, k^2}{r^2} 
    + \mathcal O(r^{-4})
    \,.
    \label{eq:energy_long_distance}
\end{equation}
Minimizing this IR energy density, for given values of $k$ and $n$,
determines the magnetic flux $\Phi$, leading to
\begin{equation}
   \Phi_B =  \Phi = (2n - k) \, \pi \,,
\label{eq:fluxval}
\end{equation}
and an IR energy density
$
    \mathcal E(r) = (\half v_c^2 + v_0^2) \, k^2  / r^2
    + \mathcal O(r^{-4})
$.

The explicit form of the radial functions is determined by minimizing
the remaining IR finite contributions to the energy.
These consist of the magnetic field energy and short distance corrections
to the scalar field kinetic and potential terms, all of which are
concentrated in the vortex core region.
Semi-explicitly,
\begin{align}
    E =
    2\pi \int r \, dr \> \Biggl[&
    \frac {h'(r)^2}{8e^2 r^2} \, (2n{-}k)^2
    +
    \frac {v_c^2 \, f_+(r)^2}{4r^2}
    \left[ (2n{-}k) (1{-}h(r)) - k \right]^2
\nonumber\\ &
    +
    \frac {v_0^2 \, k^2 f_0(r)^2}{r^2} 
    +
    \frac {v_c^2 \, f_-(r)^2}{4r^2}
    \left[ (2n{-}k) (1{-}h(r)) + k \right]^2
\nonumber\\ &
    + v_c^2\left[f_+'(r)^2+f_-'(r)^2\right] + v_0^2 f_0'(r)^2
    + \mbox{(potential terms)} \Biggr] \,.
\label{eq:tree_level_Veff}
\end{align} 
Minimizing this energy leads to straightforward but
unsightly ordinary differential equations which determine
the precise form of the radial profile functions, see Appendix~\ref{sec:EoMAppendix}.
Qualitatively,
the gauge field radial function $h(r)$ approaches its asymptotic value
of one exponentially fast on the length scale $\textrm{min}(m_A^{-1},\widetilde{m}^{-1})$, where 
$m_A = 2e v_c $ and $\widetilde{m}^2 \equiv 4\lambda_c v_c^2 + 2\epsilon v_0$.
The scalar field profile functions $f_0(r)$ and $f_\pm(r)$ approach their
asymptotic large $r$ values with $1/r^2$ corrections on the length scales set by the
corresponding masses $m_0$ and $m_c$.  

For a given non-zero winding number $k$, the above procedure
generates an infinite sequence of vortex solutions distinguished by
the value of $n$, or more physically by the quantized value of the
magnetic flux (\ref{eq:fluxval}) carried in the vortex core.
The minimal energy vortex, for a given winding number, is the one
which minimizes this flux.
For even winding numbers, this is $n = k/2$ and vanishing magnetic flux.
In such solutions, the phases of the two charged scalar fields are identical
with $\nu_\pm = - k/2$.

For odd winding number $k$
there are two degenerate solutions with
$n = (k\pm 1)/2$ and magnetic flux $\Phi = \pm \pi$.
In these solutions, the charged scalar fields have differing phase windings
with
$
    \nu_+ = -(k \mp 1)/2
$
and
$
    \nu_- = -(k \pm 1)/2 
$.
For minimal $|k| = 1$ vortices, one of the charged scalars has a
constant phase with no winding, while the other charged scalar has a
phase opposite that of $\phi_0$.
%% The magnitude of the charged
%% scalar with the constant phase must approach $|m_c^2|/(2 \lambda_c)$
%% at $r=0$.

The gauge field holonomy surrounding a vortex, far from its core,
is simply $\pm 1$ depending on whether the magnetic flux is an even
or odd multiple of $\pi$ and this, in turn, merely depends on whether the
vortex winding number $k$ is even or odd,
\begin{equation}
    \langle \Omega(C) \rangle_k = e^{i \Phi} = (-1)^k \,.
\end{equation}

\begin{figure}[t]
\centering
\includegraphics[width=.6\textwidth]{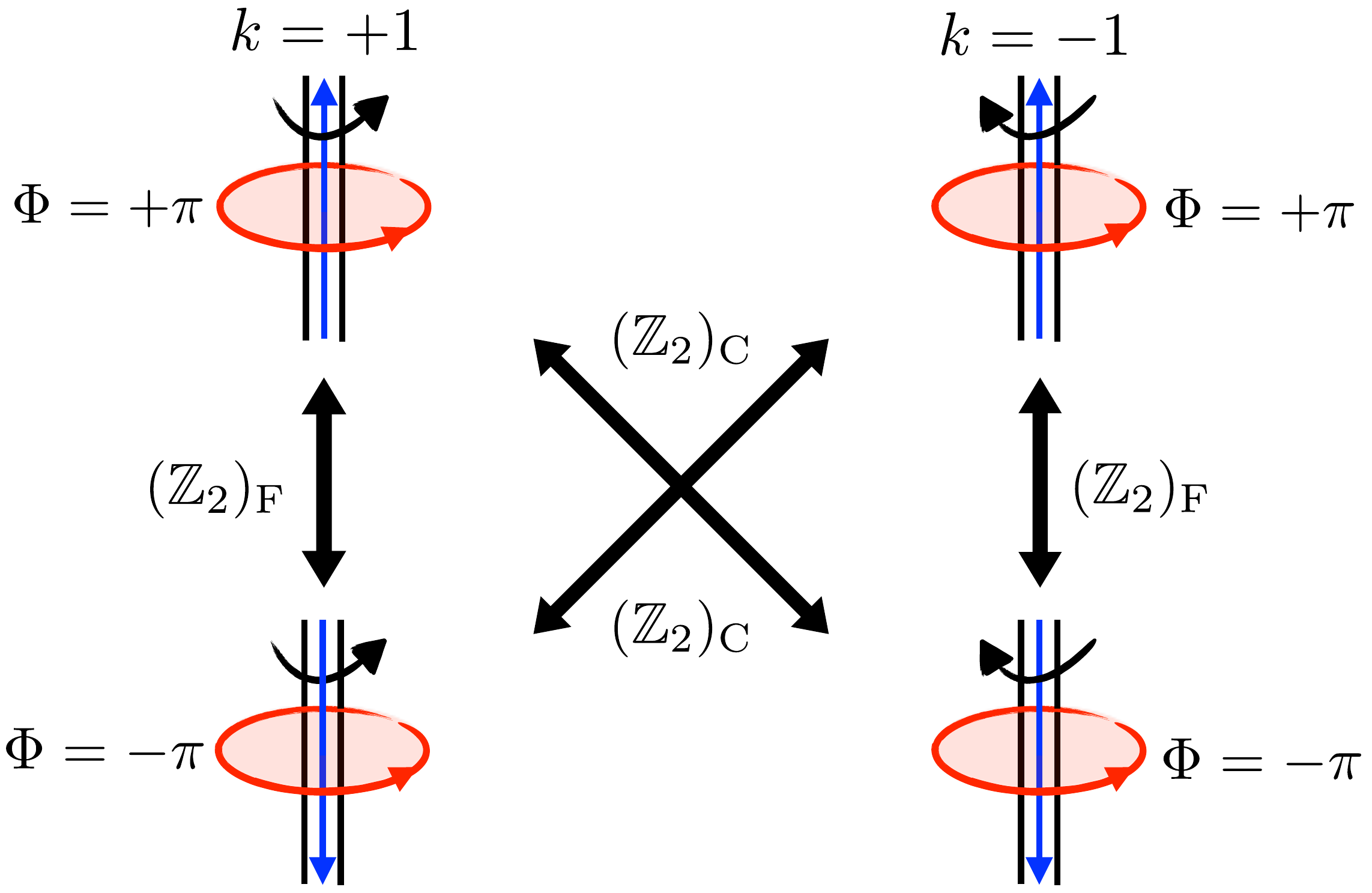}
\caption
    {%
    There are four distinct minimal energy vortex solutions, with winding number
    $k = \pm 1$ and magnetic flux $\Phi = \pm \pi$.
    The $\ZF$ and $\ZC$ discrete symmetries relate these vortices as shown.
    \label{fig:min_vortex}
    }
\end{figure}

%%%%%%%%%%%%%%

The net result is that there are four different minimal energy vortex solutions,
illustrated in Fig.~\ref{fig:min_vortex},
having $(k,\Phi) = (1,\pi)$, $(1,-\pi)$, $(-1,\pi)$, and $(-1,-\pi)$.
As indicated in the figure,
the $\ZF$ symmetry interchanges vortices with identical winding number
and opposite values of magnetic flux,
while the $\ZC$ symmetry interchanges vortices with opposite values
of both winding number and magnetic flux.
Therefore, all these vortices have identical energies.
For our purposes, the key result is that the
long distance holonomy is the same for all minimal vortices, namely
$\langle \Omega(C) \rangle_{k = \pm1} = -1$.
Consequently, we find
\begin{align}
    O_{\Omega} = -1 \, \;\; \textrm{ at tree-level}.
\label{eq:O_tree}
\end{align}

We now consider the effects of quantum fluctuations on this result.
Using standard effective field theory (EFT) reasoning,
as one integrates out
fluctuations below the UV scale $\Lambda_{\rm UV}$, the action
\eqref{eq:the_model} will receive scale-dependent corrections which
(a) renormalize the coefficients of operators appearing in the action
\eqref{eq:the_model}, and (b) induce additional operators of
increasing dimension consistent with the symmetries of the theory.
But  the result \eqref{eq:O_tree} follows directly from the leading long-distance
form \eqref{eq:energy_long_distance} of the energy density whose
minimum fixes the vortex magnetic flux equal to $\pm \pi$  for minimal
winding vortices. Because this $1/r^2$ energy density leads to a total
energy which is logarithmically sensitive to the spatial volume, short
distance IR-finite contributions to the energy cannot affect the flux
quantization condition \eqref{eq:fluxval}  in the limit of large
spatial volume. Only those corrections which modify this $1/r^2$ long
distance energy density have the potential to change the quantization
condition.

One may construct the long distance EFT as an expansion in
derivatives, with the effective expansion parameter being
the small ratio of fundamental length scales (such as
the vortex core size or Compton wavelengths of massive
excitations) to the arbitrarily large
length scale of interest.
Any term in the EFT
action with more than two derivatives will produce a contribution to
the energy density which falls faster than $1/r^2$ when evaluated on a
vortex configuration, and hence cannot
contribute to the $\mathcal{O}(1/r^2)$ long distance energy density
\eqref{eq:energy_long_distance}. Similarly,
terms with less than two derivatives also
do not contribute to the $\mathcal{O}(1/r^2)$ long distance energy
density \eqref{eq:energy_long_distance}.  Hence the only
fluctuation-induced terms that might affect the long distance
vortex holonomy are those with
precisely two derivatives acting on the charged scalar fields.
Consequently,
the portion of the effective action that controls
holonomy expectation values around vortices can be written in the form
\begin{align} \label{eq:quantum_eff_action} 
    S_{\textrm{eff}, \, U(1)\textrm{ holonomy}}
    = \int d^{3}x \, \Bigl\{ \,
	   & f_1(\phi_0, \phi_{+},\phi_{-})(D_{\mu}\phi_{+})(D^\mu\phi_+)^\dagger+f_1(\phi_0, \phi_{-},\phi_{+})(D_{\mu}\phi_{-})(D^\mu\phi_-)^\dagger  \nonumber \\
 {}+{} & f_2(\phi_0,\phi_+,\phi_-) (D_\mu\phi_+)(D^\mu \phi_+)+f_2(\phi_0,\phi_-,\phi_+) (D_\mu\phi_-)(D^\mu \phi_-)  \nonumber \\[6pt]
 {}+{} & f_3(\phi_0,\phi_+,\phi_-) (D_\mu\phi_+)(D^\mu \phi_-)^\dagger+ f_3(\phi_0,\phi_-,\phi_+) (D_\mu\phi_-)(D^\mu \phi_+)^\dagger \nonumber \\
    {}+{}  & f_4(\phi_0, \phi_{+},\phi_{-})(D_{\mu} \phi_{+})(D^{\mu} \phi_{-})
	    \Bigr\}+\textrm{h.c.}\,,
\end{align}
with coefficient functions $\{f_i\}$ depending on
the fields $\phi_0$, $\phi_{\pm}$ (but not their derivatives)
such that each term is $\UG$ and gauge invariant.
We emphasize that the
long-distance EFT (\ref{eq:quantum_eff_action})
does not rely on a weak-coupling expansion.
It is valid at long distances whenever the theory is in the Higgs phase.%
\footnote
    {%
    More precisely, the long-distance EFT (\ref{eq:quantum_eff_action})
    neglects the instanton-monopole induced potential for the dual photon
    and, as such, is valid provided the mass $m_A$ (\ref{eq:m_A})
    generated by the Higgs mechanism is large compared to the
    monopole induced photon mass $m_\gamma$ (\ref{eq:mgamma}).
    }

The $f_1$ terms represent wavefunction renormalizations which simply
modify the overall normalizations in the energy density
\eqref{eq:energy_long_distance}, and have no effect on the flux
quantization condition \eqref{eq:fluxval}.  When
evaluated on the vortex, the $f_2$ terms also have the same form as
the long distance energy density \eqref{eq:energy_long_distance}.
The $f_3$ and $f_4$ terms produce a
$1/r^2$ contribution to the vortex energy density proportional to
%The $f_2$ term produces a $1/r^2$
%contribution to the energy density of the form
%\begin{align}
%    f_2  
%    \left[ \phi_0 (D_{\mu} \phi_{+}) (D^{\mu} \phi_{-}) +\textrm{h.c.}\right]
%    \propto 
%    \frac{v_0 \, v_c^2 }{r^2} \left( n - k - \frac{\Phi}{2\pi} \right) 
%\left(-n +\frac{\Phi}{2\pi} \right) ,
%\end{align}
\begin{align}
\frac{v_c^2 }{r^2} \left( n - k - \frac{\Phi}{2\pi} \right) \left(n -\frac{\Phi}{2\pi} \right) = \frac{v_c^2 }{2r^2}\left[  \left(n -\frac{\Phi}{2\pi} \right)^2 + \left( n - k - \frac{\Phi}{2\pi} \right)^2-k^2\right]
\end{align}
Hence, up to holonomy-independent terms, 
the $f_3$ and $f_4$ terms also merely change the normalization of
the tree-level energy density \eqref{eq:energy_long_distance}.
Therefore, provided fluctuations are not strong enough to flip its
overall sign, the holonomy-dependent $1/r^2$ energy density has minima
(with respect to $\Phi$) at $\Phi = (2n-k)\pi$. In particular,
\emph{all} minimal-circulation ($k=\pm1$) vortices which minimize the
quantum-corrected long-distance energy density carry flux $\Phi = \pi$
modulo $2\pi$. If fluctuations do flip the sign in front of
Eq.~\eqref{eq:energy_long_distance}, then the energy density becomes
unbounded below as a function of $\Phi$, with no additional local
minima appearing. The EFT description \eqref{eq:quantum_eff_action}
therefore breaks down, signaling the departure from the Higgs phase.
Therefore, within the Higgs phase, the fluctuation-induced corrections
to the effective action have no effect on the flux quantization
condition \eqref{eq:fluxval}.

This shows that the minimal vortex expectation value $\langle
\Omega(C) \rangle_{1}$  at large distance remains real and negative to
all orders in perturbation theory, provided that the fluctuations are
not so large that they completely destroy the Higgs phase. The size of
quantum fluctuations in this model is controlled by the dimensionless
parameter $e^2/m_A = \mathcal O(e\lambda_c^{1/2}/|m_c|) =\mathcal
O(e^2/|m_c|)$, where we have assumed $\lambda_c^{1/2} \sim \epsilon^{1/3} \sim e$ and $g_0, g_c \ll 1$ for
simplicity, and hence this conclusion about a negative value of
$\langle \Omega(C) \rangle_{1}$ holds exactly
%  in perturbation theory 
whenever $m^2_c/e^4$ is
sufficiently negative to put the theory into the Higgs phase.

As discussed earlier, quantum fluctuations do suppress the magnitude
of holonomy expectation values leading to perimeter law exponential
decay. By construction, this size dependence cancels in our ratio
$O_{\Omega} = \langle \Omega(C)\rangle_{1}/ \langle \Omega(C)\rangle
$. Unbroken $\ZF$ symmetry
(or $\ZC$, or reflection symmetry)
in the vacuum state guarantees that the
ordinary expectation value $\langle \Omega(C)\rangle $  in the
denominator is real.  It is easy to check that it is positive at tree
level, and sufficiently small quantum fluctuations cannot make it
negative.  So $O_{\Omega}$ is determined by the phase of the vortex
state holonomy expectation value in the numerator. The net result from
this argument is that within the Higgs phase,
%%% at least when $m^2_c/e^4$ is sufficiently large and negative,
\begin{align}
    \boxed{\textrm{Higgs phase:}\;\; O_{\Omega} = -1} \,,
\label{eq:holonomy_higgs}
\end{align}
holds precisely.
The next subsection gives useful alternative perspectives on the same conclusion. 

%%%%%%%%
\subsubsection{Vortex junctions, monopoles, and vortex flux quantization}
\label{sec:vortex_junctions}
%%%%%%%%

In the preceding section we analyzed the physics of vortices using
effective field theory in the bulk $3$-dimensional spacetime.   This
analysis showed that the minimal energy vortices carry quantized
magnetic flux $\pm \pi$, and the phase of the holonomy around vortices
is quantized, leading to result \eqref{eq:holonomy_higgs}. We now reconsider the
same physical questions from the perspective of an effective field
theory defined on the vortex worldline. This will lead to a discussion
of vortex junctions, their interpretation as magnetic monopoles, a
connection between vortex flux quantization and Dirac charge
quantization, and finally to distinct logically independent arguments
for the result \eqref{eq:holonomy_higgs}.

The $(0{+}1)$ dimensional effective field theory describing
fluctuations of a vortex worldline includes two gapless modes arising
from the translational moduli representing the spatial position of
the vortex.  The vortex effective field theory must include an
additional real scalar field which may be chosen to equal
the magnetic flux $\Phi$ carried by a vortex configuration.
This field will serve as a coordinate along field configuration
paths which interpolate between distinct vortex solutions.
The field $\Phi$
appears in the 1D worldline EFT in the form
\begin{align}
    S_{\textrm{vortex EFT}}
    = \int dt
	\left[c_K \,(\partial_t \Phi)^2 + c_V \,V(\Phi) \right]
    + \cdots \,.
    \label{eq:vortex_EFT}
\end{align} 
Here $t$ is a coordinate running along the vortex worldline, $\Phi$ is
dimensionless,  $c_K$ and $c_V$ are low-energy constants with
dimensions of inverse energy and energy, respectively, and the
ellipsis represents terms with additional derivatives or couplings to
other fields on the worldline.  

The worldline potential $V(\Phi)$ in expression \eqref{eq:vortex_EFT} obeys two important
constraints.  First, since $\ZF$ symmetry acts on $\Phi$ by $\Phi \to
-\Phi$, $V(\Phi)$ is an even function.  Second, Dirac charge
quantization in the underlying bulk quantum field theory 
%implies that
%any two distinct minima $\Phi_1, \Phi_2$ of $V(\Phi)$ must obey
%$\Phi_1 - \Phi_2 \in 2\pi \mathbb{Z}$.  
further constrains the possible minima of $V(\Phi)$. To see this, suppose that
$V(\Phi)$ has a minimum at $\Phi = \Phi_{\rm min} \neq 0$.  Since $V(\Phi)$ is
an even function, it must also have a distinct minimum at $\Phi = -\Phi_{\rm
min}$. For generic values of the microscopic parameters, the potential $V$ is
finite for all finite values of $\Phi$. This means that there exists a solution
to the equation of motion for $\Phi$ in which $\Phi$ interpolates between
$-\Phi_{\rm min}$ and $\Phi_{\rm min}$ as the worldline coordinate $t$ runs from
$-\infty$ to $+\infty$.  Suppose that this tunneling event has an action which
is both UV- and IR-finite, so that it is meaningful to describe it within the worldline effective field theory.  What is its interpretation in bulk
spacetime? It has unit $\UG$ circulation at all times, but also possesses a
``junction'' at some finite time where the magnetic flux changes sign. For the tunneling event to have finite action, the azimuthal component of the electric field far from the vortex core must decay faster than $1/r$. Then the flux
of the field strength through a $2$-sphere surrounding the junction is simply
$\Phi_{\rm min} - (-\Phi_{\rm min}) = 2\Phi_{\rm min}$. Comparing this to the
Dirac charge quantization condition in \eqref{eq:flux_quant} implies that
$\Phi_{\rm min} \in \pi \mathbb{Z}$ when $\ZF$ is unbroken.\footnote{If $\ZF$
symmetry is explicitly broken, Dirac charge quantization together with the
assumption that tunneling events have finite action leads to the conclusion
that any two distinct minima $\Phi_1, \Phi_2$ of $V(\Phi)$ must satisfy $\Phi_1
- \Phi_2 \in 2\pi \mathbb{Z}$.}  These remarks imply that the worldline
tunneling events can be interpreted as monopole-instantons in the 3d bulk, and
their action must depend on the UV completion of our compact Abelian gauge
theory.\footnote{Appendix~\ref{sec:nonAbelian} describes an explicit $SU(2)$ gauge theory
which reduces to our $U(1)$ gauge theory at long distances, and where $\SI \sim
m_W/e^2$ with $m_W$ the $W$-boson mass.}

%%%%%%%%%%%%%%
\begin{figure}    
    \centering
    \includegraphics[width=.2\textwidth]{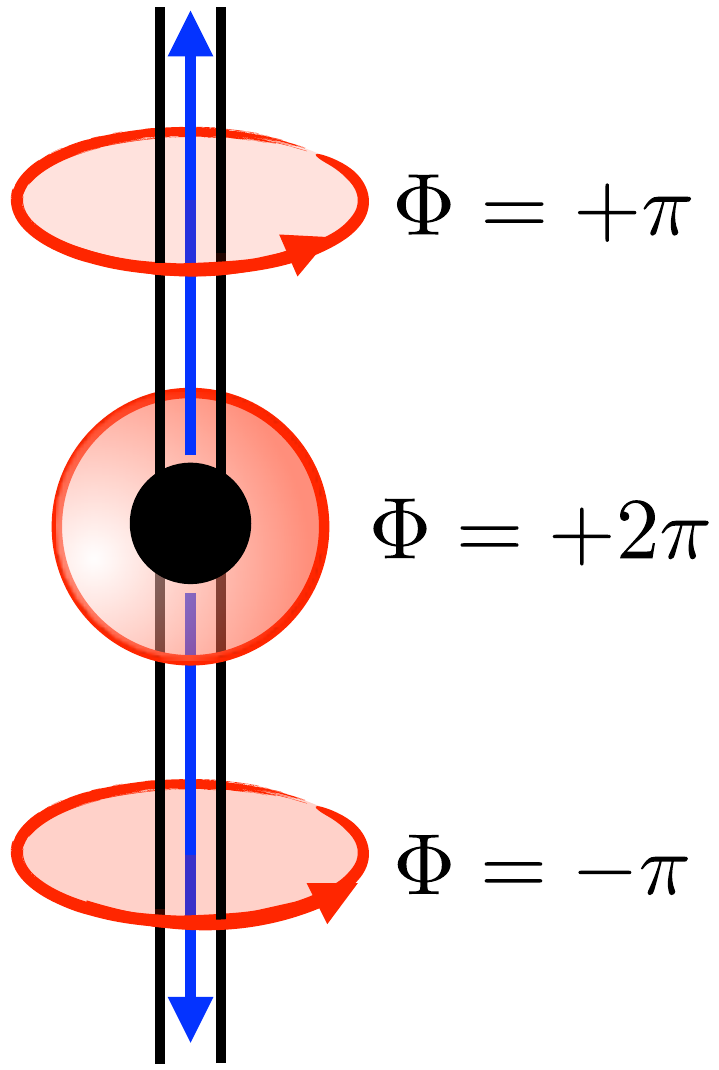}
    \caption
        {%
        A junction between the two minimal energy unit-winding vortex
        worldlines is a magnetic monopole with flux $2\pi$.
	\label{fig:single_monopole}
        }
 \end{figure}
%%%%%%%%%%%%%%

In the preceding section, we saw that in the Higgs phase
minimal-energy unit-circulation vortices carry magnetic flux $\pm\pi$
at tree-level.  The vortex flux quantization argument in the paragraph
above implies that quantum corrections cannot change this
result, again leading to result \eqref{eq:holonomy_higgs}.   We also learn that
a junction between two minimal-energy unit-circulation vortices with
flux $\pi$ and $-\pi$ can be interpreted as a magnetic monopole
carrying the minimal $2\pi$ flux consistent with Dirac charge
quantization, as illustrated in Fig.~\ref{fig:single_monopole}. This
is the Higgs phase version of a single monopole-instanton, discussed
earlier, when the $\phi_0$ condensate has unit winding.

\begin{figure}
\centering
\includegraphics[width=.7\textwidth]{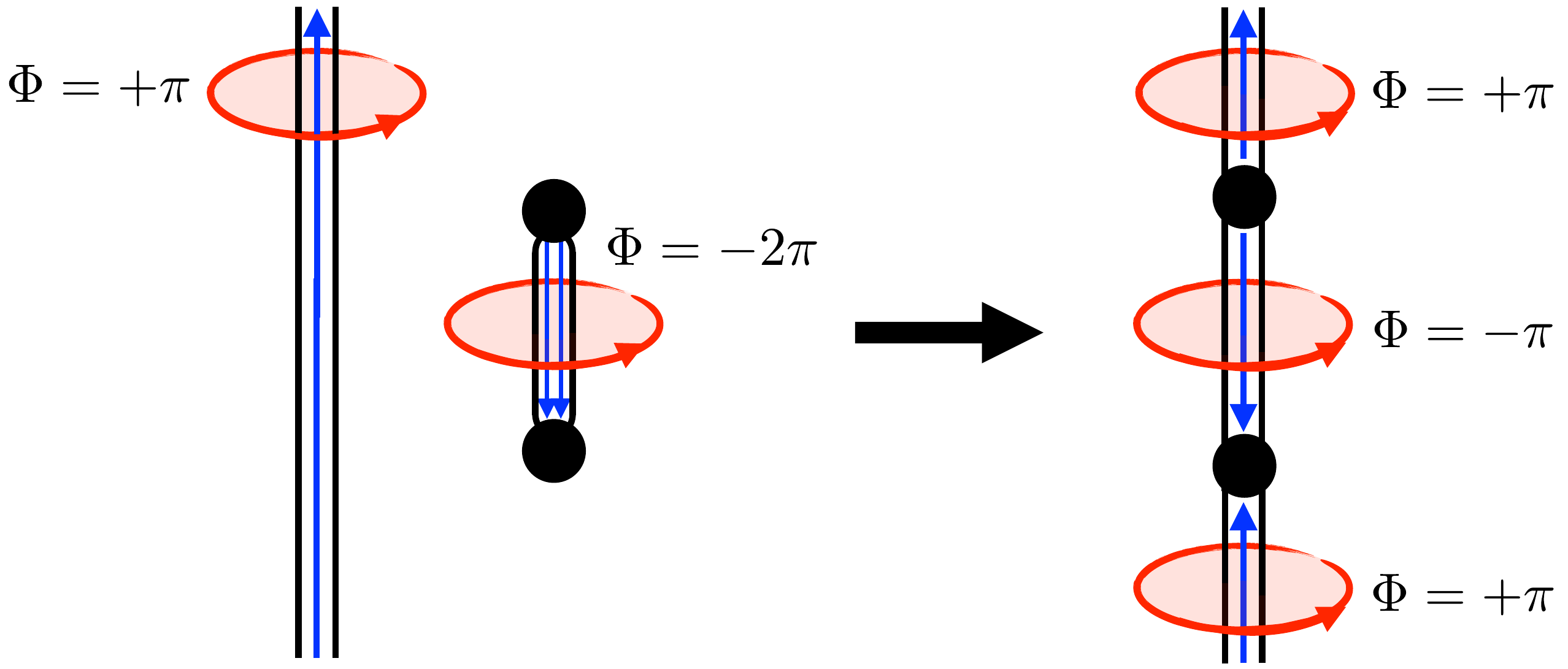}
\caption{
    Monopole--antimonopole pairs with minimal magnetic flux $2\pi$
    are confined in bulk spacetime, but such pairs are attracted
    to the worldline of a minimal global vortex where they become
    deconfined.
    \label{fig:monopole_anti_monopole}
    }
\end{figure}

As noted earlier near the end of Sec.~\ref{sec:our_model},
Higgs phase monopole--antimonopole pairs are connected by magnetic flux tubes
(which can break at sufficiently large separation due to
monopole--antimonopole pair creation). This is true in the absence of
any vortices carrying unit $\UG$ winding. But in the presence of a
unit circulation vortex, a monopole--antimonopole pair can bind to
the vortex, with the monopole and antimonopole then free to separate
arbitrarily along the vortex worldline.%
\footnote
    {%
    Deconfinement of magnetic monopoles on both local
    and semilocal vortices with and without supersymmetry has been
    extensively studied previously. In our model the vortices are global
    but the monopole deconfinement mechanism described here is essentially
    identical to previous discussions in, for example,
    Refs.~\cite{Hindmarsh:1985xc,Tong:2003pz,Shifman:2004dr,Hanany:2004ea,
    Eto:2009tr,Eto:2009kg,Cipriani:2011xp,Gorsky:2011hd,Chatterjee:2019zwx}.
    }
This is illustrated in Fig.~\ref{fig:monopole_anti_monopole}. To see
this, note that for fixed separation $L$ between monopole and
antimonopole, the action will be lowered if the monopole and
antimonopole move onto the vortex line, provided they are oriented
such that adding the monopole--antimonopole flux tube to the vortex
magnetic flux has the effect of merely flipping the sign of vortex
magnetic flux on a portion of its worldline. This eliminates the cost
in action of the length $L$ flux tube initially connecting the
monopole and antimonopole. As noted above, the $\ZF$ symmetry
guarantees that the vortex action per unit length is independent of
the sign of the magnetic flux. Once the monopole and antimonopole are
bound to the vortex worldline, there is no longer any cost in action
(neglecting exponentially falling short distance effects) to separate
the monopole and antimonopole arbitrarily.   In summary,  the
monopole--antimonopole string tension vanishes on the vortex, and
magnetic monopoles are deconfined on minimal Higgs phase vortices.%
\footnote
    {%
    Provided monopoles and antimonopoles alternate along the
    vortex worldline.
    There is a direct parallel between this phenomenon and
    charge deconfinement in 2D Abelian gauge theories
    at $\theta = \pi$, see for example Refs.~\cite{Coleman:1975pw,Coleman:1976uz,Witten:1978ka,Anber:2018jdf,Anber:2018xek,Armoni:2018bga,Misumi:2019dwq}.
    }

%%%%%%%%%%%%%%
%%%%%%%%%%%%%%

One can also regard the monopole--antimonopole pair as an
instanton--antiinstanton pair in the worldline EFT
\eqref{eq:vortex_EFT}.\footnote{The deconfinement of magnetic
monopoles on unit-circulation vortices corresponds to the fact that
the separation of an instanton--antiinstanton pair is a quasi-zero
mode.} We now argue that this perspective leads to yet another derivation
of the result \eqref{eq:holonomy_higgs}. The existence of degenerate global
minima with flux $\pm\Phi_{\rm min}$ means that the $\ZF$ symmetry is
spontaneously broken on the worldline to all orders in perturbation
theory.  But non-perturbatively, the finite-action worldline
instantons connecting these minima  will proliferate and restore the
$\ZF$ symmetry. As is familiar from double-well quantum
mechanics, the unique minimal energy vortex state will be a symmetric
linear combination of $\Phi_{\rm min}$ and $-\Phi_{\rm min}$
configurations. 

From our previous arguments we know that $\Phi_{\rm min} = \pi$,
so that both of these vortex configurations have the same $-1$ long
distance holonomy, and none of this non-perturbative physics has any
effect on the validity of the result \eqref{eq:holonomy_higgs}
regarding Higgs phase vortices.  But suppose that we did not already
know that $\Phi_{\rm min} = \pi$. The existence of finite-action
tunneling events connecting the two $\Phi$ minima would imply
that the minimal energy vortex state with a given winding number is
unique and invariant under $\ZF$. Unbroken $\ZF$ symmetry in turn
implies that the holonomy expectation value in the minimal vortex
state is purely real. Therefore, on symmetry grounds alone, our
observable $O_\Omega$ is quantized to be either $+1$ or $-1$. Our
analysis in the weakly coupled regime serves to establish that in the
Higgs phase the value is $-1$, and we again arrive at
result \eqref{eq:holonomy_higgs}.

%%%%%%%%%%%
\subsection{$O_\Omega$ in the $\UG$-broken confining regime}
\label{sec:holonomy_confining}
%%%%%%%%%%%

We now turn to a consideration of holonomies around vortices in the
$\UG$-broken confining phase.
Once again, it is useful to consider the appropriate effective
field theory deep in this regime, near the SE corner of the
phase diagram of Fig.~\ref{fig:3D_phase_diagram}.

Suppose that $m_c^2 \gg e^4$. Given the scale separation, it is useful
to to integrate out the charged fields.  The resulting effective action
retains the gauge field and neutral scalar $\phi_0$ and
has the form
\begin{align}
    S_{\rm eff} = \int d^{3}x\,
    &\left[ \frac{1}{4 e^2}\,  F_{\mu\nu}^2 + \Vm(\sigma) + |\partial_\mu \phi_0|^2 + V(|\phi_0|) 
    + \frac{a}{m_c^2} \, |\phi_0|^2 F_{\mu\nu}^2 
    + \cdots \right]\,,
\label{eq:effective_action}
\end{align}
where the ellipsis denotes higher dimension terms involving
additional powers of fields and derivatives.
The dimension five term shown explicitly, with coefficient $a$,
is the lowest dimension operator coupling the gauge and
neutral scalar fields.
This term describes ``Raleigh scattering'' processes in which
photons scatter off fluctuations in the magnitude of $\phi_0$.
Within this EFT, the $\ZF$ symmetry
simply flips the sign of the gauge field and
hence forbids all terms involving odd powers
of the gauge field strength.

When $m_0^2$ is sufficiently negative so that the
$\UG$ symmetry is spontaneously broken and $\phi_0$ condenses,
the leading effect of the $|\phi_0|^2 F^2$ coupling is merely
to shift the value of the gauge coupling by an amount
depending on the condensate $v_0 \equiv \langle \phi_0 \rangle$,
\begin{equation}
    \frac 1{e^2}
    \to
    \frac 1{e'^{\,2}}
    \equiv
    \frac 1{e^2}
    + \frac {4a \, |v_0|^2}{m_c^2} \,.
\label{eq:shift}
\end{equation}
This is a small shift of relative size $\mathcal O(e^4/m_c^2)$
within the domain of validity of this effective description.
The $\ZF$ symmetry (or parity) guarantees that the 
neutral scalar condensate cannot source the gauge field strength,
so the magnetic field 
$B \equiv \half \epsilon_{ij} F^{ij}$ ($i,j = 1,2$)
must have vanishing expectation value.

Within this $\UG$ broken phase,
there are vortex configurations in which
the condensate $\langle \phi_0 \rangle$
has a phase which winds around the vortex,
while its magnitude decreases in the vortex core,
vanishing at the vortex center. 
As far as the gauge field is concerned,
one sees from the effective action (\ref{eq:effective_action})
that the only effect this has is to modulate
the gauge coupling,
effectively undoing the shift (\ref{eq:shift}) in the vortex core.
But such coupling renormalizations, or dielectric effects,
do not change the fact that the effective action is an 
even function of magnetic field which is minimized
at $B = 0$.
In other words, even in the presence of vortices,
the neutral scalar field does not source
a magnetic field.
And consequently, both the vacuum state \emph{and}
minimal energy vortex states are invariant under the
$\ZF$ symmetry.

Once again, invariance of the both the vacuum and
vortex states under the $\ZF$ symmetry implies
that holonomy expectation values in both states are real,
and hence our observable $O_\Omega$ must be either
$+1$ or $-1$.
The Abelian gauge field holonomy is, of course,
nothing but the exponential of the magnetic flux,
$
    \Omega(C)
    =
    e^{i \oint_C A}
    =
    e^{i \int_S B}
    =
    e^{i \Phi_B}
$
(with contour $C$ the boundary of disk $S$).
The above EFT discussion shows that
deep in the confining $\UG$-broken phase
the influence of a vortex on the magnetic field
is tiny and hence $\langle \Omega(C) \rangle_1$ is
positive,
implying that $O_\Omega = +1$.
And once again, by analyticity, this result
must hold throughout the confining $\UG$-broken phase.
In summary,
\begin{align}
    \boxed{\textrm{$\UG$-broken confining phase:}\;\; O_\Omega = +1} \,,
\end{align}
is an exact result within this phase.

%%%%%%%%%%%%
\subsection{Higgs-confinement phase transition}
\label{sec:ColemanWeinberg}
%%%%%%%%%%%%

We have seen that $O_\Omega$ has constant magnitude but changes sign
between the Higgs and confining, $\UG$-broken regimes; it cannot be a
real-analytic function of $m_c^2$. Hence, there must be at least one
phase transition as a function of $m_c^2$.  A single phase transition
would be associated with an abrupt jump of $O_{\rm \Omega}$ from $-1$
to $1$ at some critical value of $m_c^2$.
%%% Indeed, thanks to the
%%% observation in Sec.~\ref{sec:vortex_junctions} that the minimal-energy
%%% unit-winding vortex state is invariant under $\ZF$, 
%%% the numerator and denominator in $O_{\Omega}$
%%% must be real, and since their magnitudes are identical by construction,
%%% unbroken $\ZF$ symmetry implies $O_{\rm \Omega} = \pm 1$.
If instead $O_{\rm \Omega}$
equals $-1$ for charged mass-squared below some value, $m_c^2 <
(m_c^2)_A$, equals $+1$ above a different value $(m_c^2)_B < m_c^2$,
and 
continuously interpolates from $-1$ to $+1$ in the intervening interval
$(m_c^2)_A < m_c^2 < (m_c^2)_B$, this would indicate the presence of
two phase transitions bounding an intermediate phase in which the
$\ZF$ symmetry is spontaneously broken.
(This follows since, as discussed above, unbroken $\ZF$ symmetry implies
that $O_\Omega$ must equal $\pm 1$.)

In much of parameter space, phase transitions in our model occur
at strong coupling and are not amenable to analytic treatment.
But the theory becomes weakly coupled
when the masses $|m_c^2|$ and $|m_0|^2$ are sufficiently large.
Specifically, we will assume that
the dimensionful couplings $|\lambda_c|$, $|\lambda_0|$
and $e^2$ are all small relative to the masses $|m_c|$ and $|m_0|$, 
the cubic coupling obeys $\epsilon \ll \textrm{min}(|m_c|^{3/2},|m_0|^{3/2})$,
and the sextic couplings are small, $g_c, g_0 \ll 1$.
If a first order transition lies within this region,
then simple analytic arguments suffice to identify
and locate the transition.

A first-order transition involving a complex scalar $\phi$ with $U(1)$
symmetry requires multiple local minima in the effective potential
viewed as a function of $|\phi|$. In four dimensions, a renormalizable
scalar potential is quartic and, as a function of $|\phi|$, has at
most a single local minimum. So to find a first-order phase transition
in a weakly coupled four-dimensional $U(1)$ invariant scalar theory
one must either be abnormally sensitive to higher order
non-renormalizable terms (and thus probing cutoff-scale physics), or
else reliant on a one-loop or higher order calculation producing
non-analytic terms like $|\phi|^2 \log |\phi|$. This is illustrated by
the classic Coleman and Weinberg analysis \cite{Coleman:1973jx}. But
in three spacetime dimensions, renormalizable scalar potentials are
sextic, and $U(1)$ invariant sextic potentials can easily have
multiple local minima. Consequently, a tree-level analysis can suffice
to demonstrate the existence of a first-order phase transition, in a
renormalizable theory, without any need to consider higher-order
corrections. 

\begin{figure}
\centering
\includegraphics[width=\textwidth]{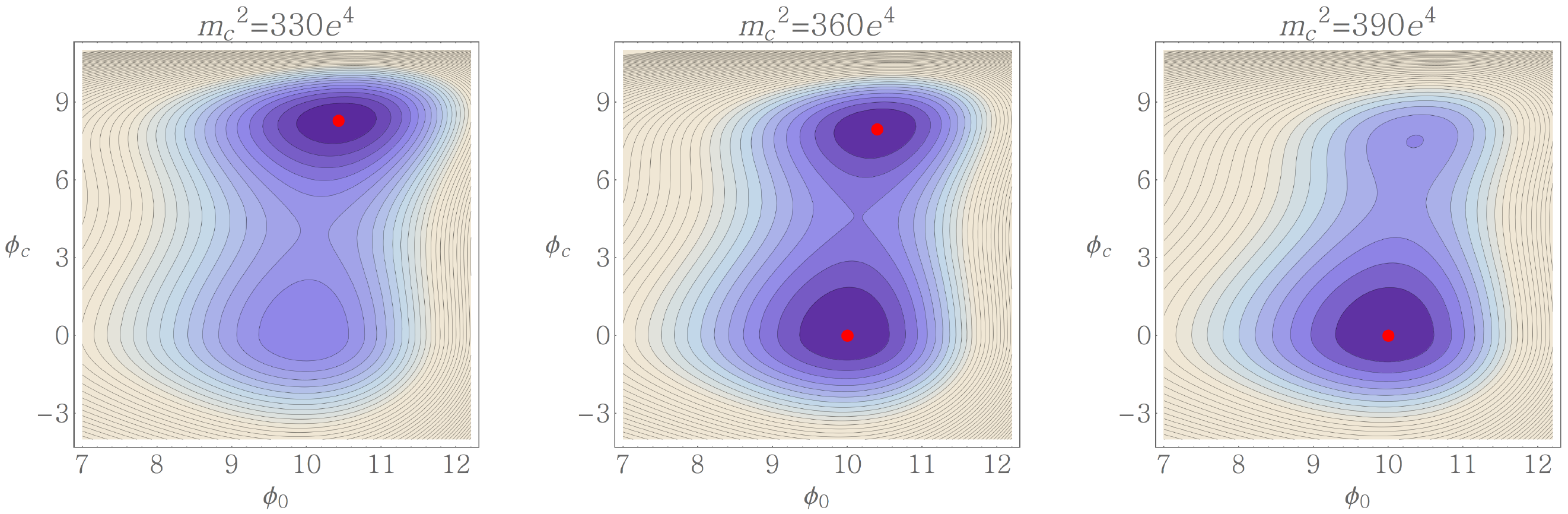}
\caption
    {Contour plots of the tree-level scalar effective potential at
    three different values of $m_c^2$ in the vicinity of the
    first-order Higgs-confinement phase transition.   We have used
    gauge and global symmetries to choose the phases of the scalar
    fields such that the potential can be interpreted as a function of
    $\phi_c \equiv \phi_{+}$ and $\phi_0$, with $\phi_- = |\phi_+|$.
    We have set $m_0^2 = -200 \, e^4$, $\epsilon = 40 \, e^3$, $\lambda_c =
    \lambda_0 = -5 \, e^2$, and $g_c = g_0 = 0.04$. Decreasing values of
    the scalar potential are colored with darker colors, and
    global minima are marked with red dots.  Note that the global
    minimum is degenerate when $m_c^2 \approx 360 \, e^4$, and the
    location of the global minimum jumps as $m_c^2$ crosses this
    value, from a point where the charged fields are condensed to one
    where they are not condensed.  This shows the presence of a strong
    first-order Higgs-confinement phase transition, with the $\UG$
    global symmetry spontaneously broken on both sides of the
    transition.
    \label{fig:contours}
    }
\end{figure}

Let us see how this works in our model. Consider the region where
$m_0^2$, $\lambda_c$, and $\lambda_0$ are all negative.  For
simplicity, let us also suppose that $e^2 \ll |\lambda_c|,$
$|\lambda_0|$, and $\epsilon \ll e^3 \ll
\textrm{min}(|m_c|^{3/2},|m_0|^{3/2}) $. In Fig.~\ref{fig:contours} we
show contour plots of the scalar potential as a function of $\phi_c
\equiv \phi_{+}$ and $\phi_0$, with $\phi_- = |\phi_+|$, as
$m_c^2/e^4$ is varied.  The figure shows that the potential has
multiple local minima with relative ordering that changes as
$m_e^2/e^4$ is varied with all other parameters held fixed. With the
parameter choices given in the caption of Fig.~\ref{fig:contours}, the
figure shows the existence of a strong first-order phase transition
between $\UG$-broken confining and $\UG$-broken Higgs states in
the regime where $m_0^2/e^4$ is large and negative and  $m_c^2/e^4$ is
large and positive. Correspondingly, the change in the derivative of
the energy density with respect to the charged scalar mass squared in
units of $e^2$, $e^{-2} \Delta(\partial \mathcal E/\partial m_c^2)$, is
large across the transition.  For the parameter values
used in Fig.~\ref{fig:contours} one finds $e^{-2}\Delta(\partial \mathcal
E/\partial m_c^2) = 2\Delta \phi_c^2/e^2 \approx 127 \gg 1$.  This behavior
is generic. The effective masses (i.e., curvatures of the potential)
at the minima are comparable to the input mass parameters, so there
are no near-critical fluctuations and the phase transition is reliably
established at weak coupling.

Finally, the analysis of the previous subsections shows that our
vortex holonomy order parameter $O_\Omega$ changes sign
across this phase transition, confirming that the 
abrupt change in this ``topological'' order parameter 
is associated with a genuine thermodynamic phase transition.

As one moves into the interior of the $(m^2_0,m^2_c)$ phase diagram,
out of the weakly-coupled periphery, we certainly expect this direct
correlation between a jump in our vortex order parameter and a
thermodynamic phase transition to persist. But one may contemplate
whether this association could cease to apply at some point in the
interior of the $\UG$ spontaneously broken domain.
In general, a line of first order phase transitions which is not
associated with any change in symmetry realization can have a critical
endpoint (as seen in the phase diagram of water). Could our model have
such a critical endpoint, beyond which the first order transition
becomes a smooth cross-over as probed by any local observable? If so,
there would necessarily remain some continuation of the phase
transition line across which our topological observable $O_\Omega$
continues to flip sign, but all local observables remain smooth. What
would be necessary for such a scenario to take place?

First, note that the magnetic flux carried by vortices can change in
steps of $2\pi$ due to alternating monopole-instanton fluctuations
appearing along the vortex worldline, but such processes do not affect
the sign of the holonomy around a vortex.  At the transition between
the Higgs and confining phases the magnetic flux carried by
minimal-winding vortices changes by $\pi$ (modulo $2\pi)$.  It is very
tempting to expect such a sudden change in the vortex magnetic flux to
imply non-analyticity in the IR-finite core energy of a vortex, or
equivalently the vortex fugacity. Whenever the $\UG$ symmetry is
spontaneously broken, the equilibrium state of the system will contain
a non-zero density of vortices and antivortices due to quantum
fluctuations. If the minimal vortex energy is non-analytic this will
in turn induce non-analyticity in the true ground state energy
density.  (This argument ceases to apply only when the vortex density
reaches the point where vortices condense, thereby restoring the $\UG$
symmetry.) In other words, if non-analyticities in vortex magnetic
flux imply non-analyticity in the vortex energy, our vortex holonomy
observable functions as a useful order parameter, identifying
thermodynamically distinct gapless phases.

There is a possible loophole in the above argument: what if the change
in vortex magnetic flux is caused by a level crossing between 
vortices of flux $\pi$ and $0$ (mod $2\pi$)?\footnote{We are grateful to N.~Seiberg for
useful discussions on this issue.}
Such a level crossing could produce non-analyticy in our vortex
holonomy observable without being associated with non-analyticity
in the ground state energy or other thermodynamic observables.
However, for such a level crossing to be possible, a (metastable)
unit-winding vortex with flux $0$ (mod $2\pi$) would need to exist
in the Higgs phase and become degenerate with the flux $\pi$ (mod $2\pi$)
unit-winding vortex as one varies parameters. Our analysis of the quantum
effective action for the vortex holonomy shows that,
within the domain of validity of the effective action
\eqref{eq:quantum_eff_action},
there simply are no static solutions describing unit-winding vortices 
with flux equal to $0$ mod $2\pi$ in the Higgs phase.
The quantum effective action 
\eqref{eq:quantum_eff_action} is a valid long distance description
throughout the Higgs regime relying, essentially, only on a large
ratio of the distance scale of interest to microscopic scales.
However,
Eq.~\eqref{eq:quantum_eff_action} does not take into account
monopole-instanton effects, so it necessarily ceases to be applicable in
a transition region between the confining and Higgs regimes where
the Higgs mass scale $m_A \sim e^2 v_c^2$ becomes comparable to the
monopole-induced photon mass scale $m_{\gamma}^2 \sim ({\mu_{\rm
UV}^3}/{e^2}) \, e^{-\SI}$.
This region in parameter space can be
made arbitrarily small by increasing $\SI$. For the level-crossing
scenario to take place, one would need to envision that as we go from
the Higgs regime toward the confining regime, a flux $2\pi$
minimal-winding vortex has to appear with a higher energy than a
$\pi$-flux minimal-winding vortex, and then cross it in energy, all
within this arbitrarily small region. Moreover, this phenomenon would
have to take place \emph{only} in the strongly-coupled region of
parameter space, because it certainly does not happen in the
weakly-coupled domain, illusgrated in Fig.~\ref{fig:contours}, where
we have shown the existence of a first order phase transition.  So while we cannot
absolutely rule out this level-crossing scenario, in our view it requires
enough conspiracies to seem very far-fetched.

This concludes our arguments for the presence of at least one phase
transition curve separating the SE and W regions of
Fig.~\ref{fig:3D_phase_diagram}.

%%%%%%%%%%%%
\subsection{Explicit breaking of flavor permutation symmetry} 
\label{sec:broken_permutations}
%%%%%%%%%%%%

We now generalize our model to include
operators which break the
$\ZF$ symmetry explicitly.
The simplest such term is just a mass perturbation giving
the two charged fields $\phi_+$ and $\phi_-$ distinct masses
$m_+$ and $m_-$.
Let 
\begin{align}
    m^2_{\rm avg} &\equiv  \frac{1}{2} (m^2_{+}+m^2_{-})\,,
    \quad
    \Delta \equiv \frac{1}{e^4}(m^2_{+} - m^2_{-}) \,,
\end{align}
denote the average mass squared and a measure of their difference,
respectively. We will examine the dependence of physics on $m^2_{\rm
avg}/e^{4}$ with $\Delta >0$ held fixed.   

If $\Delta$ is sufficiently large then
there are two seemingly different regimes where no global symmetries
are spontaneously broken:
one where no scalar fields are condensed, and another where
only $\phi_{-}$ is condensed.
The latter regime is not a distinct phase as condensation of the
charged field $\phi_-$, by itself, does not imply a non-vanishing
expectation value of any physical order parameter.
In fact, these two regimes are smoothly
connected to each other and are trivial in the sense that they have a
mass gap and a vacuum state which is invariant under all global
symmetries. 

The more interesting regimes of the model are those with spontaneously broken
$\UG$ symmetry.  The cubic term in the action $\epsilon \phi_{0}
\phi_{+} \phi_{-} + \textrm{h.c.}$ ensures that there is no regime
where $\phi_0$ and only one of the two charged fields are condensed.
Hence we only need to consider two regimes with spontaneously broken
$\UG$ symmetry: one where all scalar fields are condensed, another
where only the neutral scalar $\phi_0$ is condensed.

%%%%%%%%
\subsubsection{Higgs regime}
%%%%%%%%
Consider the Higgs regime where $-m_{\rm avg}^2 \gg e^4$ and all
scalars are condensed. The tree-level long-distance
energy density that determines the holonomy around a
$\UG$ vortex of winding number $k$
is given by an obvious generalization of Eq.~\eqref{eq:energy_long_distance},
\begin{equation}
    \mathcal E(r)
    =
	\frac{v_+^2}{r^2}\left(n-k - \frac \Phi{2\pi}\right)^2 + \frac{v_-^2}{r^2}\left(-n + \frac \Phi{2\pi}\right)^2
    +
    \frac {v_0^2 \, k^2}{r^2} 
    + \mathcal O(r^{-4})
    \,.
\end{equation}
Due to the explicit breaking of $\ZF$, the magnitudes of the charged
scalar expectation values $v_+$ and $v_-$ are no longer equal;
let us denote their average by $v_{\rm avg}$. For given values of $k$ and
$n$, minimizing the above energy density yields 
\begin{equation}
    \Phi = \left(2n-k\frac{v_+^2}{v_{\rm avg}^2}\right)\pi,
\end{equation}
and $\mathcal E = \left(\frac{1}{2}\frac{v_+^2v_-^2}{v_{\rm
avg}^2}+v_0^2\right)k^2/r^2 + \mathcal O(r^{-4})$. Due to the explicit breaking
of $\ZF$, there are no longer two degenerate minimal-winding vortices
at tree-level. Suppose $v_-^2 < v_+^2$ without loss of generality. Then
the unique minimal energy unit-winding vortex (corresponding to
$k=1,n=1$) carries magnetic flux 
\begin{equation}
    \Phi = \frac{v_-^2}{v_{\rm avg}^2} \, \pi \,,
\label{eq:brokensymflux}
\end{equation}
which is no longer quantized in units of $\pi$.
This means that the holonomy encircling a vortex,
$\langle \Omega(C)\rangle_1 = e^{i\Phi}$,
is no longer real.

The ordinary holonomy expectation value in the denominator
of $O_\Omega$ necessarily remains real and positive due to
the continuing presence of unbroken $\ZC$ symmetry.
Consequently, in this tree-level analysis,
our nonlocal order parameter $O_\Omega$ is a non-trivial phase which
differs from both $-1$ and $+1$.
Small quantum corrections cannot bring the vortex magnetic flux
(\ref{eq:brokensymflux}) to $0$, so this conclusion must hold
generically throughout the phase which extends inward from the
weakly coupled regime. 
In particular,
\begin{align}
    \boxed{\textrm{Higgs phase without $\ZF$ symmetry:} \;\;O_{\Omega} \neq 1} \,.
\label{eq:O_Omega_HiggsWithoutZP}
\end{align}
Following the analysis in the next subsection, we will see that one can
actually interpret the condition \eqref{eq:O_Omega_HiggsWithoutZP} as a
gauge-invariant criterion defining the Higgs phase.

%%%%%%%
\subsubsection{$\UG$-broken confining regime}
%%%%%%%%

Now consider the regime where neither charged scalar field is condensed.
When $m_{\rm avg}^2$ is large (compared to other scales)
one may integrate out both charged fields and the effective 
description of the theory is given by Eq.~\eqref{eq:effective_action}, with $m_c$
now defined as the mass of the lightest charged field, $m_c = \min(m_+,m_-)$,
plus additional higher dimension operators which are no
longer forbidden by the $\ZF$ symmetry.
Writing out the lowest dimension such term explicitly, we have
\begin{align}
\label{eq:broken_effective_action}
    S_{\rm eff} = \int d^{3}x &
    \left[
	\frac{1}{4e^2} \, F_{\mu\nu}^2
	+ \Vm(\sigma)
	+|\partial_\mu \phi_0|^2 + V(|\phi_0|)
	+  \frac{a}{m_c^2} \, |\phi_0|^2  F_{\mu\nu}^2
	+  \frac{b}{m_c^2} \, S_{\mu\nu} F^{\mu\nu}
	+ \cdots
    \right],
\end{align}
where the ``polarization''
$
    S_{\mu\nu} \equiv
    \frac{i}{2}\big[
	(\partial_\mu\phi_0^\dagger)(\partial_\nu\phi_0)
	-(\partial_\nu\phi_0^\dagger)(\partial_\mu\phi_0)
	\big]
$.
To examine the effect of this new dimension-5 $\ZF$-odd term $S \cdot F$
in the presence of vortices,
it will prove helpful to integrate by parts and rewrite it as
direct coupling between the gauge field and a current,
$\int d^3x \> A_\mu J^\mu_{\rm eff}$,
with the current built out of gradients of the neutral scalar $\phi_0$,
\begin{equation}
\label{eq:bound_current}
    J^{\mu}_{\rm eff} = \frac{2b}{m_c^2} \, \partial_{\nu} S^{\mu\nu}  \,.
\end{equation}
This current is automatically conserved,
$\partial_{\mu} J^{\mu}_{\rm eff} = 0$,
as required by gauge invariance.%
\footnote
    {%
    Alternatively,
    one might be tempted to eliminate this term, which induces
    mixing between $S_{\mu\nu}$ and $F_{\mu\nu}$,
    by making a suitable redefinition of the gauge field.
    But for our purposes such a field redefinition is unhelpful
    as it complicates the evaluation of holonomies,
    effectively introducing a current-current interaction between
    the $\UG$ current and the current associated with a heavy
    electrically-charged probe particle used to measure the
    holonomy.
    } 

Now consider the minimal vortex configuration where
the neutral scalar has a spatially varying magnitude and phase,
$\phi_0 = v_0 \, f_0(r)\, e^{i\theta}$.
This induces a non-zero antisymmetric $S_{\mu \nu}$ with
\begin{align}
    S_{r \theta} = v_0^2\, \frac{ f_0(r) f_0'(r)}{r} \,.
\label{eq:Svortex}
\end{align}
This polarization is localized on the vortex core
(with an $\mathcal O(r^{-4})$ power-law tail).
The associated current $J^\mu_{\rm eff}$ has an
azimuthal component,
$
    J^{\theta}_{\rm eff}(r)
    =
    \frac{2b\,v_0^2}{m_c^2} \, \partial_{r} \big[ {f_0(r)f_0'(r)}/r \big]
$.
As in any solenoid,
this current sources a magnetic field which is also localized 
within the vortex core,
i.e., $r \lesssim |m_0|^{-1}$, 
up to an $\mathcal O(r^{-4})$ tail.

What does all of this mean for holonomies around vortices?  There are
several distinct physical length scales in the $\UG$-broken confining
phase. Recall that the non-perturbative monopole-instanton induced
contribution to the action depends on the classical action $\SI$ of a
monopole-instanton and the scale $\mu_{\rm UV}$ which is set by the inverse
length scale of the monopole core,
$S_{\rm monopole} =\int d^3x\, \Vm(\sigma)= -\int d^3x\,
\mu_{\rm UV}^3 \, e^{-\SI} \cos(\sigma)$. This term is responsible for linear
confinement with a string tension
$T \sim  e^2 m_\gamma \sim \sqrt{e^2\mu_{\rm UV}^3}\, e^{-\SI/2}$.
Suppose that $T^{1/2} \ll m_c$, as is the case in the weakly coupled
portion of the phase. The possibility of charged scalar pair
production implies that sufficiently long strings can break. The
string-breaking length,
\begin{equation}
    L_{\rm br} \equiv 2 m_c/T \,,
\end{equation}
characterizes the length scale beyond which string-breaking effects
cannot be neglected. Hence, linear confinement and area law behavior
for Wilson loops only holds for intermediate distance scales between
$T^{-1/2}$ and $L_{\rm br}$.

For our purposes, the quantity of primary interest is the holonomy 
for a circular contour around
a vortex when the contour radius $r$ exceeds
the largest intrinsic scale of the theory, $r \gg L_{\rm br}$.
However, let us work up to this case by considering holonomies
calculated on circles of progressively increasing size.

Consider a circular contour $C$ with a unit-winding $\phi_0$ vortex at
its center. To begin, suppose that the radius $r$ of the contour $C$
is large compared to the coherence length $\xi \sim 1/|m_0|$ but small
compared to $T^{-1/2}$, the inverse dual photon mass. Then confinement
and monopole effects can be ignored, and a calculation of the magnetic
flux using Eqs.~\eqref{eq:broken_effective_action}--\eqref{eq:Svortex}
gives 
\begin{align}
    \langle \Omega(C) \rangle_1 = e^{-2\pi r \mu'} \, e^{i \Phi}
    \,, \qquad \xi \ll r \ll T^{-1/2} \,,
\label{eq:small_circle}
\end{align}
where $\mu'$ is a scheme dependent renormalization scale and the flux
is given by
\begin{align}
\label{eq:vme} 
\Phi = 2\pi b\left(\frac{e\, v_0}{m_c}\right)^2 \,.
\end{align}
So, for contours encircling vortices in this ``inner'' distance regime
(but still far outside the vortex core),
we find that 
\begin{align}
    \frac{\langle \Omega(C) \rangle_1}{\langle \Omega(C) \rangle}
    = e^{i \Phi} \,,  
    \qquad \xi \ll r \ll T^{-1/2} \,.
\end{align}

Next, suppose that the contour radius satisfies $T^{-1/2} \ll r \ll
L_{\rm br}$. The dual photon mass term is important in this regime. To
compute the behavior of the Wilson loop, we recall the usual
prescription of Abelian duality, see e.g.~%
Ref.~\cite{Polyakov:1976fu,Unsal:2008ch}: an electric Wilson loop
along a contour $C$ maps to a configuration of the dual photon with a
$2\pi$ monodromy on curves that link $C$. A very large Wilson loop in
the $x$-$y$ plane can be described by a configuration of $\sigma$
which, well inside the loop, is purely $t$-dependent, with $\sigma$
vanishing as $t \to \pm\infty$ while having a $2\pi$-discontinuity at
$t=0$. In the Abelian dual description, the effective action
\eqref{eq:broken_effective_action} then takes the form
\begin{align}
    S_{\rm eff}
    =
    \int d^{3}x\, &\left[
	|\partial \phi_0|^2
	+ V(|\phi_0|)
%%%	+ \frac{1}{2}\left(\frac{e}{2\pi}\right)^2
	+ \frac{e^2}{8\pi^2} \,
	    (\partial\sigma)^2
	    \Bigl( 1 - \frac {4a \, e^2}{m_c^2} \, |\phi_0|^2 \Bigr) \right.
	- \mu_{\rm UV}^3 \, e^{-\SI} \cos(\sigma)
\nonumber\\
    &\left.\vphantom{[|\partial \phi_0|^2}
%%	- \frac{2a\, e^2 }{m^2}\left(\frac{e}{2\pi}\right)^2|\phi_0|^2
%%	    (\partial\sigma)^2
	+  \frac{ib\, e^2}{2\pi m_c^2} \, \epsilon^{\mu\nu\rho} \, S_{\mu\nu}
		\, \partial_{\rho} \sigma
	+ \cdots \right] \,.
\label{eq:dual_effective_action}
\end{align}
On the vortex configuration, the final $b$ term becomes
\begin{equation}
    \frac{ib\, e^2}{2\pi m_c^2} \, 
	\epsilon^{\mu\nu\alpha} \, S_{\mu\nu} \,
	\partial_\alpha\sigma
    =
    \frac{ib}{\pi} \, \left(\frac{ev_0}{m_c}\right)^2 \frac{ f_0(r) f_0'(r)}{r}
	\,{\partial_t\sigma} \,.
\end{equation}
Since the vortex configuration is time-independent,
the integral of this term only receives
a contribution from the $2\pi$ discontinuity in $\sigma$ at $t = 0$.
Evaluating the effective action \eqref{eq:dual_effective_action} on this
solution gives a result for the holonomy expectation value of
\begin{align}
    \langle \Omega(C) \rangle_1  = e^{-2\pi r \mu'} e^{-T \pi r^2} e^{i \Phi}
    \,, \qquad  T^{-1/2} \ll r \ll L_{\rm br} \,,
\end{align}
showing area-law decrease in magnitude together with the same phase 
\eqref{eq:vme} appearing in smaller holonomy loops.
Of course, without a vortex the $b$ term vanishes and
the holonomy expectation shows pure area-law decrease with no phase,
\begin{align}
    \langle \Omega(C) \rangle  = e^{-2\pi r \mu'} e^{-T \pi r^2}
    \,, \qquad  T^{-1/2} \ll r \ll L_{\rm br} \,.
\end{align}
Consequently, for this ``intermediate'' range of circle sizes we again find
\begin{align}
    \frac{\langle \Omega(C) \rangle_1}{\langle \Omega(C) \rangle} = e^{i \Phi}
    \,, \qquad  T^{-1/2} \ll r \ll L_{\rm br} \,.
\end{align}

Now we are finally ready to consider the most interesting regime of
holonomy contours, those with $r \gg L_{\rm br}$.
First, consider the unconstrained vacuum expectation value.
Due to the presence of heavy dynamical charged excitations,
Wilson loop expectation values 
contain a sum of area-law and perimeter-law contributions,
but the perimeter-law contribution dominates in the long-distance regime,
\begin{align}
    \langle \Omega(C) \rangle
    = e^{-2\pi r \mu'} \big( e^{-T \pi r^2} + e^{-2\pi r m_c } \big)
    \sim e^{-2\pi r(m_c + \mu') }
    \,, \qquad L_{\rm br} \ll r \,,
\label{eq:vachol}
\end{align}
(Here, irrelevant prefactors are neglected.) Physically, this Wilson
loop expectation describes a process where a unit test charge and
anticharge are inserted at some point, separated, and then recombined
after following semicircular worldlines (in Euclidean space) forming
two halves of the contour $C$. The second perimeter-law term arises
from contributions in which dynamical charges of mass $m_c$ are
pair-created and dress the test charge and anticharge to create two
bound gauge-neutral ``mesons''. These mesons have physical
size of order $\ell_{\rm
meson} \sim \textrm{min}(T^{-1/2},(e^2 m_c)^{-1/2})$, and experience
no long range interactions.\footnote{When $m_c \gg T^{1/2}/e^2$ the dressed test
charges are analogous to charmed $B$ mesons in QCD, and can be described
as $2{+}1$D Coulomb bound states, see e.g. Ref.~\cite{Aitken:2017ayq}. }
Once the loop size exceeds $L_{\rm br}$,
pair creation of dynamical charges of mass $m_c$ and the associated
meson formation becomes the dominant process.

Finally, suppose that this very large contour $C$ encircles a minimal
vortex. Then the area-law contribution to the holonomy expectation
acquires the phase $\Phi$, in exactly the same manner described above.
In contrast, the perimeter-law contribution arises from fluctuations
of the charged fields within distances of order of $\ell_{\rm
meson}$ from any point on the contour $C$. The amplitude for such
screening fluctuations, and consequent meson formation, must be
completely insensitive to the presence of a vortex very far away at
the center of the loop. Consequently, in the presence of a vortex the
two different contributions to the holonomy expectation value have
different phases,
\begin{align}
    \langle \Omega(C) \rangle_1 =
    e^{-2\pi r \mu'} \big( e^{-T \pi r^2} e^{i\Phi} + e^{-2\pi r m_c }  \big)
    \,.
\label{eq:vorhol}
\end{align}
Once again, in the long distance regime, $r \gg L_{\rm br}$ the
string-breaking or perimeter-law term dominates.

Combining the vortex holonomy expectation (\ref{eq:vorhol})
with the vacuum expectation (\ref{eq:vachol}),
we find that their ratio, in the long distance regime,
equals 1 up to exponentially small corrections,
\begin{align}
    \frac{\langle \Omega(C) \rangle_1}{\langle \Omega(C) \rangle}
    &= 
    1 + \mathcal O\Big(e^{-T \pi r^2 (1-L_{\rm br}/r)} \left(e^{i \Phi}-1\right)\Big) \,.
%%%    + \cdots\nonumber\\  
%%%    &= 1+ \exp\left[-TL^2\left(1-\frac{L_{\rm br}}{2L}\right) \right] \left(e^{i \Phi}-1\right) + \cdots
\end{align}
Hence, the large $r$ limit defining our vortex observable $O_\Omega$
exists and yields the simple result:
\begin{align}
    \boxed{\textrm{$\UG$-broken confining phase without $\ZF$ symmetry:}
    \;\;O_{\rm \Omega} = +1} \,.
\end{align}
Being strictly constant (i.e., with no dependence
whatsoever on microscopic parameters), this result must hold exactly
throughout the phase connected to the weakly-coupled confining
$\UG$-broken regime. \emph{Any} deviation from $O_\Omega = +1$ must
signal a phase transition.

\subsubsection{Summary}

Let us take stock of what we have learned about the relation between
the Higgs and confining $\UG$-broken regimes in the absence of the $\ZF$
symmetry.  So long as the $\UG$ global symmetry is spontaneously broken,
there is no way to distinguish the Higgs and confining regimes
within the Landau paradigm using local order parameters.
But our vortex holonomy order parameter \emph{does} 
%continue to
distinguish them!
Consider the theory with large positive $m_{\rm avg}^2$, in its
regime where $\UG$ is spontaneously broken
due to the dynamics of the neutral scalar sector,
and imagine progressively decreasing $m_{\rm avg}^2/e^4$.
Initially, for large positive  $m_{\rm avg}^2/e^4$,
the gauge field holonomy calculated on arbitrarily large circles
around $\UG$ vortices is trivial, dominated by perimeter-law contributions,
and our order parameter $O_{\Omega} = +1$.
But once $m_{\rm avg}^2/e^4$ decreases sufficiently, the
charged scalars condense. Then the holonomy around vortices acquires
a non-trivial phase, with $O_{\Omega}$ first deviating from $1$ at some
critical value of $m_{\rm avg}^2/e^4$.
The same reasoning as in Sec.~\ref{sec:ColemanWeinberg}
implies that this non-analytic behavior in $O_\Omega$ should also signal
a genuine phase transition.

%%%%%%%%%%%%
\section{QCD and the hypothesis of quark-hadron continuity}
\label{sec:QCD}
%%%%%%%%%%%%

%%% In this section we discuss the implications of our analysis for QCD.
%%% In Sec.~\ref{sec:QCD_review}, we review the Higgs phase of 4d QCD and
%%% the Schafer-Wilczek conjecture of quark-hadron continuity and its
%%% status in the literature to date.  Then in
%%% Sec.~\ref{sec:QCD_discontinuity} we explain how this conjecture is
%%% ruled out by a generalization of the analysis of the preceding
%%% sections. 

%%%%%%%%%
%%% \subsection{The Higgs phase of QCD and the Sch\"afer-Wilczek conjecture}
%%% \label{sec:QCD_review}
%%%%%%%%%

A central topic in strong interaction physics is understanding the
phase structure of QCD as a function of baryon number density, or
equivalently as a function of the chemical potential $\mu_B$
associated with the $U(1)_B$ baryon number symmetry. (For reviews see,
for example, Refs.~\cite{Alford:2007xm,Baym:2017whm}.) At low
(nuclear) densities, or small $\mu_B$, it is natural to describe the
physics in terms of nucleons, while at large $\mu_B$ a description in
terms of quark matter is appropriate thanks to asymptotic freedom. Are
``confined'' nuclear matter and ``deconfined'' quark matter sharply
distinct phases of matter, necessarily separated by at least one phase
transition, or might they be smoothly connected, similar to the gas
and liquid phases of water?  

Following Sch\"afer and Wilczek \cite{Schafer:1998ef},
we focus on the behavior of QCD with three flavors of quarks having
a common mass $m_q$, so that there is a vector-like $SU(3)$ flavor symmetry.
We ignore the weak, electromagnetic, and gravitational interactions.
Some readers may wonder why it is especially interesting to consider
the limit of QCD with $SU(3)$ flavor symmetry.
Physically there are, of course, six quark flavors in the
Standard Model.  But the three heaviest quark flavors (charm, bottom
and top) are so heavy that it is an excellent approximation to ignore
them entirely when considering the possible continuity between nuclear
matter and quark matter.
The three lightest quark flavors (up, down and strange) have distinct
masses in nature, so there is no exact global $SU(3)$ symmetry acting
on the light quark fields.
However, in practice the strength of $SU(3)$ flavor symmetry
breaking is not terribly large, since none of the three lightest
quarks are heavy compared to the strong scale $\Lambda_{\rm QCD}$. So
one motivation to study the $SU(3)$ flavor symmetric limit of QCD is
that the physics is simplest in this limit, and at the same time it is
a useful starting point for much phenomenology.  

There is also a more theoretical justification for focusing on the
$SU(3)$ flavor symmetric limit.  Suppose that the up and down quarks are
approximately degenerate in mass, but $SU(3)$ flavor symmetry is
broken because the strange quark is heavier, as is the case in nature.
In dense QCD, the effective strength of $SU(3)$-flavor breaking effects
due to unequal quark masses depends on the mass differences relative to $\mu_B$.
At sufficiently large $\mu_B$, or high density,
$SU(3)$ flavor breaking effects are negligible and
one is always in the so-called CFL regime, described below.
%
%% When the ratio between the strange quark mass and lightest quark mass
%% scale becomes very large, so that the flavor $SU(3)$ symmetry is badly
%% broken, it is already known that super-dense quark matter and nuclear
%% matter are separated by a phase transition.
%
However, when the strange quark mass is made large enough
compared to the light quark mass scale, one can show
reliably that at intermediate values of $\mu_B$
the theory lies in a different regime called 2SC.
The 2SC regime is known to be separated by phase
transitions from \emph{both} nuclear matter and high density CFL regimes,
because the realizations of global symmetries in the 2SC phase differ
from those in both confined nuclear matter and the CFL phase
\cite{Rischke:2000cn,Alford:2007xm}.
The open issue is to understand
what happens to the phase structure of QCD near the $SU(3)$ flavor limit.
%%% This is the reason the $SU(3)$ flavor symmetric point is the focus of
%%% the Sch\"afer-Wilczek conjecture on quark-hadron
%%% continuity~\cite{Schafer:1998ef} (reviewed below) and also the focus of
%%% our analysis.

Let us briefly review what is known about the
behavior of the $SU(3)$ flavor symmetric QCD as a function
of $\mu_B$. There is a critical value of $\mu_B$, which we denote by
$\muBs$, at which the baryon number density $n_B$ jumps from zero
to a finite value known as the nuclear saturation density,
$n_B^{\rm sat}$.\footnote{For physical values of quark masses, 
$n_B^{\rm sat} \sim 0.17 \, \textrm{fm}^{-3}$ and $\muBs \sim 920
\, \textrm{MeV}$.}  For $\mu_B$ above but close to $\muBs$, the
ground state of QCD may be thought of as modestly compressed
nuclear matter,
by which we mean that a description in terms of 
interacting nucleon quasiparticles is useful.
It is believed that $U(1)_B$ is spontaneously broken for any
$\mu_B > \muBs$ due to condensation of dibaryons,
so $SU(3)$-symmetric nuclear matter is a superfluid
(see, e.g., Refs.~\cite{Dean:2002zx,Gandolfi:2015jma}).
In real nuclear matter neutron pairs condense, while in
$SU(3)$ symmetric QCD it is flavor singlet $H$-dibaryons
which condense.
Nuclear matter should be regarded as a ``confined phase'' of QCD,
with quark confinement defined in the
same heuristic fashion as at zero density.
(The infamous difficulties of making the notion of confinement precise in
theories like QCD are reviewed in,
e.g., Ref.~\cite{Greensite:2016pfc}.)

In contrast, when $\mu_B \gg \muBs$ it becomes
natural to describe the system in terms of interacting quarks
rather than interacting nucleons.
Cold high density quark matter is known to
feature ``color superconductivity.''
Attractive gluon mediated interactions between quarks near the Fermi
surface lead to quark pairing and condensation,
analogous to phonon-induced Cooper pairing of electrons in conventional
superconductors.
The condensing diquarks in $SU(3)$ flavor-symmetric
three-color QCD have the quantum numbers of 
color-antifundamental scalar fields with charge $2/3$ under $U(1)_B$.
The condensation of these diquark fields spontaneously breaks $U(1)_B$
to $\mathbb{Z}_2$.  At the same time, the color $SU(3)$ gauge group is
completely Higgsed, while the flavor $SU(3)$ symmetry is unbroken. The
unbroken symmetry transformations consist of common
global $SU(3)$ rotations in color and flavor space, and
as a result the high density regime of three-flavor QCD is called the
``color-flavor-locked'' (CFL) phase.
The term ``color superconductivity'' for this phase is something
of a misnomer as there are no physically observable
macroscopic persistent currents or related phenomena analogous
to those present in real superconductors.
It is far better to think of this phase as a baryon superfluid in which
the $SU(3)$ gauge field is fully Higgsed.

Consequently, as $\mu_B$ is increased from $\muBs$ to values that are
very large compared to $\textrm{max}(\Lambda_{QCD},m_q)$,
the ground state of flavor symmetric QCD evolves from a
confining regime with spontaneously broken baryon number symmetry to
a Higgs regime which also has spontaneously broken $U(1)_B$.
The realization of all conventional global symmetries is
identical between the low and high density regimes.
One may also confirm that `t Hooft anomalies match and the pattern
of low energy excitations in the different regimes may be
smoothly connected
\cite{Schafer:1998ef,Rajagopal:2000wf,Wan:2019oax}.

So a natural question is whether there is a phase transition between
the nuclear matter and quark matter regimes of flavor-symmetric QCD
\cite{Schafer:1998ef}. If one can argue that such a phase transition
is required, then ``confined'' nuclear matter and ``Higgsed'' or
``deconfined'' quark matter become sharply distinct phases of QCD, and
one would obtain some insight into the meaning of the loosely defined
term ``confinement'' in QCD.

This question was the subject of the well-known conjecture by
Sch\"afer and Wilczek \cite{Schafer:1998ef}. Based on the matching
symmetry realizations and other points noted above, they argued that
no phase transition is required between the Higgsed (quark matter) and
confined (nuclear matter) regimes of $SU(3)$ flavor symmetric QCD, a
conjecture known as ``quark-hadron continuity.'' 
It should be noted that this conjecture is more general than its name
suggests. The arguments in favor of this conjecture do not rely on the
existence of fermionic fundamental representation matter fields, and
apply just as well to gauge theories with fundamental scalar fields
and analogous symmetry structures.
The Sch\"afer-Wilczek conjecture can be summarized as the statement
that if one considers a gauge theory with gauge group $G$, 
fundamental representation matter,
a $U(1)$ global symmetry,
and parameters that allow one to interpolate between 
a ``confining'' regime where the $U(1)$ global symmetry is
spontaneously broken, and
a regime where the gauge group $G$ is completely Higgsed and the $U(1)$ global
symmetry is also spontaneously broken,
then these regimes are smoothly connected (i.e., portions of a single phase)
at zero temperature.%
\footnote
    {%
    This may sound similar to the Fradkin-Shenker-Banks-Rabinovici
    theorem~\cite{Fradkin:1978dv,Banks:1979fi} but, as discussed in the
    introduction, the Fradkin-Shenker-Banks-Rabinovici theorem does not apply
    in situations where the Higgs field is charged under global symmetries,
    while the Sch\"afer-Wilczek conjecture concerns precisely such situations.
    }

Apart from its intrinsic theoretical interest,
the status of quark-hadron continuity
is also of experimental interest, at least to the extent that the
flavor symmetric limit of QCD is a decent approximation to QCD
with physical quark masses.
If phase transitions between nuclear matter and
quark matter do occur, then
the interiors of neutron stars may reach densities where
the equation of state and transport properties
are strongly affected by such transitions, leading to signatures that
might be detectable via multi-messenger observations of neutron stars
\cite{Lin:2005zda,Sagert:2008ka,Lattimer:2015nhk,Alford:2015gna,Han:2018mtj,Most:2018eaw,McLerran:2018hbz,
Bauswein:2018bma,Christian:2018jyd,Xia:2019pnq,Gandolfi:2019zpj,
Chen:2019rja,Alford:2019oge,Han:2019bub,Christian:2019qer,
Chatziioannou:2019yko,Annala:2019puf,Chesler:2019osn,Fischer:2020xjl,Zha:2020gjw}.

%%%%%%%%%%%%%%%
\subsection{Status of the Sch\"afer-Wilczek conjecture}
%%%%%%%%%%%%%%%

In the two decades since Sh\"afer and Wilczek hypothesized
quark-hadron continuity in flavor symmetric QCD,
based on compatible symmetry realizations and other necessary
but not sufficient correspondences,
their conjecture has reached the status of a highly plausible folk theorem.
The expectation of quark-hadron continuity
has been used as the starting point for a large number of further
conjectures and developments, see e.g., Refs.~\cite{Hatsuda:2006ps,McLerran:2007qj,
Yamamoto:2007ah,Alford:2019oge,Baym:2019iky,Nishimura:2020odq,
Hirono:2018fjr,BitaghsirFadafan:2018uzs,Schmitt:2010pf,Buballa:2003qv,
Fukushima:2013rx,Fukushima:2015bda,Schafer:1999pb,Schafer:2000tw,Masuda:2012ed,Kovensky:2020xif}.

Recently, however, three of the present authors argued that
a change in particle-vortex statistics between the Higgs regime
(quark matter) and the confined regime (nuclear matter)
should be interpreted as compelling evidence for invalidity
of the Sch\"afer-Wilczek conjecture \cite{Cherman:2018jir}.%
\footnote
    {%
    For other examinations of vortices in dense quark matter,
    see also Refs.~\cite{Eto:2009kg,Eto:2009tr,Cipriani:2012hr,
    Chatterjee:2015lbf,Chatterjee:2018nxe,Hirono:2019oup}.
    }
We showed that color holonomies around minimal circulation $U(1)_B$
vortices have non-trivial phases of $\pm 2\pi/3$ in high density quark matter,
noted that these holonomies should have vanishing phases
in the nuclear matter regime, and used this sharp change in the physics
of topological excitations to argue that the nuclear matter
and quark matter regimes of dense QCD will be separated by a phase transition.  

Subsequent work by other authors \cite{Hirono:2018fjr,Alford:2018mqj}
offered some objections to the arguments in our
Ref.~\cite{Cherman:2018jir}. Let us address these objections, starting
with Ref.~\cite{Hirono:2018fjr} by Hirono and Tanizaki. Changes in
particle-vortex statistics are a commonly used diagnostic for phase
transitions in \emph{gapped} phases of matter, see e.g.,
Refs.~\cite{doi:10.1080/00018739500101566,Hansson:2004wca}. In gapped
phases, changes in particle-vortex statistics are connected to changes
in intrinsic topological order, which in turn can be related to
changes in the realization of higher-form global
symmetries~\cite{Gaiotto:2014kfa}. Reference~\cite{Hirono:2018fjr}
tacitly assumed that these statements also hold in gapless systems,
and misinterpreted our work \cite{Cherman:2018jir} as proposing that
the zero temperature high density phase of QCD is topologically
ordered. Reference~\cite{Hirono:2018fjr} then argued that this is not
the case by discussing the realization of a putative low-energy
``emergent'' higher-form symmetry in a gauge-fixed version of $N_c=3$
Yang-Mills theory coupled to fundamental Higgs scalar fields. Besides
relying on a non-manifestly gauge invariant approximate description to
suggest some higher form symmetry, this discussion missed the central
points of Ref.~\cite{Cherman:2018jir} for two reasons. First,
Ref.~\cite{Cherman:2018jir} already explicitly emphasized that the CFL
phase of QCD is not topologically ordered according to the standard
definition of that term, so arguing that the CFL phase does not have
topological order in no way contradicts the analysis of
Ref.~\cite{Cherman:2018jir}. Second, while Ref.~\cite{Hirono:2018fjr}
agreed with us that in the flavor-symmetric limit, CFL quark matter
features non-trivial color holonomies around $U(1)_B$ vortices, it did
not address the key question of how this could be consistent with the
expected behavior of color holonomies in the nuclear matter regime.
Without addressing this crucial question, one cannot conclude that
quark-hadron continuity remains a viable scenario in QCD.

Reference~\cite{Alford:2018mqj} by Alford, Baym, Fukushima, Hatsuda
and Tachibana accepted the main result of Ref.~\cite{Cherman:2018jir},
namely that in the flavor-symmetric limit color holonomies around
vortices take sharply different values in the nuclear matter and quark
matter phases.   But Ref.~\cite{Alford:2018mqj} argued that the
hadronic and color-superconducting regimes may nevertheless be smoothly
connected.  Alford et al. considered (straight)
minimal-circulation vortices in a setting where the density varies
along the direction of a vortex, and argued that a single superfluid vortex in the color-superconducting
regime can connect to a single superfluid vortex in the hadronic regime.%
\footnote
    {%
    The argument for this statement in Ref.~\cite{Alford:2018mqj} is
    very simple: one can consider a gedanken situation involving a
    rotating bucket of density stratified quark/nuclear matter when
    the quantized superfluid circulation equals unity on every
    cross-section of the bucket. There must then be a single minimal
    circulation vortex threading both phases and crossing the
    interface between them.
}  
This was interpreted as evidence against a ``boojum''\footnote{A boojum is a junction or special defect
at points where vortices pass through the interface between distinct
superfluid phases.} of the sort
discussed in Refs.~\cite{Cipriani:2012hr,Chatterjee:2018nxe} in which
three vortices in the quark matter phase must join in order to pass
into the hadronic phase.  We agree that there is no reason for a
boojum  at the interface between quark matter and nuclear matter to
necessarily involve multiple vortices joining together. Instead, given
the behavior of the color holonomy, it is entirely consistent for the
interface to be a genuine boundary between distinct thermodynamic
phases, with minimal-energy boojums involving just one minimal
circulation vortex on either side, with the behavior of the color
gauge fields changing sharply at the interface.

In our view, the key limitations of our work in Ref.~\cite{Cherman:2018jir}
were that we could not explicitly compute expectation values of
color holonomies in the superfluid nuclear matter regime
and demonstrate that they have trivial phases,
nor could we give a proof that a change in the behavior of
gauge field holonomies around vortices must be associated with a
bulk thermodynamic phase transition.
(Although we did give physical arguments for this which we believe
are convincing.)

In the preceding sections of the present paper, we have analyzed a
3D model which was deliberately constructed to be analogous to dense QCD,
and to which Sch\"afer and Wilczek's continuity conjecture applies and
predicts that no phase transition separates the $\UG$-broken Higgs and
confining regimes.
This allowed us to examine both of these earlier limitations in
the context of this instructive model, and find that continuity does
\emph{not} hold. The Higgs and confining $\UG$-broken regimes of
the 3D theory are distinct phases of matter characterized by a
novel order parameter.

%%%%%%%%%
\subsection{Higgs versus confinement in 4D gauge theory}
\label{sec:QCD_discontinuity}
%%%%%%%%%

In earlier sections we focused on our 3D Abelian model because
this provided the simplest setting in which to examine the
issue of Higgs-confinement continuity within superfluid
(or spontaneously broken $U(1)$) phases, with good theoretical
control in both regimes.
It is, of course, of interest to understand how the relevant
physics might change when one turns to 4D gauge theories
which are more QCD-like.

To that end, we now
consider an $SU(3)$ gauge theory coupled to three antifundamental
representation scalar fields, as well as an additional gauge-neutral
complex scalar field $\phi_0$.  We will build a model with
$SU(3)$ flavor symmetry, and write the charged scalar fields as a $3\times 3$
matrix $\Phi$ which transforms in the bifundamental representation of
$SU(3)_{\rm flavor} \times SU(3)_{\rm gauge}$,
%%% If we denote its matrix elements by $\Phi^I_{\ A}$,
%%% with $I$ a flavor index and $A$ color,
%%% then $\Phi$ transforms as
\begin{align}
    \Phi \to F \, \Phi \, C^\dagger \,,
%%%    \Phi^I_{\ A} \to F^I_{\ J} \, \Phi^J_{\ B} (C^\dagger)^B_{\ A},
    \quad F \in SU(3)_{\rm flavor}, \,\, C \in SU(3)_{\rm gauge}\,.
\end{align} 
We also assume the theory has a $U(1)$ global symmetry, which acts as
\begin{equation}
    \UG: \Phi \to e^{2i\alpha/3}\,\Phi, \quad
	\phi_0 \to e^{2i\alpha} \, \phi_0\, ,
\end{equation}
and assume that there exist (or could exist) heavy `baryon' test particles
with unit charge under the $\UG$ global symmetry.  Since $\UG$ phase
rotations which lie within $\mathbb Z_3$ coincide with the action of
$SU(3)$ gauge transformations, the faithfully acting $U(1)$ global
symmetry is $\UG/\mathbb{Z}_3$.

The action defining this model is given by
\begin{align}
    \label{eq:4dScalarQCD}
    S = \int d^{4}x \, &\left[
	\frac{1}{2g^2} \, \tr F_{\mu\nu}^2
	+ \tr  (D_{\mu} \Phi)^{\dag} D^{\mu} \Phi
	+ |\partial_{\mu}\phi_0|^2
	+ m_{\Phi}^2 \, \tr \Phi^{\dag}\Phi 
	+ m_{0}^2 \, |\phi_0|^2
	\right.
\nonumber\\
    &\left.\vphantom{[\frac{1}{2g^2}\tr F_{\mu\nu}^2}
	+ \lambda_0|\phi_0|^4 + \lambda_\Phi \, \tr (\Phi^\dagger\Phi)^2
	+ \epsilon\, (\phi_0^\dagger \det \Phi + \textrm{h.c}) + \cdots
	\right] .
\end{align}
%%%$ where $\tr$ denotes the usual trace over color indices, while $\Tr$
%%%$ denotes a trace over both color and flavor indices.
As usual, $D_{\mu}
\Phi = \partial_{\mu} \Phi + i \Phi A_{\mu}$ is the covariant
derivative in the antifundamental representation, and the ellipsis denotes possible further
scalar self-interactions which are invariant under
the chosen symmetries. The field strength $F_{\mu \nu} \equiv F^a_{\mu
\nu} t^a$, with Hermitian $SU(3)$ generators satisfying $\tr t^a t^b =
\half \delta^{ab}$. 

This 4D model is very similar to the scalar part of the effective
field theory that describes high-density three-color QCD in the CFL
quark matter regime~\cite{Alford:2007xm}, with $\UG/\mathbb{Z}_3$ playing the role
of $U(1)_B$ in QCD. The matrix-valued scalar $\Phi$ represents three
color-antifundamental diquark fields, so that $\det\Phi$ has the
quantum numbers of flavor-singlet dibaryons, which are condensed in
both the CFL phase and the $SU(3)$-symmetric nuclear matter phases. Due to the $\epsilon$ coupling between
the gauge-neutral scalar $\phi_0$ and $\det \Phi$, one can think of
$\phi_0^\dagger$ as a (dynamical) source for flavor-singlet dibaryons.
Explicitly introducing the neutral scalar $\phi_0$ allows
the model \eqref{eq:4dScalarQCD} to describe both the Higgs regime and a regime
where dibaryons are light, but the gauge and charged scalar fields can
be integrated out.

Of course, the effective action for dense QCD in the CFL regime is
rotation-invariant but not Lorentz invariant, and also includes heavy
fermionic excitations, in contrast to the purely bosonic
Lorentz-invariant theory defined by Eq.~\eqref{eq:4dScalarQCD}. These
differences are not relevant to our discussion, and we expect the
phase structure of the model \eqref{eq:4dScalarQCD} to mimic the phase
structure of QCD with approximate $SU(3)$ flavor symmetry.

% We are specifically interested in the regime
% where the $\UG$ symmetry is spontaneously broken with non-vanishing
% expectation values for  $\phi_0$ and $\det \Phi$.

Consider the Higgs regime of the model \eqref{eq:4dScalarQCD} where
(in gauge-fixed language) $\Phi$ has an expectation value of
color-flavor locked form, $\langle \Phi \rangle =  v_{\Phi} \,
\mathbf{1}_3$, and there is a residual unbroken $SU(3)_{\rm global}$
symmetry acting as $\Phi \to U \Phi U^{\dag}$, with $U \in SU(3)$. The
$\UG$ global symmetry is spontaneously broken implying, as always, the
existence of vortex topological excitations. To describe a straight
``superfluid'' vortex, using cylindrical coordinates with $r=0$ at the
center of the vortex, one may fix a gauge in which the vortex
configuration has $\Phi$ diagonal and $A_\mu$ taking values in the
Cartan subalgebra, 
\begin{subequations}
\label{eq:QCDvortex}
\begin{align}   
    \phi_0(r,\theta) &= v_0 \, f_0(r)\, e^{ik\theta} \,,
\\[5pt]
    \Phi(r,\theta) &= v_{\Phi} \>
    \mathrm{diag}
    \left(
	f_1(r)\, e^{i(n+k)\theta} ,\>
	f_2(r)\, e^{i(m-n)\theta} ,\>
	f_3(r)\, e^{-im\theta}
    \right) ,
%%%    \begin{pmatrix} f_1(r)\, e^{i(n+k)\theta} & 0 & 0 \\ 0 &  f_2(r)\, e^{i(m-n)\theta} & 0 \\
%%%    0 & 0 &  f_3(r)\, e^{-im\theta}
%%%    \end{pmatrix},
\\
    A_{\theta}(r) &= \frac{a\, h_8(r)}{2\pi r} \, t_8
		    + \frac{b\, h_3(r)}{2\pi r} \, t_3 \,.
\end{align}
\end{subequations}
Here $k,m,n \in \mathbb Z$, with $k$ the vortex winding number,
$t_8 \equiv \frac{1}{2\sqrt{3}}\,\textrm{diag}(1,1,-2)$ and
$t_3 \equiv \frac{1}{2}\,\textrm{diag}(1,-1,0)$
are the usual diagonal $SU(3)$ generator matrices,
and the radial profile functions $\{ f_i \}$ and $\{ h_i \}$
approach $1$ as $r\to \infty$.
%%% and vanish as $r\to 0$.
%%% LGY: $f_3(0)$ needn't vanish if $m = 0$
Minimizing the long-distance energy density of the
vortex configuration determines the gauge field asymptotics.
One finds,
\begin{equation}
    a = -\tfrac{2\pi}{\sqrt{3}} \, (k+3m) \,,\quad
    b = -2\pi(k+2n-m) \,.
\end{equation}
The minimal energy vortex with unit circulation ($k=1$) 
corresponds to $n=m=0$
(with physically equivalent forms related by Weyl reflections),
in which case
\begin{equation}
    a = -\tfrac{2\pi}{\sqrt{3}} \,,\quad
    b = -2\pi \,,
\label{eq:minvals} 
\end{equation}
and
\begin{equation}
    \Phi(r,\theta) = v_\Phi \,\textrm{diag}
    \left(f(r)\, e^{i\theta},\> g(r),\> g(r) \right),\quad
    A_\theta(r) = \frac{h(r)}{3r} \,\textrm{diag}(-2,1,1) \,.
\label{eq:minimalQCDvortex}
\end{equation}
Here, we have set $f_1(r) = f(r)$, $f_2(r) = f_3(r) = g(r)$,
and $h_3(r) = h_8(r) = h(r)$.
The minimal-energy vortex configuration
\eqref{eq:minimalQCDvortex} preserves an $SU(2) \times U(1)$ symmetry
(cf., Ref.~\cite{Auzzi:2003fs}).
Hence, these minimal energy unit-winding vortices have zero
modes associated with the moduli space
\begin{align}
    \frac{SU(3)}{SU(2) \times U(1)} =
\mathbb{CP}^2 \,.
\end{align} 
Consequently, the worldsheet effective field theory for a vortex contains a
$\mathbb{CP}^2$ non-linear sigma model~\cite{Eto:2009kg,Gorsky:2011hd,Eto:2011mk}.
But the $\mathbb{CP}^2$ model in two spacetime dimensions
(with vanishing topological angle $\theta$) 
has a mass gap and a unique ground state angle,
see e.g., Refs.~\cite{Campostrini:1992ar,Bonati:2019owp}.
So, despite the appearance of the classical configuration \eqref{eq:QCDvortex},
the $SU(3)_{\rm global}$ symmetry is unbroken both in the vacuum and
in the presence of vortices.

Now consider the behavior of our vortex holonomy order
parameter in this theory.
The gauge field holonomy is now a path-ordered exponential around some
contour $C$,
$\Omega(C) \equiv \mathcal P (e^{i\int_C A})$,
and defines an $SU(3)$ group element.
The natural non-Abelian version of our vortex order parameter involves
gauge invariant traces of holonomies,
\begin{align}
    O_{\Omega}
    \equiv
    \lim_{r \to \infty} 
    \frac{\langle \tr \Omega(C) \rangle_1}{\langle \tr \Omega (C) \rangle} \,,
\label{eq:QCD_order_parameter}
\end{align}
where in the numerator the circular contour $C$ encircles a minimal
vortex in the same direction as the circulation of the $\UG$ current.%
\footnote
    {%
    Once again, the numerator is defined by a constrained functional
    integral with a prescribed vortex world-sheet, with the
    size of that world-sheet and the minimal separation between
    the vortex world-sheet and the holonomy contour $C$ scaling
    together as the contour radius $r$ increases.
    }
Both expectations in the 
ratio \eqref{eq:QCD_order_parameter} have perimeter-law dependence on
the size of the contour $C$ arising from quantum fluctuations on
scales small compared to $r$,
but this geometric factor cancels by construction in the ratio.
Unbroken charge conjugation symmetry implies that the denominator
is real, and it must be positive throughout any phase connected
to a weakly coupled regime.
So as in our earlier Abelian model, the behavior of $O_{\Omega}$
is determined by the phase of the vortex expectation value in the numerator.

A trivial calculation (identical to that in Ref.~\cite{Cherman:2018jir})
shows that at tree-level, far from the vortex,
\begin{align}
   \frac{1}{3} \left\langle \tr \Omega(C) \right\rangle_1^{\rm tree}
    = e^{2\pi i /3} \,.
\end{align} 
demonstrating that $O_\Omega = e^{2\pi i/3}$ at tree-level. An
effective field theory argument, analogous to that given in section
\ref{sec:Higgs_holonomy_Abelian} (see also
Appendix~\ref{sec:nonAbelian}), shows that this result is unchanged
when quantum fluctuations are taken into account, as long as they are
not so large as to restore the spontaneously broken $\UG$ symmetry. To
see this, consider the form of the effective action generated by
integrating out fluctuations on scales small compared to $r$. Only
terms in the effective action with two derivatives acting on the
charged scalar field $\Phi$ can contribute to the $\mathcal O(1/r^2)$
holonomy-dependent part of the energy density, and hence affect the
gauge field asymptotics (\ref{eq:minvals}) which determines the
expectation value of holonomies far from the vortex core.
Traces of operators containing a single covariant
derivative, such as $\Tr \Phi^\dagger D_\mu \Phi$, are independent of
the gauge field far from the vortex core. Consequently, the portion
of the effective action which controls the holonomy expectation value
far from a vortex may be written in the form
\begin{align}
    S_{\textrm{eff}, \, SU(3) \textrm{ holonomy}}
    = \int d^{4}x \, \Big\{ &
     \Tr \!\big[
	 f_1(\phi_0,\Phi) (D_{\mu} \Phi)^\dagger
	 f_2(\phi_0,\Phi) (D^\mu\Phi) \big]
\nonumber\\
%     & +
%	 \Tr\!\big[f_3(\Phi,\phi_0) \Phi^\dagger (D^\mu\Phi)\big] \,
%	 \Tr\!\big[f_4(\Phi,\phi_0)(D_{\mu} \Phi)^\dagger \Phi\big]
%\nonumber\\[3pt]
     & +
	 \epsilon^{ABC}\epsilon_{IJK} \,
	 f_3(\phi_0,\Phi)^I_{\ A}
	 (D_\mu\Phi)^J_{\ B}
	 (D^\mu\Phi)^K_{\ C}
   \Big\}+ \textrm{h.c.}\,,
\label{eq:pert}
\end{align}
where $A,B,C$ are color indices and $I,J,K$ are flavor indices. The
three coefficient functions $\{ f_i\} $ depend on the fields $\phi_0$
and $\Phi$, but not on their derivatives, only in combinations which
are are invariant under $\UG$. The function $f_1$ is a color adjoint
and flavor singlet (like $\Phi^\dagger\Phi$), $f_2$ is color singlet
and flavor adjoint (like $\Phi\Phi^\dagger$), and $f_3$ is
antifundamental in color and fundamental in flavor (like
$\phi_0^\dagger \Phi$). Plugging in the configuration
\eqref{eq:QCDvortex}, one can easily verify that both terms in
\eqref{eq:pert} have extrema, with respect to the asymptotic gauge
field coefficients $a$ and $b$, at the same location
\eqref{eq:minvals} regardless of the form of the functions $\{f_i\}$.
Therefore small quantum corrections do not perturb the gauge field
asymptotics far from a vortex, and hence cannot shift the phase of the
vortex holonomy expectation $\left\langle \tr \Omega(C)
\right\rangle_1$ away from $2\pi/3$. Hence, we learn that 
\begin{align}
\boxed{
\textrm{$\UG$-broken Higgs phase:}\;\; O_{\Omega} = e^{2\pi i/3} } \,,
\label{eq:QCD_holonomy_higgs}
\end{align}
holds exactly throughout the phase connected to the
weakly coupled Higgs regime.

Alternatively, when $m_{\Phi}^2 \gtrsim \Lambda^2$, with $\Lambda$ 
the strong dynamics scale of the theory,
we can recycle the arguments of
Sec.~\ref{sec:broken_permutations} to understand the behavior of
$O_{\Omega}$.  In this regime, due to the presence of heavy
dynamical charged excitations, the expectation values of large
fundamental representation Wilson loops are (exponentially) dominated
by a perimeter-law contribution. Physically, a Wilson loop describes a
process where a fundamental representation test particle and
antiparticle are inserted at some point, separated and then recombined
as they traverse the contour $C$. The perimeter law behavior arises
from configurations in which dynamical fundamental representation
excitations of mass $m_{\Phi}$ are pair-created and dress the test charge and
anticharge to create two bound gauge-neutral ``mesons.'' These mesons
have physical size of order $\ell_{\rm meson} \sim
\textrm{min}\left(\Lambda^{-1}, (\alpha_s m_{\Phi})^{-1} \right)$, and experience no
long range interactions. Once the Wilson loop size exceeds the string
breaking scale $\sim 2 m_{\Phi}/\Lambda^2$, pair creation of dynamical
charges of mass $m_{\Phi}$ and the associated meson formation becomes
the dominant process contributing to fundamental Wilson loop
expectation values.

The perimeter law contribution to large fundamental representation
Wilson loop expectation values arises from fluctuations of the
gauge-charged fields within distances of order of $\ell_{\rm meson}$ from
any point on the contour $C$. The amplitude for such screening
fluctuations, and consequent meson formation, must be completely
insensitive to the presence of a vortex very far away at the center of
the loop.  This means that the holonomy expectations in the 
numerator and denominator of the vortex observable
\eqref{eq:QCD_order_parameter} will be identical (up to exponentially
small corrections vanishing as $r \to \infty$), leading to the conclusion that 
\begin{align}
    \boxed{\textrm{$\UG$-broken confining phase:}\;\;  O_{\Omega} = 1 } \,.
\label{eq:QCD_holonomy_conf}
\end{align}
Once again, the differing results 
\eqref{eq:QCD_holonomy_higgs}
and
\eqref{eq:QCD_holonomy_conf},
each strictly constant within their respective domains,
implies that $O_{\Omega}$ cannot be a real-analytic function of $m_{\Phi}^2$.
Adapting the arguments in Sec.~\ref{sec:ColemanWeinberg} regarding the
impact of abrupt changes in the properties of vortex loops on the
ground state energy,
we see that $O_{\Omega}$ functions as an order
parameter that distinguishes the $\UG$-broken Higgs and $\UG$-broken
confining phases of this four-dimensional $SU(3)$ gauge theory
with $SU(3)$ flavor symmetry.%
\interfootnotelinepenalty=10
\footnote
{%
Further evidence that changes in our non-local order parameter signal
genuine phase transitions in non-Abelian gauge theories
may be gained by considering other calculable examples.
One such case is described in Appendix~\ref{sec:nonAbelian}.
A different example which is closer to the model discussed in
this section consists of a version of the theory \eqref{eq:4dScalarQCD}
in three spacetime dimensions, with gauge group $SU(2)$ and two flavors
of $SU(2)$ antifundamental scalar fields, with a
global flavor symmetry containing an $SU(2)$ factor.
Generalizing the analysis in Sec.~\ref{sec:ColemanWeinberg}
to this non-Abelian model, we have checked
that there is a set of parameters (essentially identical to the ones
in Sec.~\ref{sec:ColemanWeinberg}) for which the phase transition
between the $\UG$-broken confining and Higgs regimes is strongly
first-order as a function of the mass of the antifundamental scalars.
The fact that the transition is strongly first-order allows the existence
of the phase transition to be reliably established despite the fact that
the gauge sector is strongly coupled within the $\UG$-broken confining
phase.  It is easy to check in this example that our vortex observable
$O_{\Omega}$ jumps from $+1$ to $-1$ across the transition,
and serves as an order parameter distinguishing distinct phases,
even when the transition is no longer strongly first-order.
Finally, it is easy to check that these statements generalize to
$N = N_f >2$ gauge theories.
}

Finally, if the $SU(3)$ flavor symmetry of this theory is explicitly broken
by a small perturbation, a simple generalization of the analysis
leading to the gauge field asymptotics \eqref{eq:minimalQCDvortex}
implies that the phase of $O_{\Omega}$ will now deviate slightly from $2\pi /3$.
But in the $\UG$-broken confined phase, $O_{\Omega}$ remains exactly $1$ due
to the confinement and string breaking effects discussed above. This
implies that the $\UG$-broken Higgs and $\UG$-broken confining regimes
of our 4D $SU(3)$ scalar theory \eqref{eq:4dScalarQCD}
must remain separated by a quantum phase transition even when
the $SU(3)$ flavor symmetry is explicitly broken.
Most importantly, essentially the same argument applies to dense QCD.

Before leaving this section, we note that one may consider
our original 3D model (\ref{eq:the_model}), or the 4D non-Abelian
generalization (\ref{eq:4dScalarQCD}), with the addition of
a non-zero chemical potential for the $\UG$ symmetry.
Such a chemical potential explicitly breaks charge conjugation
symmetry, just like the baryon chemical potential in dense QCD.
In our earlier discussion we used unbroken charge conjugation
symmetry to conclude that the ground state expectation value
of the holonomy must be real.
But, as noted in footnote \ref{fn:reflect}, for a reflection-symmetric
holonomy contour (such as a circle), reflection symmetry is an equally good
substitute.
Consequently, all of our arguments demonstrating that the
phase of the holonomy encircling a vortex at large distance
serves as an order parameter distinguishing ``confining''
and ``Higgs'' superfluid phases go through without
modification in the presence of a non-zero chemical potential.

In summary, we have shown that consideration of our new order
parameter implies that there is a phase transition between nuclear
matter and quark matter in dense QCD near the $SU(3)$ flavor limit.
This means that the confining nuclear matter regime of QCD (at least
with approximate $SU(3)$ flavor symmetry) has a sharp definition as a
phase of QCD where the expectation values of color holonomies around
superfluid vortices are positive, while quark matter --- a Higgs
regime --- can be defined as the phase of QCD where these holonomy
expectation values become complex.   Given the notorious difficulties
in giving a sharp definition for confining and Higgs regimes in gauge
theories with fundamental representation matter (see Ref.~\cite{Greensite:2016pfc}
for a review), this is a satisfying result in the theory of strong
interactions.   Our results are also encouraging for observational searches for
evidence of quark matter cores in neutron stars, see
e.g.~Refs.~\cite{Lattimer:2015nhk,Alford:2015gna,Han:2018mtj,McLerran:2018hbz,
Bauswein:2018bma,Christian:2018jyd,Xia:2019pnq,Gandolfi:2019zpj,
Chen:2019rja,Alford:2019oge,Han:2019bub,Christian:2019qer,
Chatziioannou:2019yko,Annala:2019puf,Chesler:2019osn}, because our
results imply that hadronic matter and quark matter must be separated
by a phase transition as a function of density.

\section{Conclusion}
\label{sec:conclusion}
%%%%%%%%%%

We have explored the phase structure of gauge theories with
fundamental representation matter fields and a $U(1)$ global symmetry.  Motivated by the
physics of dense QCD, we considered both Higgs and confining portions of
the phase diagram
in which the $U(1)$ global symmetry is spontaneously broken,
and hence the theory is gapless due to the presence of a
Nambu-Goldstone boson.
These two regimes cannot be distinguished by conventional
local order parameters probing global symmetry realizations,
nor do they naturally fit into more modern classification schemes
based on topological order and related concepts.
Nevertheless, using a novel
vortex order parameter introduced in Sec.~\ref{sec:vortices_and_holonomies},
we found that $U(1)$-broken confining and Higgs regimes
are sharply distinct phases of matter
separated by at least one phase transition in parameter space,
as illustrated in Fig.~\ref{fig:3D_phase_diagram}.
In Secs.~\ref{sec:our_model} and \ref{sec:vortices_and_holonomies}
(and Appendix~\ref{sec:nonAbelian}) we examined instructive 
parity-invariant Abelian (and non-Abelian) gauge theories 
in three spacetime dimensions
illustrating this physics.
Then in Sec.~\ref{sec:QCD} we considered related theories with a
$U(1)$ global symmetry in four spacetime dimensions, and
explained how our considerations serve to rule out the
Sch\"afer-Wilczek conjecture of quark-hadron continuity in cold dense
QCD.

Why are these results interesting?  First, we have added to the
toolkit of techniques for diagnosing phase transitions in gauge
theories, and shown that it predicts previously unexpected phase
transitions in theories with fundamental representation matter fields.
Second, our analysis implies a phase transition between
quark matter and nuclear matter in dense QCD near the $SU(3)$ flavor
limit, with possible implications for observable properties of neutron stars.
Third, our analysis provides a sharp distinction 
between a confined nuclear matter regime of QCD and
dense quark matter.
In other words, it provides sharp answers to some basic questions
about strong dynamics:
\begin{itemize}
    \item ``What is the confined phase of QCD?''  Our work shows that
this question has a sharp answer when the $U(1)_B$ baryon number symmetry is
spontaneously broken.  The confined phase of QCD with spontaneously
broken $U(1)_B$ symmetry can be defined as the phase of QCD
where the expectation values of color holonomies around
minimal-circulation superfluid vortices are positive.
    \item ``What is cold quark matter?'' Our analysis shows that cold
    quark matter can be defined as the phase of QCD where the
    expectation values of color holonomies around minimal-circulation
    superfluid vortices have non-vanishing phases.
\end{itemize}
Our results raise a number of other interesting questions that we hope can
be addressed in future work.  These include:
\begin{itemize}
\item
    What is the nature of the point in
    Fig.~\ref{fig:3D_phase_diagram} where the three different phase
    transition curves intersect?
\item
    What can be said in general about the order of the phase
    transition(s) separating $U(1)$-broken confining and Higgs phases
    in the theories we have considered? As discussed in
    Sec.~\ref{sec:ColemanWeinberg}, for some ranges of parameters
    there is a single first order phase transition. Is this always the
    case, or is there a range of parameters where the transition
    becomes second-order? How does the answer depend on the spacetime
    dimension? These issues are of more than just theoretical
    interest, because the properties of the
    nuclear to quark matter phase transition(s) in dense QCD can have
    observational impacts for the physics of neutron stars.
\item
    Relatedly, when the transition is first order
    what is the physics on an interface separating coexisting phases?
    This is also directly connected to potential neutron star phenomenology.
\item
    What happens to the phase structure of the class of theories
    we have considered, in both three and four spacetime dimensions,
    at non-zero temperature?

\item
    How should the modern classification of the phases of matter be
    generalized when considering transitions between gapless regimes?
    Is there a natural embedding of the constructions in this paper
    into some more general framework?  In Appendix~\ref{sec:gaugingU1}
    we gauge the $\UG$ symmetry of our 3D Abelian model and show that
    the resulting gapped theory (which flows to TQFTs at long
    distances) has a phase transition analogous to the
    Higgs-confinement phase transition studied in the body in the
    paper. But we also argue that, by itself, this cannot be used to
    infer the existence of a phase transition in the original model
    with a global $\UG$ symmetry.

\item 
    Can our construction be generalized to gauge theories where the
    $\UG$ global symmetry is explicitly broken to a discrete subgroup
    $\mathbb{Z}_k$? Such theories would contain domain walls, and the behavior of gauge field holonomies around domain wall junctions could be used to identify phase transitions. 

\item
    Are there condensed matter systems which realize the physics of
    $U(1)$-broken Higgs-confinement phase transitions?
\end{itemize}

\section*{Acknowledgments}

We are especially grateful to Fiona Burnell for extensive discussions
and collaboration at the initial stages of this project. We are also
grateful to M.~Alford, F.~Benini, S.~Benvenuti, K.S.~Damle, L.~Fidkowski,
D.~Harlow, Z.~Komargodski, S.~Minwalla, E.~Poppitz, N.~Seiberg, Y.~Tanizaki and
M.~\"Unsal for helpful discussions and suggestions during the long
gestation of this paper. AC acknowledges support from the University
of Minnesota. TJ is supported by a UMN CSE Fellowship. SS acknowledges
the support of Iowa State University startup funds. LY acknowledges
support from the U.S. Department of Energy grant DE-SC\-0011637.

\appendix

\section{Higgs phase vortex profile functions}
\label{sec:EoMAppendix}

Recall that the vortex configuration in our 3D model has the form
given by Eqs.~(\ref{eq:ansatz}) and (\ref{eq:nupm}), repeated here:
\begin{equation}
    \phi_0 = v_0\, f_0(r)\, e^{ik\theta}, \quad
    \phi_+ = v_c\, f_+(r)\, e^{i(n-k)\theta}, \quad
    \phi_- = v_c\, f_-(r)\, e^{-in\theta}, \quad
    A_\theta = \frac{\Phi\, h(r)}{2\pi r}, 
\end{equation}
with the radial profile functions $h$, $f_\pm$, and $f_0$ all approaching
1 as $r\to\infty$. The equation of motion for
the gauge field profile $h(r)$ is
\begin{equation}
    \frac{\Phi}{2\pi}
    \left(\frac{d^2h}{dr^2}-\frac{1}{r}\frac{dh}{dr}\right)
    =
    -2e^2v_c^2
    \left[ \left(n-k-\frac{\Phi\, h}{2\pi}\right) \, f_+^2
	+  \left(n-\frac{\Phi\, h}{2\pi}\right) \, f_-^2\right],
\label{eq:gauge} 
\end{equation}
while the scalar field profile functions obey 
\begin{subequations}
\begin{align}
    \left[\frac{d^2f_+}{dr^2}+ \frac{1}{r}\frac{df_+}{dr} - \frac{f_+}{r^2}\left(n-k-\frac{\Phi\, h}{2\pi}\right)^2\right]
    &=
    -|m_c^2| \, f_+ + 2\lambda_c v_c^2 \, f_+^3
    -\epsilon v_0 \, f_- f_0 \,,
\label{eq:scalar1}
\\
    \left[\frac{d^2f_-}{dr^2}+ \frac{1}{r}\frac{df_-}{dr} - \frac{f_-}{r^2}\left(n-\frac{\Phi\, h}{2\pi}\right)^2\right]
    &=
    -|m_c^2| \, f_- + 2\lambda_c v_c^2 \, f_-^3
    -\epsilon v_0 \, f_+ f_0 \,,
\label{eq:scalar2}
\\
    \left[\frac{d^2f_0}{dr^2}+ \frac{1}{r}\frac{df_0}{dr} - \frac{f_0}{r^2}\left(k\right)^2\right]
    &=
    -|m_0^2| \, f_0 +2\lambda_0 v_0^2 \, f_0^3
    -\epsilon \frac{v_c^2}{v_0} \, f_+ f_- \,.
\label{eq:scalar3}
\end{align}
\end{subequations}
As discussed in section \ref{sec:Higgs_holonomy_Abelian}
(cf. Eq.~\eqref{eq:fluxval})
the minimal energy solution has $\Phi = (2n-k)\pi$.
Inserting this value and examining the resulting large $r$ asymptotic
behavior of the profile functions,
one finds that $h(r)$ equals 1 up to exponentially falling corrections.
This will be demonstrated below.
Neglecting such exponentially small terms, the scalar profile
functions satisfy
\begin{subequations}\label{eq:larger}%
\begin{align}
    \left[\frac{d^2f_\pm}{dr^2}+ \frac{1}{r}\frac{df_\pm}{dr}
    - \frac{k^2}{4} \frac{f_\pm}{r^2} 
    \right]
    &=
    -|m_c^2| \,f_\pm
    + 2\lambda_c v_c^2 \, f_\pm^3
    -\epsilon v_0 \, f_\mp f_0 \,,
\\
    \left[\frac{d^2f_0}{dr^2}+ \frac{1}{r}\frac{df_0}{dr}
    - k^2 \,\frac{f_0}{r^2}\right]
    &=
    -|m_0^2| \, f_0
    +2\lambda_0 v_0^2 \, f_0^3
    -\epsilon \frac{v_c^2}{v_0} \, f_+ f_- \,.
\end{align}
\end{subequations}
Demanding that the scalar profile functions $\{ f_i \}$
approach 1 as $r \to \infty$
and requiring that the resulting right-hand sides of Eq.~(\ref{eq:larger})
vanish determines the condensate magnitudes $v_0$ and $v_c$.
One finds $v_c^2 = (|m_c^2| + \epsilon v_0)/(2\lambda_c)$
with $v_0$ the positive solution of the cubic equation
\begin{equation}
    4\lambda_0 \lambda_c \, v_0^3
    -(2\lambda_c |m_0^2| + \epsilon^2) \, v_0
    - \epsilon |m_c^2|
    = 0 \,.
\end{equation}
One may then verify that the resulting scalar profile functions have
the asymptotic forms
\begin{equation}
    f_\pm(r) = 1 - \frac{\ell_c^2}{r^2} + \mathcal O(r^{-4}) \,,\qquad
    f_0(r) = 1 - \frac{\ell_0^2}{r^2} + \mathcal O(r^{-4}) \,,
\label{eq:power_law_falloff}
\end{equation}
with
\begin{align}
    \ell_c^2
    =
    \frac{k^2}{4 v_c^2} \,
    \frac{6\lambda_0v_0^2- |m_0^2|+4\epsilon v_0}{2\lambda_c(6\lambda_0v_0^2- |m_0^2|)-\epsilon^2}
    \,, \qquad
    \ell_0^2
    = \frac{k^2}{4v_0} \,
    \frac{\epsilon+ 8\lambda_c v_0}{2\lambda_c(6\lambda_0 v_0^2-|m_0^2|)-\epsilon^2} \,.
\end{align}
The large $r$ power-law tails in scalar field profile functions 
are characteristic features of global vortices.
But since we are in a Higgs phase,
the gauge field should have exponential fall-off.   
To verify this, 
we parallel the treatment of Ref.~\cite{Eto:2009kg}
and rewrite the coupled equations in terms of
sums and differences of the charged field profiles.
Let%
\begin{subequations}
\begin{align}
    h=1+ H \,,\qquad
    f_+ =1+ F+\frac{ G}{2} \,,
\\
    f_{0} = 1+ F_0 \,,\qquad
    f_- =1+ F-\frac{ G}{2} \,,
\end{align}
\end{subequations}
and then linearize the field equations in the deviations
$F_0$, $F$, $G$ and $H$.
The leading behavior of $F$ and $F_0$ can be read off from
Eq.~\eqref{eq:power_law_falloff}.
The linearized equations for $G$ and $H$ do not involve $F$ or $F_0$
and read
\begin{subequations}\label{eq:GH}
\begin{align}
    \left[\frac{d^2}{dr^2}+\frac{1}{r}\frac{d}{dr} - \widetilde{m}^2-\frac{k^2}{4r^2}\right] G &= \frac{k\,(2n{-}k)}{r^2} \, H\, ,
\\
    (2n{-}k)\left[\frac{d^2}{dr^2}-\frac{1}{r}\frac{d}{dr} -m_A^2 \right]H &=  k\, m_A^2 \, G \,,
\end{align} 
\end{subequations}
where $m_A^2 \equiv 4 e^2 v_c^2$ and
$\widetilde{m}^2 \equiv 4\lambda_c v_c^2 + 2\epsilon v_0$.
Taking, for simplicity, $k = n = 1$, one may check that
the two independent homogeneous solutions have the asymptotic forms
\begin{subequations}
\begin{align}
    G_{\rm I}(r)
    &\sim
    m_A^2 \, (m_A r)^{-3/2} \, e^{-m_A r} \,, 
    &H_{\rm I}(r)
    &\sim (\widetilde m^2 {-} m_A^2) \, (m_A r)^{1/2} \, e^{-m_A r} \,,
\label{eq:mAAsymptotics}
\\
    G_{\rm II}(r)
    &\sim
    (\widetilde m^2 {-} m_A^2) \,
    (\widetilde m r)^{-1/2} \, e^{-\widetilde{m} r}\,,
    &H_{\rm II}(r) 
    &\sim   m_A^2 \,
    (\widetilde{m} r)^{-1/2} \, e^{-\widetilde{m} r}  \,.
\label{eq:tildeMAsymptotics}
\end{align}
\end{subequations}
The most general solution is a linear combination of
solutions I and II.
Depending on whether $m_A$ or $\widetilde m$ is smaller,
either solution I or solution II dominates at large distance.
In either case, the gauge field profile function approaches
its asymptotic value far from the vortex exponentially fast.

%%%%%%%%%%%%
\section{Embedding in a non-Abelian gauge theory}
\label{sec:nonAbelian}
%%%%%%%%%%%%

%%%In this appendix we discuss an embedding of our Abelian model into a
%%%non-Abelian gauge theory.  Such an embedding makes the discussion of
%%%monopole-instantons more explicit, since they are manifestly
%%%finite-action field configurations in the continuum model we will
%%%discuss.  This non-Abelian model also illuminates some of the
%%%implications of our analysis for QCD. 

Consider a parity-invariant $SU(2)$ gauge theory
containing one real adjoint representation scalar field $\zeta$, one
fundamental representation scalar $\Phi$, and one $SU(2)$-singlet
complex scalar field $\phi_0$.
We take the action of the theory to be 
 \begin{align}
    S = \int d^3&x\,
    \Bigl[ \frac{1}{2g^2} \,
	\tr \mathcal{F}_{\mu\nu}^2
	+ |D_\mu\Phi|^2
	+ \tr (D_{\mu}\zeta)^2
	+ |\partial_{\mu} \phi_0|^2
	+ m^2_{\Phi} |\Phi|^2
	+  m_{0}^2 |\phi_0|^2
\nonumber \\ &{}
	+ \lambda_{\Phi} |\Phi|^4
	+ \lambda_{0} |\phi_0|^4
	+\lambda_{\zeta} \tr\!\left(\zeta^2{-}\tfrac 14 {v_{\zeta}^2}\right)^2 
	+\varepsilon\left(\phi_0 \, \Phi^Ti\sigma_2\zeta \Phi
	+ \text{h.c.}\right) 
	+ \cdots \Bigr],
\label{eqn:UVaction}
\end{align}  	 
where the covariant derivatives 
$D_{\mu} \Phi \equiv \partial_{\mu} \Phi - i \mathcal{A}_{\mu} \Phi$ and
$D_{\mu}\zeta \equiv \partial_{\mu}\zeta - i[\mathcal{A}_{\mu},\zeta]$,
the gauge field $\mathcal{A}_{\mu} \equiv \mathcal{A}^a_{\mu} \, t^a$,
and the $SU(2)$ generators obey $\tr t^a t^b = \half \delta^{ab}$.
The couplings $\varepsilon$, $\lambda_{\Phi}$, $\lambda_0$, and $\lambda_{\zeta}$
are assumed real and positive, and
the ellipsis stands for further scalar potential terms consistent with the
symmetries imposed below.

In addition to parity (and gauge and Euclidean invariance),
we assume the theory has a $\UG$ global symmetry acting as
\begin{equation}
    \UG: \  \Phi \to e^{-i\alpha} \Phi, \ \phi_0 \to e^{2i\alpha}\phi_0 \,.
\label{eqn:UGnonAbelian} 
\end{equation}
The $\mathbb{Z}_2$ subgroup generated by the $\alpha = \pi$ sign flip is part of
the $SU(2)$ gauge symmetry, so the faithfully-acting global symmetry
is $\UG/\mathbb{Z}_2$. We also assume the existence of a discrete global
symmetry we will call $\ZF$, acting as
\begin{align}
\ZF: 
\begin{pmatrix}
    \zeta \\[2pt]
    \Phi \\[2pt]
    \mathcal{A}_\mu
\end{pmatrix}
\to 
\begin{pmatrix}
    - U_4 \, \zeta \, U_4^{\dag} \\[2pt]
    -i U_4 \, \Phi \\[2pt]
    U_4 \, \mathcal{A}_\mu \, U_4^{\dag}
\end{pmatrix},
\label{eq:nonAbelianZ2}
\end{align}
where
$
    U_4 \equiv
    \left( \begin{smallmatrix} 0 & i \\ i & 0 \end{smallmatrix} \right)
    \in SU(2)$.
This transformation leaves the action (\ref{eqn:UVaction}) invariant
and acts as a $\mathbb{Z}_2$ symmetry on gauge invariant observables. 

We consider this model when $v_{\zeta}^2 \gg g^2$,
leading to Higgsing of the $SU(2)$ gauge group down to a $U(1)$
Cartan subgroup.
Choosing, for convenience, a gauge where $\zeta = v_{\zeta} \sigma_3/2$,
one sees that the ``color'' components $\mathcal A^1$ and $\mathcal A^2$ become
massive while $\mathcal{A}^3_{\mu}$ remains massless.
Writing $\Phi \equiv \big(\begin{smallmatrix} \phi_+ \\ \phi_- \end{smallmatrix}\big)$,
the component fields  $\phi_{\pm}$ transform with charge $\pm 1/2$ under
the unbroken $U(1)$ gauge group.
To write the resulting low energy theory, below the scale
$m_W \equiv g v_\zeta$,
in the most convenient form let
$e \equiv g/2$ and  $A_{\mu} \equiv \frac{1}{2}\mathcal{A}^{3}_{\mu} $.
This makes $A_{\mu}$ an Abelian gauge field with coupling $e$ interacting
with fields $\phi_\pm$ having charges $\pm 1$.
The final term in the action \eqref{eqn:UVaction} with coefficient $\varepsilon$
becomes
\begin{align}
    \varepsilon \left(\phi_0\Phi^Ti\sigma_2\zeta \Phi  +\text{h.c.}\right) 
    &=
    \varepsilon \, \phi_0 
	\left( \phi_+,\,  \phi_-  \right)
	\begin{pmatrix} 
	    0 & -v_{\zeta}/2 \\ 
	    -v_{\zeta}/2 & 0
	\end{pmatrix}
	\begin{pmatrix}
	    \phi_+ \\ 
	    \phi_- 
	\end{pmatrix} 
    + \text{h.c.}
\nonumber\\ &=
    -\varepsilon v_\zeta
    \left( \phi_0 \phi_+\phi_- + \text{h.c.}\right) ,
\label{eqn:UVcoupling}
\end{align}
after setting $\zeta$ to its expectation value.
If one now identifies $\epsilon \equiv \varepsilon v_{\zeta}$ and
$m_{c}^2 \equiv m_{\Phi}^2$,
then the resulting low-energy description of this non-Abelian  model,
now involving the fields $\phi_{\pm}$, $\phi_0$, and $A_{\mu}$,
precisely coincides with our original Abelian model (\ref{eq:the_model}).

Now consider the behavior of the non-Abelian model (\ref{eqn:UVaction})
as the mass parameters $m_{\Phi}^2$ and $m_0^2$ are varied.
To begin, suppose that $m_{\Phi}^2 \gg g^2 v_{\zeta}^2$,
so that the fundamental scalar field $\Phi$ is not condensed.
If $m_{0}^2$ is sufficiently negative, the neutral scalar
$\phi_0$ will condense and the $\UG$ symmetry will be spontaneously broken;
otherwise $\UG$ will be unbroken.
In either case, the $SU(2)$ adjoint Higgs mechanism
leads to the existence of stable finite-action
monopole-instantons~\cite{tHooft:1974kcl,Polyakov:1974ek}
whose stability is guaranteed by $\pi_2(SU(2)/U(1)) = \mathbb{Z}$.
The Abelian magnetic flux
(defined on scales large compared to $m_W^{-1}$)
through a spacetime surface $M_2$
has an $SU(2)$ gauge-invariant definition
\begin{align}
    \Phi_B(M_2)
    \equiv
    \frac{1}{|v_{\zeta}|}\int_{M_2} \tr\left( \zeta \mathcal{F} \right)\,,
\label{eq:nonAbelianflux}
\end{align}
where $\mathcal{F}$ is the two-form field strength.
We have normalized the flux $\Phi_B$ (not to be confused with the field $\Phi$)
so that when written in terms of the 2-form field strength
$F$ of the Abelian gauge field $A_{\mu}$, the flux has the conventional form
$\Phi_B = \int_{M_2} F$.
Note that $\Phi_B$ is odd under the $\ZF$ global symmetry.
The minimal magnetic monopole-instantons in $SU(2)$ gauge
theory have $\Phi_B(S^2) = \pm 2\pi$, where $S^2$
is a spacetime two-sphere surrounding the center of the
monopole-instanton. If the center of the monopole is at $r=0$,
then as $r\to \infty$, then at large distance from the monopole the
$SU(2)$ gauge field and adjoint scalar approaches the asymptotic forms
\begin{align}
    (\mathcal{A}^{\mu})^{a} \to \frac{ \epsilon^{a\mu \nu} \, \hat r_{\nu} }{r}
    \,,\qquad
    \zeta^{a} \to v_{\zeta}\, \hat r^a  \,,
\end{align}  
in ``hedgehog'' gauge, with $\hat r_{\mu}$ a radial unit vector.
The action of a monopole-instanton has the form
\begin{align}
    \SI = \frac{4\pi v_{\zeta}}{g} \,
    f\!\left(\frac{\lambda_{\zeta}}{g^2}\right),
\end{align}
where the dimensionless and monotonically increasing function $f$ varies
between  $f(0) = 1$ and $f(\infty) = 1.787$~\cite{Bogomolny:1975de,Prasad:1975kr,PhysRevD.24.999}.
%% (When $\lambda_{\zeta} = 0$, the action of the monopole satisfies a
%% Bogomol'nyi-Prasad-Sommerfield
%% bound~\cite{Bogomolny:1975de,Prasad:1975kr}.)
The associated monopole operator,
characterizing the effect of a monopole on long distance physics,
has the form $e^{i\sigma} e^{-\SI}$,
with $\sigma$ the magnetic dual of the Abelian field strength $F$.
As discussed in section \ref{sec:our_model},
these monopole-instantons generate a potential
of the form $e^{-\SI} \cos(\sigma)$
for the dual photon of $A_{\mu}$,
leading to a mass gap for the low energy Abelian gauge field and confinement of
heavy test charges.

Alternatively, if $-m^2_{\Phi}\gg g^4$ then the fundamental representation
scalar $\Phi$ will condense, leading to complete Higgsing of the
$SU(2)$ gauge symmetry, along with spontaneous breaking of the global
$\UG$ symmetry, regardless of the value of $m_{0}^2$.
Monopole-instantons are now confined by magnetic flux tubes,
and have negligible effect on long distance physics.
All components of the $SU(2)$ gauge field acquire mass via the
Higgs mechanism.

The question remains: are the confining (via Polyakov mechanism)
and fully Higgsed regimes, both with spontaneous $\UG$-breaking,
smoothly connected?
All of the analysis of section \ref{sec:vortices_and_holonomies}
generalizes in a straightforward fashion to this non-Abelian model,
and shows that the answer is no.
To see this, one may consider the natural generalization of our
previous vortex holonomy observable (\ref{eq:order_parameter}) which
replaces the Abelian gauge field holonomy with the trace of the
non-Abelian holonomy,%
\footnote
    {%
    Alternatively, one might consider writing the Abelian holonomy
    as the exponential of the magnetic flux through a surface spanning
    the holonomy contour, and then insert the definition (\ref{eq:nonAbelianflux})
    of the Abelian flux in terms of the underlying non-Abelian field strength.
    However, this generalization is undesirable as it converts the original
    line operator into a surface operator, for which one can no longer
    argue that the magnitude of the expectation value, in the large $r$ limit,
    must be independent of the presence of a vortex piercing the surface.
    With this generalization, the ratio of vortex and ordinary expectation
    values of the surface operator need not be a pure phase.
    }
\begin{align}
    O_{\Omega}^{SU(2)}
    \equiv
    \lim_{r \to \infty}
    \frac{\langle \tr \Omega(C) \rangle_1}{\langle \tr \Omega (C) \rangle} \,.
\label{eq:SU2_holonomy_order_parameter}
\end{align}
Just as in our original Abelian model,
the holonomy expectation values in  numerator and denominator will have
perimeter law decay of their magnitudes, but this size dependence cancels in the
ratio by construction.
The denominator is guaranteed to be positive in weakly coupled regimes
because charge conjugation (or reflection) symmetry requires it to be real,
it is positive at tree level, and hence
small quantum corrections cannot turn it negative.
So the ratio of expectations is determined by the phase of the
vortex expectation value in the numerator.

One can easily uplift the entirety of the analysis in section
\ref{sec:vortices_and_holonomies} to this non-Abelian setting.
In the Higgs phase, a unit-winding vortex configuration has the form
\begin{align}
       \Phi(r,\theta)
        = 
        v_{\Phi} 
        \begin{pmatrix}
            f_{+}(r) \, e^{i (n-1) \theta }  \\
            f_{-}(r) \, e^{ - i n \theta}
           \end{pmatrix}, \quad
        \phi_{0}(r,\theta) = v_0 \, f_0(r) \, e^{i \theta } \,, \quad
        \mathcal{A}_{\theta}(r) &= \frac{a\, h(r)}{2\pi r} \frac{\sigma_3}{2} \,.
\label{eq:SU2_ansatz}
\end{align}
The resulting long-distance energy density, 
generalizing Eq.~\eqref{eq:energy_long_distance}, is
\begin{equation}
    \mathcal E(r)
    =
    \frac {v_{\Phi}^2}{r^2}
    \left[
	\left(n-1 - \frac{a}{4\pi}\right)^2 + \left(-n + \frac{a}{4\pi}\right)^2
    \right]
    +
    \frac{v_0^2 \, k^2}{r^2} 
    + \mathcal O(r^{-4})
    \,.
\label{eq:SU2_energy_long_distance}
\end{equation}
The minimum lies at $a = 2\pi(2n{-}1)$,
leading to the tree-level result,
\begin{align}
    \label{eq:SU2_holonomy_tree_level}
    \half \langle \tr \Omega(C) \rangle_1
    =
    - 1 \,,
\end{align}
and a phase of $\langle \tr \Omega(C) \rangle_1$, at long distance, equal to $\pi$.

To see that the phase of $\langle \tr \Omega(C) \rangle_1$ must remain
at $\pi$ even when quantum fluctuations are taken into account, one can
adapt the effective field theory argument at the end of
Sec.~\ref{sec:Higgs_holonomy_Abelian}.  Integrating out fluctuations
generates corrections to the tree-level effective action. The only terms in
the effective action that can affect expectation value of holonomies
along contours far from the vortex core are those with exactly
two derivatives acting on $\Phi$, because only such operators 
can affect the  $\mathcal{O}(1/r^2)$ holonomy-dependent part of the energy density.
Given the symmetries of our $SU(2)$ model,
all such terms may be written in the form
\begin{align}
    \label{eq:eff_action_SU2}
    S_{\textrm{eff}, \, SU(2) \textrm{ holonomy}}
    = \int d^{3}x \, \bigg\{
     & D_{\mu} \Phi^\dagger f_1(\phi_0,\zeta,\Phi) D^\mu\Phi
 \nonumber\\ & {}
    +  (D_{\mu} \Phi)^{T} i\sigma_2\, f_2(\phi_0,\zeta,\Phi) D^{\mu} \Phi \bigg\} + \textrm{h.c.} 
\end{align}
%The functions $f_1$ and $f_2$, depending on the indicated fields 
%but not their derivatives, must be constructed so as to be
%invariant under the $\UG$ symmetry, transform in the adjoint representation
%of the $SU(2)$ group,
%and be conjugated by $U_4$
%under the $\ZF$ symmetry \eqref{eq:nonAbelianZ2}.
The functions $f_1$ and $f_2$, depending on the
indicated fields but not their derivatives, transform in the adjoint
(or singlet) representation of the $SU(2)$ group. The function $f_1$
is invariant under the $\UG$ symmetry, and is conjugated by $U_4$
under the $\ZF$ symmetry \eqref{eq:nonAbelianZ2} (like
$\Phi^\dagger\Phi$ or $\Phi\, \Phi^\dagger$). The function $f_2$ has
charge $+2$ under $\UG$, and transforms as $f_2 \to -U_4 f_2
U_4^\dagger$ under $\ZF$ (like $\phi_0\Phi\, \Phi^\dagger\zeta$ or $
i\sigma_2(\Phi^\dagger)^T \Phi^\dagger$). 

Just as in
Sec.~\ref{sec:Higgs_holonomy_Abelian}, one may verify that both terms,
in the presence of a unit-winding vortex,
have a minimum at the value
$a = 2\pi(2n{-}1)$ for the asymptotic coefficient of the gauge field.
Therefore, small quantum corrections cannot shift the phase of the holonomy
and we learn that:
\begin{align}
    \boxed{ \textrm{Higgs phase:}\;\; O_{\Omega}^{SU(2)} = -1} \,.
\label{eq:SU2_holonomy_higgs}
\end{align}

On the other hand, in the $\UG$-broken confining regime when $m_{\Phi}^2 \gg g^4$,
one may reapply the arguments of Sec.~\ref{sec:holonomy_confining} to show that:
\begin{align}
    \boxed{\textrm{$\UG$-broken confining phase:} \;\;O_{\Omega}^{SU(2)} = 1} \,.
\end{align}
So, as claimed,
$O_{\Omega}^{SU(2)}$ serves as an order parameter that distinguishes the
$\UG$-broken confining and Higgs phases in this $SU(2)$ gauge theory. 

%%%%%%%%%%%%
\section{$U(1) \times U(1)$ gauge theory and topological order}
\label{sec:gaugingU1}
%%%%%%%%%%%%

Gauging the global $\UG$ symmetry of our model (\ref{eq:the_model}),
by adding a second dynamical gauge field minimally coupled to the
conserved current associated with the $\UG$ symmetry, converts the
model into a $U(1) \times U(1)$ gauge theory. This process has the
effect of converting the massless Nambu-Goldstone boson associated
with spontaneous breaking of the global $\UG$ symmetry into the
longitudinal component of a massive gauge field, thereby producing a
mass gap in the $U(1) \times U(1)$ gauge theory. Superfluid systems
(gapless due to global symmetry breaking) and superconducting systems
(gapped due to the Meissner effect) are related in precisely this
manner.

The $U(1) \times U(1)$ gauge theory produced by gauging the $\UG$
symmetry of our model (\ref{eq:the_model}) no longer has any
continuous global symmetries and is expected to have a non-vanishing
mass gap at generic points within its parameter space.  This makes it
easier to analyze than the gapless models considered in the body of
the paper.\footnote
{%
We thank Z.~Komargodski for urging us to pursue the calculations
described in this appendix.}
The long-distance physics of the $U(1)
\times U(1)$ gauge theory can be described by topological quantum
field theories (TQFTs) at generic points in parameter space.    The
phase diagram of the $U(1) \times U(1)$ gauge theory produced by
gauging the $\UG$ symmetry of our model (\ref{eq:the_model}) turns out
to be very similar to the phase structure of our original model
\eqref{eq:the_model}.%

Let $X$ and $Y$ denote the gauge fields associated each of the $U(1)$ factors
of the gauge group, which we henceforth denote as
$\UX \times \UY$.
Let $F_X$ and $F_Y$ denote the corresponding field strengths,
and $e_X$ and $e_Y$ the gauge couplings of the two different gauge fields.
The charge assignments of the scalar fields are:
\begin{align}
\begin{array}{c|ccc}
 & \phantom{+}\phi_+ &  \phantom{+}\phi_- & \phantom{+}\phi_0 \\ \hline
\UX & +1 & -1 & \phantom{+}0 \\
\UY & -1 & -1 & +2
\end{array}
\label{eq:chargeTable2}
\end{align}
We assume the standard magnetic flux quantization condition
holds for both $F_{X}$ and $F_{Y}$,
\begin{align}
    \int_{S^2} F_X = 2\pi k_X \,, \;\;     
    \int_{S^2} F_Y= 2\pi k_Y \,, \qquad k_{X},k_{Y} \in \mathbb{Z}\,, 
\end{align}
and assume that finite action monopole-instantons
preclude the existence of any magnetic $U(1)$ global symmetries. 
%
% \footnote
%     {%
%     This Abelian model can be embedded in a natural way in an $SU(3)$
%     gauge theory coupled to one adjoint scalar $\zeta$ and one fundamental
%     scalar $\Phi$.
%     }
We also assume that the fundamental representation Wilson loop operators
$\Omega_X = e^{i \int_C X}$ and $ \Omega_Y = e^{i \int_C Y}$ are 
genuine line operators in the sense of Ref.~\cite{Kapustin:2014gua}.

The action is a simple extension of the original model (\ref{eq:the_model}),
\begin{align}
\label{eq:the_model_XY}
    S = & \int d^3x \, \left[
	\frac{1}{4e_X^2}F_X^2
	+ \frac{1}{4e_Y^2}F_Y^2
	+ |D\phi_+|^2 + |D\phi_-|^2 + |D\phi_0|^2
	- \epsilon \, (\phi_+\phi_-\phi_0 + \text{h.c.})
    \right.
\nonumber \\ &\qquad \left.\vphantom{\frac{1}{4}}
	+ m_+^2 |\phi_+|^2 + m_-^2|\phi_-|^2 + m_0^2|\phi_0|^2
	+ \lambda_+|\phi_+|^4 + \lambda_-|\phi_-|^4 + \lambda_0|\phi_0|^4
	+ \cdots \right.
\nonumber\\ & \qquad \left. \vphantom{\,\frac{1}{4}}
	 + \Vm(\sigma_X) + \Vm(\sigma_Y) \right] \,.
\end{align}
The $\Vm(\sigma_X)$ and $ \Vm(\sigma_Y)$ terms describe the effects of
monopole-instantons for the $X$ and $Y$ gauge fields, respectively.
The cubic $\epsilon$ term ensures that there is no global $U(1)$ symmetry
despite the presence of three charged scalar fields and only two gauge
bosons. 

Our model enjoys Euclidean (or Lorentz) invariance, including reflection
and time-reversal symmetry.
We do not assume any discrete flavor symmetry permuting the different
scalar fields, nor any symmetry interchanging the two $U(1)$ subgroups.\footnote{A nearly identical model to \eqref{eq:the_model_XY} with a $\mathbb{Z}_2$ flavor permutation symmetry was studied in Ref.~\cite{Carroll_1998}.}
Given the charge assignments (\ref{eq:chargeTable2}),
the $\mathbb{Z}_2$ transformation $(-1,-1) \in \UX \times \UY$
acts trivially on all three scalar fields $\phi_{+}$, $\phi_{-}$ and $\phi_{0}$.
Consequently, the theory has a 1-form symmetry,
which we denote by $(\mathbb{Z}_2)^{(1)}_{XY}$,
which acts on topologically non-trivial Wilson loops as:
\begin{align}
    (\mathbb{Z}_2)^{(1)}_{XY}: \;\Omega_X \to - \Omega_X \,, 
    \qquad \Omega_Y \to - \Omega_Y.
\label{eq:Z2XY}
\end{align}   

Now consider the resulting phase diagram. When all three scalars have
large positive masses, they can be integrated out resulting in a pure
$\UX\times\UY$ gauge theory, which has a mass gap and a unique vacuum
thanks to the Polyakov mechanism.  If only one of the fields
$\phi_{\pm}$ is condensed, then there is again a mass gap and a unique
vacuum thanks to a combination of the Higgs and Polyakov mechanisms.
Other portions of the phase diagram can be mapped out by considering:
(i) the regime where $\phi_0$ is condensed but the $\phi_{\pm}$ fields
are not condensed, and (ii) the regime where all three scalar fields,
$\phi_{\pm}$ and $\phi_0$, are condensed. So long as the cubic
coupling $\epsilon \neq 0$,  there is no separate regime where the
fields $\phi_{+}$ and $\phi_{-}$ are condensed, but $\phi_0$ is not.
Nor is there a regime where, e.g., $\phi_0$ and $\phi_{+}$ are
condensed but $\phi_{-}$ is not. 

%%%%%%%%%%%%
\subsection*{$\phi_0$ condensed phase}
%%%%%%%%%%%%

Suppose that the $\phi_{\pm}$ fields have large positive masses,
so that they may be integrated out.
The $\UX$ gauge field will be gapped, as usual, thanks to the Polyakov mechanism.   
Condensation of $\phi_0$ causes Higgsing of the $\UY$ gauge field,
showing that this regime is (generically) gapped.  Let us call the
regime where only the $\phi_0$ field is condensed the Y regime.
Despite the fact that it is gapped, the Y regime is
not become completely trivial in the deep infrared because the
emergent gauge group at long distances is $\mathbb{Z}_2$, and the
resultant physics is described by a non-trivial topological quantum
field theory (TQFT).%
\footnote
    {%
    See, e.g., Refs~\cite{Maldacena:2001ss,Hansson:2004wca}.
    Reference~\cite{Gaiotto:2014kfa}
    explains that the TQFT associated with a $\mathbb{Z}_2$ gauge theory
    can be viewed as an effective field theory describing a spontaneously
    broken $\mathbb{Z}_2$ 1-form symmetry.  Here, this one-form $\mathbb{Z}_2$
    symmetry acts by flipping the sign of $\Omega_Y$.
    }

Before discussing the TQFT description, let us consider the physics of
the system through a more direct approach. 
After integrating out $\phi_{+}$ and $\phi_{-}$, the resulting effective action is
\begin{align}
\label{eq:SeffGauged}  
    S_{\text{eff}} =
    \int d^3x \,&\left[
	\frac{1}{4e_X^2} \, F_X^2
	+ \frac{1}{4e_Y^2} \, F_Y^2 
	+ |D \phi_0|^2 + V(|\phi_0|) \right.
\nonumber\\ &\left.{}
	+ \frac{c_X}{m^2} \, |\phi_0|^2F_X^2
	+ \frac{c_Y}{m^2} \, |\phi_0|^2F_Y^2
	+ \frac{b_X}{m^2} \, \mathcal{S}_{\mu\nu}F^{\mu\nu}_X  
	+ \frac{b_Y}{m^2} \, \mathcal{S}_{\mu\nu}F^{\mu\nu}_Y 
	+\cdots \right],
\end{align}
where $m= \min(m_+,m_-)$,
$D_\mu \equiv \partial_\mu - 2i Y_\mu$,
and
$
    \mathcal S_{\mu\nu} \equiv
    \frac{i}{2}[(D_\mu\phi_0)(D_\nu\phi_0^\dagger)-(D_\nu\phi_0)(D_\mu\phi_0^\dagger)]
$.
A minimal vortex configuration has the usual form,
written as 
\begin{align}
    \phi_0(r,\theta) &= v_0 \, f_0(r)\, e^{i\theta}\,,
    \qquad
    Y_\theta(r) = {\Phi_Y\, h(r)}/(2\pi r) \,, 
\label{eq:Bvortex} 
\end{align}
where $v_0$ is the $\phi_0$ vacuum expectation value
and the radial functions $f_0$ and $h$ interpolate between 0 and 1 as
$r$ goes from 0 to $\infty$.
The asymptotic gauge field coefficient $\Phi_Y$ is determined by minimizing
the long-distance energy density,
\begin{equation}
    \mathcal E(r) =
    \frac{v_0^2}{r^2} \left(1  -\frac{\Phi_Y}{\pi}\right)^2 + \mathcal O(r^{-4}),
\end{equation}
which must vanish to prevent a logarithmic IR divergence in the vortex
energy, implying that $\Phi_Y = \pi$.  One can prove that small
quantum corrections cannot shift $\Phi_Y$ away from this value by
using the effective field theory analysis in
\ref{sec:Higgs_holonomy_Abelian}.  
Hence, if $C$ is a large circular contour centered on the vortex, then
the holonomy $\Omega_Y(C)$ has a phase of $\pi$.
More physically, this means that a test particle with unit charge under $\UY$
picks up a phase of $\pi$ when it moves around a unit-circulation vortex.  

Next let us consider the behavior of $X$ holonomies.  Consider a
test particle with unit charge under $U(1)_X$ and zero charge under
$U(1)_Y$.  What is the phase acquired by such a test particle when it
encircles the $\phi_0$ vortex?   The answer is not immediately obvious
when the coupling $b_X$ is non-zero.
(A non-zero value for $b_X$ may appear in the
absence of any flavor permutation symmetry which also flips the sign of
$F_X$.) In the presence of a winding-$k$ vortex, the antisymmetric
tensor $\mathcal S_{\mu\nu}$ is non-vanishing with
\begin{align}
    \mathcal S_{r\theta} = \frac{f(r)f'(r)}{r} \left[k- \frac{\Phi_Y}{\pi} \, h(r) \right].
\label{eq:Srtheta} 
\end{align}
When $k=1$, we know that $\Phi_Y = \pi$. When $b_X \neq 0$, the source
\eqref{eq:Srtheta} corresponds to an azimuthal $J_X$ current
encircling the vortex which, in turn, generates a magnetic field $B_X
= \frac{1}{2}\epsilon_{ij} F_X^{ij}$ localized on the vortex core.
Just as in Sec.~\ref{sec:broken_permutations}, due to confinement, the
extent to which this matters depends on size of the loop with which
one probes the system.  Consider a spatial disk $D$ with a $\phi_0$
vortex at its center and the $\Omega_X$ holonomy calculated along the
boundary of $D$. When the radius of $D$ is small compared to the
string breaking scale, the magnetic flux through $D$ is non-zero, and
$O_{\Omega_X} = e^{i\Phi}$ with $\Phi \propto b_X$.  But for
holonomies on large contours (with radius $r \gg L_{\rm br})$, string
breaking effects remove the sensitivity to $b_X$, and we find
$O_{\Omega_X} =1$. So the expectation values of $X$ holonomies that
encircle $\phi_0$ vortices on contours $C = \partial D$ are positive
in the limit of large contour radius.

The information about holonomies around vortices is encoded into the
TQFT description of the infrared limit of the system. The action for
the topological field theory describing the Y regime, which we denote
as $\textrm{TQFT}_{Y}$, can be written using the $K$-matrix formalism
\cite{PhysRevLett.65.1502,PhysRevB.42.8145,PhysRevB.46.2290,
doi:10.1080/00018739500101566,Lu:2012dt,Lu:2013jqa,Delmastro:2019vnj}.
In the regime we are considering, $\phi_0$ vortex excitations cost
finite energy and can be viewed as one type of probe excitation, while
test particles with unit charge under $\UY$ are another. Let $J_Y$
denote the conserved $\UY$ current, and $J_{V}$ the topologically
conserved vortex number current. These currents couple to two
different one-form gauge fields, $a^i$, with $i = 1,2$, each obeying
$\int_{M_2} d a^i \in 2\pi\mathbb{Z}$ for any closed 2-surface $M_2$.
Physically, we can identify $a^1_{\mu} = Y_{\mu}$, while $a^2_{\mu}$
arises in the derivation of the TQFT description as a Lagrange
multiplier that enforces the condition that almost everywhere
$\epsilon^{\mu\nu\rho}\, \partial_{\nu} \partial_{\rho} \phi_0 = 0$
while allowing $\oint dx_{\mu} \, \partial^{\mu} \phi_0 \in 2\pi
\mathbb{Z}$.  Then a description of this $\mathbb Z_2$ TQFT is
provided by the action
\begin{align}
    S_{\textrm{TQFT}_{Y}}
    = \int d^3x\,\left[
	 \frac{i}{4\pi} (K_Y)_{ij} \, \epsilon^{\mu\nu\rho} \, a^{i}_{\mu} \, \partial_{\nu} a^{j}_{\rho}
	+ a^1_{\mu}\, J_Y^{\mu} + a^2_{\mu}\, J_V^{\mu} \right],
\label{eq:phi0_TQFT}
\end{align} 
where the $K$-matrix and its inverse (times $2\pi$) are given by
\begin{align}
    K_Y &= \begin{pmatrix}
	\phantom{,}0 & \phantom{+}2 \phantom{,}\\	
	\phantom{,}2 & \phantom{+}0\phantom{,}
	\end{pmatrix}
    ,\qquad
    2\pi K_Y^{-1} = \begin{pmatrix}
	\phantom{,}0 & \phantom{+}\pi \phantom{,}\\	
	\phantom{,}\pi & \phantom{+}0\phantom{,}
	\end{pmatrix} .
\label{eq:Kmtx}
\end{align}
The matrix element $(2\pi K_Y^{-1})_{ij}$ gives
the phase that an excitation of type $i$
picks up under braiding around one of type $j$.  
This TQFT has $|\det K_Y|^{g} = 4^g$ ground states on compact spatial
manifolds of genus $g$.  Note that the $X$ gauge field does not appear
in the TQFT description at all.   In this way the TQFT \eqref{eq:phi0_TQFT} is
implicitly consistent with the above result 
that all $X$ holonomies have trivial phases in the long-distance limit.

%%%%%%%%%%%%
\subsection*{$\phi_0$, $\phi_{+}$, $\phi_{-}$ condensed phase}
%%%%%%%%%%%%

Now consider the regime where all three scalar fields $\phi_{0}$,
$\phi_{+}$, and $\phi_{-}$ are condensed, and  the $X$ and $Y$ gauge
fields are both Higgsed.  We will call this the XY regime.  The
infrared physics of the XY regime be described by a topological
field theory which we denote as
$\textrm{TQFT}_{XY}$.

% The XY regime is
% weakly coupled when the Higgs mass scales $m_X, m_Y$ are large
% compared to $e_X^2, e_Y^2$ and the other dimensionful interaction
% parameters, and the dimensionless sextic couplings are small.   

Before discussing the TQFT description, it is again useful to explore
the physics of vortices directly. Consider the weakly coupled corner
of the parameter space of the XY regime, and suppose that there is a vortex where $\phi_+$ winds
by $2\pi n_+$ and $\phi_-$ winds by $2\pi n_-$ on contours encircling
the vortex core. Such a field configuration has the form:
\begin{subequations}
\begin{align}
    \phi_+(r,\theta) &= v_+\, f_+(r)\, e^{in_+\theta}\,,
    &X_\theta(r) &= {\Phi_X\, g(r)}/(2\pi r)\,,
\\
    \phi_-(r,\theta) &= v_-\, f_-(r)\, e^{in_-\theta}\,,
    & Y_\theta(r) &= {\Phi_Y\, h(r)}/(2\pi r)\,, 
\\
    \phi_0(r,\theta) &= v_0\, f_0(r)\, e^{-i(n_++n_-)\theta}\,.
\end{align}
\end{subequations}
The radial functions $f$, $g$ and $h$ approach $1$ as $r\to\infty$.
The long-distance energy density of this vortex configuration is 
\begin{align}
\label{eq:all_condensed_gauged}
    \mathcal E(r) 
    = & \frac{1}{r^2} \left[ 
	 v_+^2 \Big(n_+ - \frac{\Phi_X{-}\Phi_Y}{2\pi}\Big)^{\!2}
	+ v_-^2 \Big(n_-+\frac{\Phi_X{+}\Phi_Y}{2\pi}\Big)^{\!2}
	+v_0^2\Big(n_+{+}n_- +\frac{\Phi_Y}{\pi}\Big)^{\!2}
    \right]
    + \mathcal O(r^{-4})
\end{align} 
and, for given values of $n_+$ and $n_-$, $ \mathcal E(r) $ is
minimized when the $X$ and $Y$ magnetic fluxes $\Phi_X,\Phi_Y$ are
\begin{align}
    \Phi_X &= (n_+-n_-)\pi\,, \;\;  \Phi_Y = -(n_++n_-)\pi \,. 
    \label{eq:TQFT_XY_flux}
\end{align} 
Since the winding numbers $n_\pm$ are integers,
these two fluxes are identical modulo $2\pi$.
The vortices with minimal winding and minimal magnetic flux correspond to:
\begin{equation}
\begin{array}{c|cccc}
 & \phantom{+}n_+ & \phantom{+}n_- & \phantom{+}\Phi_X & \, \Phi_Y \phantom{+} \\ \hline
 V_+  & \phantom{+}1 & \phantom{+}0 & \phantom{+}\pi & \, -\pi\phantom{+} \\
 V_-  & \phantom{+}0 & \phantom{+}1 & -\pi &  \, -\pi\phantom{+}
\end{array}
\label{eq:vortices2}
\end{equation}
The result (\ref{eq:vortices2}) shows that test particles with unit
charge under $\UY$ pick up a phase of $-\pi$ when encircling either
$V_{+}$ or $V_{-}$, while test particles with unit charge under $\UX$
pick up a phase of $\pm \pi$ when encircling a $V_\pm$ vortex. 
% This
% agrees with the phases encoded in the $K$-matrix \eqref{eq:Kmtxy}.   

The discussion in the body of this paper implies that the XY and Y
regimes cannot be smoothly connected.  Both phases contain vortices
with flux $\Phi_{Y} = \pi \textrm{ mod } 2\pi$, but in the XY regime
any vortices that have $\Phi_{Y} = \pi \textrm{ mod } 2\pi$ also have
$\Phi_{X} =   \pi \textrm{ mod } 2\pi$.  By comparison, vortices in
the Y regime only carry Y flux.   The X flux carried by vortices
changes non-analytically as we go from one regime to the other.
%%% If we vary parameters so that the system interpolates between the Y and
%%% XY regimes along a vortex with minimal $Y$ flux, then there is a
%%% boojum --- an $X$ magnetic monopole --- pinned to the $Y$ vortex at
%%% the interface between the regimes, with X flux being localized on
%%% one side of the phase boundary and spread out on the other.    
We can repeat the logic in Sec.~\ref{sec:ColemanWeinberg} to
argue that such changes are associated with non-analyticities in
thermodynamic observables, showing that the Y and XY regimes are
distinct phases of matter. The phase transition between the Y and XY
regimes  is the parallel of the confinement to Higgs phase transition
discussed in the main part of this paper.  

We now consider the TQFT description of the long-distance physics
of the XY regime.
Given the charge assignments (\ref{eq:chargeTable2}), the unbroken
part of the gauge group in the XY regime is generated by the $\mathbb
Z_2$ transformation $(-1,-1) \in \UX\times\UY$, and the 1-form
symmetry $(\mathbb{Z}_2)^{(1)}_{XY}$, acting as shown in
Eq.~\eqref{eq:Z2XY}, is spontaneously broken.
%%% The long distance
%%% physics of the XY regime is thus necessarily described by a TQFT,
%%% which we denote as $\textrm{TQFT}_{XY}$.
We can derive the appropriate TQFT describing this regime
directly from the original model~\eqref{eq:the_model_XY}
in the weakly coupled corner of parameter space of the XY regime. By
virtue of being topological, the resulting effective action will
furnish a valid description of the physics even away from the weak
coupling limit.  We follow a procedure similar to that in Sec.~3.3
of Ref.~\cite{Hansson:2004wca}. In the weak coupling and long distance
limits, we can freeze the moduli of the scalar fields to their vacuum
expectation values because all physical fluctuation modes around the
expectation values are gapped and can be integrated out.  Let us
denote the phases of the three scalar fields by $\varphi_0$, $\varphi_+$,
and $\varphi_-$. Since we are interested in the low energy form of the
effective action, we note that minimizing the cubic $\epsilon$ term in
model \eqref{eq:the_model_XY} implies that $\varphi_0 +\varphi_+ +
\varphi_-=0$.  With all this taken into account, the relevant part
of the effective action becomes just
\begin{align}
\label{eq:L0}
\mathcal L_{\textrm{St\"uckelberg }} &=
    v_+^2 \, (\partial_\mu\varphi_+ -X_\mu{+}Y_\mu)^2 
    + v_-^2 \,(\partial_\mu\varphi_- +X_\mu{+}Y_\mu)^2
%\nonumber\\ &{}
    + v_0^2\, (\partial_\mu\varphi_+ +\partial_\mu\varphi_-{+}2Y_\mu)^2 \,,
\end{align}
where $v_0$, $v_+$, and $v_-$ are the magnitudes of the expectation
values of $\phi_0$, $\phi_+$, and $\phi_-$, respectively. We have
dropped the Maxwell terms because we are interested in length scales
which are large compared to $1/e_X^2$ and $1/e_Y^2$.
%%% We also dropped the scalar
%%% potential terms because their role is simply to determine the values
%%% of $v_{0},v_+, v_{-}$.
The long distance TQFT is obtained by
dualizing $\varphi_+$ and $\varphi_-$ and taking the low energy limit
$v_0,v_+,v_- \to \infty$. To this end, we introduce one-form Lagrange
multiplier fields $a^+$ and $a^-$ satisfying $\int_{M_2}da^\pm  \in
2\pi\mathbb{Z}$ for any closed 2-manifold $M_2$. Then the Lagrangian,
\begin{align}
\mathcal L_{\textrm{dual}} = \mathcal L_{\textrm{St\"uckelberg }}
+ \frac{i}{2\pi}\epsilon^{\mu\nu\alpha} \, a^+_\mu \, \partial_\nu \partial_\alpha\varphi_+
+ \frac{i}{2\pi}\epsilon^{\mu\nu\alpha} \, a^-_\mu \, \partial_\nu \partial_\alpha\varphi_- \,,
\end{align}
enforces
$\epsilon^{\mu\nu\rho}\partial_\nu\partial_\rho\varphi_\pm = 0$
almost everywhere while allowing
$\oint dx_\mu \, \partial^\mu \varphi_\pm \in 2\pi \mathbb{Z}$.
Using the resulting equations of motion one finds
\begin{align}
    \partial^\mu\varphi_\pm =
    \pm X^\mu - Y^\mu
    -\frac{i}{4\pi}
    \frac{(v_{\mp}^2 + v_0^2) \, \epsilon^{\mu\nu\rho} \,
	\partial_\nu a^{\pm}_\rho
	- v_0^2 \, \epsilon^{\mu\nu\rho} \, \partial_\nu a^\mp_\rho}
	{v_0^2v_+^2+v_0^2v_-^2+v_+^2v_-^2} \,,
\end{align}
and
\begin{align}
    \mathcal L_{\textrm{dual}} =
    \frac{i}{2\pi} \, \epsilon^{\mu\nu\rho}
	\left[
	    (X_\mu-Y_\mu) \, \partial_\nu a^+_\rho
%%%    +\frac{i}{2\pi} \, \epsilon^{\mu\nu\rho} \,
	    -(X_\mu+Y_\mu) \, \partial_\nu a^-_\rho
	\right] + \mathcal O(v_i^{-2}) \,.
\end{align}
This Lagrangian (when the $v_i \to \infty$)
describes a topological field theory,
$\textrm{TQFT}_{XY}$.   To write the result  in a more
useful form, denote the set of gauge fields by $\{a^i_\mu\} =
\{X_\mu, Y_\mu, a^+_\mu, a^-_\mu\}$. We also introduce a set of
currents $\{J^i_\mu \}$, which are respectively the $U(1)_X$ and
$U(1)_Y$ Noether currents and the topological vortex number currents
associated with $\phi_+$ and $\phi_-$ vortices.  Then the action for
$\textrm{TQFT}_{XY}$ can be written as
\begin{align}
    S_{\textrm{TQFT}_{XY}} =
     \int d^3x \,\left[\frac{i}{4\pi}
	(K_{XY})_{ij} \, \epsilon^{\mu\nu\rho} \,  a^i_\mu \partial_\nu a^j_\rho
	+ a_\mu^i \, (J^{\mu})_i\right] \, , 
\end{align} 
with $K$-matrix (and its inverse)
\begin{align}
    K_{XY} = \begin{pmatrix}
    \phantom{+}0  & \phantom{+}0  & \phantom{+}1 &        -1\,\, \\
    \phantom{+}0  & \phantom{+}0  & -1 & -1\,\, \\
    \phantom{+}1  & -1  & \phantom{+}0 & \phantom{+}0\,\, \\
    -1            & -1  & \phantom{+}0 & \phantom{+}0\,\,
    \end{pmatrix},
    \qquad
    2\pi K_{XY}^{-1} = \begin{pmatrix}
    \phantom{+} 0   & \phantom{+}0    & \phantom{+}\pi  & -\pi\,\, \\
    \phantom{+} 0   & \phantom{+}0    & -\pi  &-\pi \,\, \\
    \phantom{+}\pi  & -\pi  &\phantom{+}0     & \phantom{+}0 \,\, \\
    - \pi           &  -\pi & \phantom{+}0    & \phantom{+}0 \,\,
    \end{pmatrix}.
    \label{eq:Kmtxy}
\end{align}
This TQFT has $|\det K_{XY}|^g = 4^g$ ground states on spatial
manifolds of genus $g$ in accordance with expectations from the
spontaneously broken $(\mathbb{Z}_2)^{(1)}_{XY}$ symmetry. 

It is possible to find an interesting relation between $\textrm{TQFT}_{Y}$ and
$\textrm{TQFT}_{XY}$.
Suppose we add a new spectator field $\chi$ which has charge $+1$ under
a new gauge field $Z_\mu$, obeying standard flux quantization
conditions, and assume that the microscopic theory admits
finite action $Z$ monopole events. We further suppose that $\chi$
is condensed in the Y regime, and has a large positive mass squared
in the XY regime.  Then the $K$-matrix of 
$\textrm{TQFT}_{XY}$ remains unchanged, but the $K$-matrix of the Y regime is enlarged and becomes
\begin{align}
     \tilde{K}_{Y} =   \begin{pmatrix}
           \phantom{+}0  & \phantom{+}2  & \phantom{+}0 &        \phantom{+}0\,\, \\
           \phantom{+}2  & \phantom{+}0  & \phantom{+}0 & \phantom{+}0\,\, \\
           \phantom{+}0  & \phantom{+}0  & \phantom{+}0 & \phantom{+}1\,\, \\
           \phantom{+}0           & \phantom{+}0  & \phantom{+}1 & \phantom{+}0\,\,
           \end{pmatrix} .
\end{align}
The action of the Y regime is now $ \int d^{3}x \,
\left[\tfrac{i}{4\pi} \epsilon^{\mu\nu\rho}a^i_\mu (\tilde{K}_{Y})_{ij}\, \partial_\nu a^j_\rho + a^i_{\mu}
J^{\mu}_i\right]$,
with the set of gauge fields $\{a^i_\mu\} = \{Y_\mu,a^0_\mu,Z_\mu,a^\chi_\mu\}$. Here $a^0_\mu$ and $a^\chi_\mu$ are auxiliary gauge fields
that couple to the $\phi_0$ and $\chi$ vortex currents, respectively.
Then one can verify the congruence relation $G^T K_{XY} \, G =
\tilde{K}_{Y}$ where $G \in GL(4,\mathbb{Z})$ is the matrix
\begin{align}
    G =  \begin{pmatrix}
        -1            & \phantom{+}0  & \phantom{+}0 & \phantom{+}0\,\, \\
        \phantom{+}1  & \phantom{+}0  & -1           & \phantom{+}0\,\, \\
        \phantom{+}0  & -1            & \phantom{+}0 & \phantom{+}0\,\, \\
        \phantom{+}0  & \phantom{+}1  & \phantom{+}0 & \phantom{+}1\,\,
        \end{pmatrix} .
\end{align}
This shows that the extended set of 
gauge fields in the Y regime are related to those in the XY regime by
the change of basis $a_{\rm Y} = G \,  a_{\rm XY}$, or explicitly
\begin{align}
    \begin{pmatrix}
        Y \\
        a^0 \\
        Z \\
        a^\chi
    \end{pmatrix}
    \to
    \begin{pmatrix}
        -X \\
        -a^{+} \\
        -X{-}Y \\
        a^{+}{+}a^{-}
    \end{pmatrix} .
\label{eq:observable_mapping}
\end{align}
We emphasize that the existence of the relation \eqref{eq:observable_mapping}
 does not contradict our assertion above that the Y and XY regimes
 cannot be smoothly connected and must be separated by a phase boundary.%
 \footnote
{%
     For an analogous example, recall that the high and low
     temperature regimes of the 2D Ising model on a square lattice
     are related by Kramers-Wannier duality. 
     Nevertheless, the high and low temperature regimes are distinct
     phases separated by a phase transition at the self-dual point.  
}

%%%%%%%%%%%%%%%%%%%%
\subsection*{Phase transitions and un-gauging limits}
%%%%%%%%%%%%%%%%%%

Having just seen the $U(1) \times U(1)$ gauge theory has distinct
gapped phases which are necessarily separated by phase transitions, in
a completely parallel fashion with what we inferred by direct
calculations in the original $U(1)$ gauge theory with gapless phases,
it is natural to ask whether our direct study in the original $U(1)$
gauge theory was really necessary. In other words, can one presume
that distinct phases present after one weakly gauges a continuous
global symmetry survive the limit of sending the coupling of the
artificially introduced gauge field back to zero? Alternatively, if
two regimes can be smoothly connected in a given theory with a
dynamical gauge field, does this necessarily remain true in the un-gauging limit?
We argue that the answer to both of these questions is no. 

For  systems with discrete symmetry groups, it has been established
that phase transitions detected by changes in particle-vortex
statistics in gauged models imply phase transitions in the parallel
un-gauged models, see e.g.,
Refs.~\cite{Levin:2012yb,Wang:2014xba,Wang:2014wka}. It may be
tempting to assume that the same should be true with continuous
symmetries, and in some simple examples this parallel between phases
in gauged and un-gauged models does hold. For instance, the phase
structure of a theory of a single parity-invariant complex scalar
field in three spacetime dimensions with a $U(1)$ global symmetry does
not change when the $U(1)$ symmetry is gauged (provided there are no monopoles), thanks to
particle-vortex duality~\cite{Peskin:1977kp,Dasgupta:1981zz}. Naively
one might take this example as part of a general pattern, and guess
that phase transitions in the $U(1)_{X}\times U(1)_{Y}$ gauge theory
necessarily imply phase transitions in $U(1)_{X}$ gauge theory
obtained via the ``un-gauging'' limit $e_{Y} \to 0$.

But it is not correct to presume, in general, that the phase structure
of a theory with a continuous global symmetry must be identical to
that of the gauged version of the theory. It is quite possible that as
the gauge coupling is sent to zero, a phase boundary appears between
two regimes which were smoothly connected in the gauged model.
Similarly, non-analyticities present in thermodynamic functions of the
gauged model may disappear when the gauge coupling is sent to zero.%
\footnote
    {%
    This subtlety does not arise in the case of gauged
    discrete symmetries studied in
    Refs.~\cite{Levin:2012yb,Wang:2014xba,Wang:2014wka}, because gauging
    discrete symmetries introduces neither local degrees of freedom nor
    continuous coupling constants. 
    }

It is easy to find examples illustrating the above scenarios. First,
consider again a compact $U(1)$ gauge theory with a single complex scalar
with charge $+1$ in three spacetime dimensions. Unlike the discussion above, suppose we specify a
UV completion of the theory that \emph{does} admit finite-action
monopole-instanton events with minimal magnetic flux. Then the Higgs
and confining regimes of the gauge theory are smoothly connected.
However, if we un-gauge the $U(1)$ symmetry, we are left with the XY
model in 3D, which has $U(1)$ global symmetry.   The $U(1)$ symmetry
broken and unbroken regimes of the XY theory are the limits of the
Higgs and confining regimes of the gauge theory.  But these regimes
are separated by a phase boundary in the  3D XY model.
    
It is also possible for phase boundaries of a gauge theory to disappear in  the ungauging limit.  To see an example of this
consider four free massless Dirac fermions in 4D spacetime. Such a
system has a global symmetry that includes $SU(4)_L \times SU(4)_R$.
If we gauge the vector-like $SU(2)_V$ subgroup of this symmetry,
introducing a gauge coupling $g$, we obtain two-color two-flavor
massless QCD. It has a $SU(2)_L \times SU(2)_R$ global symmetry, and
an $SU(2)_A$-breaking phase transition as a function of temperature.
The critical temperature has a non-perturbative dependence on $g$ due
to dimensional transmutation. The phase transition exists for any
non-zero value of the $SU(2)$ gauge coupling  $g$, but disappears at
$g=0$ where the theory becomes free. Another example is given by $N_c
N_f$ free massless Dirac fermions in 4D spacetime when one gauges an
$SU(N_c)$ subgroup of the vector-like global symmetry, yielding
massless QCD with $N_f$ massless quark flavors. In the large $N_c$
limit with $N_f/N_c$ and $g^2 N_c$ fixed, the model is known to go
through at least two quantum phase transitions as a function of
$N_f/N_c$ when $g^2 N_c$ is fixed a non-zero, see e.g.
Refs.~\cite{Caswell:1974gg,Banks:1981nn}. But there are no such phase
transitions at $g = 0$.

Similar concerns apply to the continuous Abelian gauge theories we
focused on in this paper. In the $\UX \times \UY$ gauge theory
\eqref{eq:the_model_XY}, one can infer the existence of phase
transitions from the behavior of gauge field holonomies whose values
become quantized on large distance scales. Holonomy quantization only
holds on distance scales large compared to \emph{all} relevant length
scales. If one is interested in the behavior of the system with
generic values of its physical parameters, then this is not a problem.
But the limit of vanishing gauge coupling, $e_Y \to 0$, is a highly
non-generic limit. In this limit, in Higgs phases of the theory, the
physical mass of a gauge boson goes to zero. In other words, the gauge
boson Compton wavelength diverges, and consequently the length scale
on which holonomy quantization holds also diverges. More formally,
there is non-uniformity between the limit $e_Y \to 0$ and large
distance limit implicit in defining the vortex holonomy. And this
means that one cannot simply infer the phase structure of the theory
exactly at $e_Y = 0$ from an analysis with $e_Y >0$. In particular,
the only way to demonstrate that non-analyticities inferred by
analysis of the gauged,
gapped theory survive to become corresponding observable
non-analyticities in the original gapless theory is to roll up one's
sleeves and examine the gapless theory of interest. This, of course,
was the main goal of this paper.

%\newpage
\bibliographystyle{utphys}
\bibliography{small_circle}

 \end{document}